\documentclass[%
reprint,
superscriptaddress,
 amsmath,amssymb,
 aps,
]{revtex4-2}
\usepackage{placeins}
\usepackage{graphicx}
\usepackage{dcolumn}
\usepackage{bm}
\usepackage{subfigure}
\usepackage{xcolor}
\usepackage{float}
\usepackage{lineno}
\usepackage{xspace}
\usepackage{cleveref}
\usepackage{soul}
\usepackage{multirow}
\usepackage[subrefformat=parens]{subcaption}
\usepackage[normalem]{ulem} 
\usepackage{nicefrac}
\usepackage{lineno}
\usepackage{booktabs}
\usepackage{siunitx}

\begin{document}

\preprint{}

\title{Measurement of muon neutrino induced charged current interactions without charged pions in the final state using a new T2K off-axis near detector WAGASCI-BabyMIND}

\providecommand{\wb}{WAGASCI-BabyMIND\xspace}
\providecommand{\ch}{CH\xspace}
\providecommand{\water}{H$_{2}$O\xspace}
\providecommand{\signal}{CC0$\pi^{\pm}$\xspace}
\providecommand{\sideband}{CC1$\pi^{\pm}$\xspace}

\providecommand{\nue}{$\nu_{e}$\xspace}
\providecommand{\nueb}{$\overline{\nu}_{e}$\xspace}
\providecommand{\numu}{$\nu_{\mu}$\xspace}
\providecommand{\numub}{$\overline{\nu}_{\mu}$\xspace}
\providecommand{\nutau}{$\nu_{\tau}$\xspace}
\providecommand{\nutaub}{$\overline{\nu}_{\tau}$\xspace}
\providecommand{\nubm}{$\overline{\nu}$-mode}
\providecommand{\nuebar}{$\overline{\nu}_{e}$\xspace}
\providecommand{\numubar}{$\overline{\nu}_{\mu}$\xspace}
\providecommand{\nubar}{$\overline{\nu}$\xspace}

\providecommand{\RefFig}[1]{Fig.~\ref{#1}}
\providecommand{\RefFigs}[2]{Figs.~\ref{#1}, \ref{#2}}
\providecommand{\RefManyFigs}[2]{Figs.~\ref{#1} to \ref{#2}}

\providecommand{\RefTab}[1]{Tab.~\ref{#1}}
\providecommand{\RefEq}[1]{Eq.~\ref{#1}}
\providecommand{\RefSec}[1]{Sec.~\ref{#1}}

\providecommand{\xsec}{cross section\xspace}
\providecommand{\Xsec}{Cross section\xspace}
\providecommand{\xsecs}{cross sections\xspace}
\providecommand{\Xsecs}{Cross sections\xspace}
\providecommand{\xsecunit}{$10^{-39} \mathrm{cm^{2}/nucleon}$\xspace}

\providecommand{\dwmfull}{WallMRD\xspace}
\providecommand{\dwmsfull}{WallMRDs\xspace}
\providecommand{\dbmfull}{BabyMIND detector\xspace}

\providecommand{\pmfull}{Proton Module detector\xspace}
\providecommand{\dpm}{PM\xspace}
\providecommand{\duwg}{WAGASCI UP\xspace}
\providecommand{\ddwg}{WAGASCI DOWN\xspace}
\providecommand{\dwgs}{WAGASCIs\xspace}
\providecommand{\dwg}{WAGASCI\xspace}
\providecommand{\dwm}{WM\xspace}
\providecommand{\dwms}{WMs\xspace}
\providecommand{\dbm}{BabyMIND\xspace}

\providecommand{\Pmu}{$p_{\mu}$\xspace}
\providecommand{\Tmu}{$\theta_{\mu}$\xspace}
\providecommand{\CTmu}{$\cos{\theta_{\mu}}$\xspace}

\providecommand{\masseV}{eV/$\mathrm{c}^{2}$\xspace}
\providecommand{\massMeV}{MeV/$\mathrm{c}^{2}$\xspace}
\providecommand{\massGeV}{GeV/$\mathrm{c}^{2}$\xspace}
\providecommand{\mmassMeV}{\mathrm{MeV}/\mathrm{c}^{2}}
\providecommand{\mmassGeV}{\mathrm{GeV}/\mathrm{c}^{2}}

\newcommand{\INSTHD}{\affiliation{University Autonoma Madrid, Department of Theoretical Physics, 28049 Madrid, Spain}}
\newcommand{\INSTFE}{\affiliation{Boston University, Department of Physics, Boston, Massachusetts, U.S.A.}}
\newcommand{\INSTD}{\affiliation{University of British Columbia, Department of Physics and Astronomy, Vancouver, British Columbia, Canada}}
\newcommand{\INSTGA}{\affiliation{University of California, Irvine, Department of Physics and Astronomy, Irvine, California, U.S.A.}}
\newcommand{\INSTI}{\affiliation{IRFU, CEA, Universit\'e Paris-Saclay, F-91191 Gif-sur-Yvette, France}}
\newcommand{\INSTGB}{\affiliation{University of Colorado at Boulder, Department of Physics, Boulder, Colorado, U.S.A.}}
\newcommand{\INSTFH}{\affiliation{Duke University, Department of Physics, Durham, North Carolina, U.S.A.}}
\newcommand{\INSTJA}{\affiliation{E\"{o}tv\"{o}s Lor\'{a}nd University, Department of Atomic Physics, Budapest, Hungary}}
\newcommand{\INSTEF}{\affiliation{ETH Zurich, Institute for Particle Physics and Astrophysics, Zurich, Switzerland}}
\newcommand{\INSTIG}{\affiliation{VNU University of Science, Vietnam National University, Hanoi, Vietnam}}
\newcommand{\INSTIE}{\affiliation{CERN European Organization for Nuclear Research, CH-1211 Gen\'eve 23, Switzerland}}
\newcommand{\INSTEG}{\affiliation{University of Geneva, Section de Physique, DPNC, Geneva, Switzerland}}
\newcommand{\INSTHJ}{\affiliation{University of Glasgow, School of Physics and Astronomy, Glasgow, United Kingdom}}
\newcommand{\INSTJG}{\affiliation{Ghent University, Department of Physics and Astronomy, Proeftuinstraat 86, B-9000 Gent, Belgium}}
\newcommand{\INSTDG}{\affiliation{H. Niewodniczanski Institute of Nuclear Physics PAN, Cracow, Poland}}
\newcommand{\INSTCB}{\affiliation{High Energy Accelerator Research Organization (KEK), Tsukuba, Ibaraki, Japan}}
\newcommand{\INSTIB}{\affiliation{University of Houston, Department of Physics, Houston, Texas, U.S.A.}}
\newcommand{\INSTED}{\affiliation{Institut de Fisica d'Altes Energies (IFAE) - The Barcelona Institute of Science and Technology, Campus UAB, Bellaterra (Barcelona) Spain}}
\newcommand{\INSTJC}{\affiliation{Institut f\"ur Physik, Johannes Gutenberg-Universit\"at Mainz, Staudingerweg 7, 55128 Mainz, Germany}}
\newcommand{\INSTHH}{\affiliation{Institute For Interdisciplinary Research in Science and Education (IFIRSE), ICISE, Quy Nhon, Vietnam}}
\newcommand{\INSTEI}{\affiliation{Imperial College London, Department of Physics, London, United Kingdom}}
\newcommand{\INSTGF}{\affiliation{INFN Sezione di Bari and Universit\`a e Politecnico di Bari, Dipartimento Interuniversitario di Fisica, Bari, Italy}}
\newcommand{\INSTBE}{\affiliation{INFN Sezione di Napoli and Universit\`a di Napoli, Dipartimento di Fisica, Napoli, Italy}}
\newcommand{\INSTBF}{\affiliation{INFN Sezione di Padova and Universit\`a di Padova, Dipartimento di Fisica, Padova, Italy}}
\newcommand{\INSTBD}{\affiliation{INFN Sezione di Roma and Universit\`a di Roma ``La Sapienza'', Roma, Italy}}
\newcommand{\INSTEB}{\affiliation{Institute for Nuclear Research of the Russian Academy of Sciences, Moscow, Russia}}
\newcommand{\INSTHI}{\affiliation{International Centre of Physics, Institute of Physics (IOP), Vietnam Academy of Science and Technology (VAST), 10 Dao Tan, Ba Dinh, Hanoi, Vietnam}}
\newcommand{\INSTJD}{\affiliation{ILANCE, CNRS – University of Tokyo International Research Laboratory, Kashiwa, Chiba 277-8582, Japan}}
\newcommand{\INSTHA}{\affiliation{Kavli Institute for the Physics and Mathematics of the Universe (WPI), The University of Tokyo Institutes for Advanced Study, University of Tokyo, Kashiwa, Chiba, Japan}}
\newcommand{\INSTID}{\affiliation{Keio University, Department of Physics, Kanagawa, Japan}}
\newcommand{\INSTIF}{\affiliation{King's College London, Department of Physics, Strand, London WC2R 2LS, United Kingdom}}
\newcommand{\INSTCC}{\affiliation{Kobe University, Kobe, Japan}}
\newcommand{\INSTCD}{\affiliation{Kyoto University, Department of Physics, Kyoto, Japan}}
\newcommand{\INSTEJ}{\affiliation{Lancaster University, Physics Department, Lancaster, United Kingdom}}
\newcommand{\INSTII}{\affiliation{Lawrence Berkeley National Laboratory, Berkeley, California, U.S.A.}}
\newcommand{\INSTBA}{\affiliation{Ecole Polytechnique, IN2P3-CNRS, Laboratoire Leprince-Ringuet, Palaiseau, France}}
\newcommand{\INSTFC}{\affiliation{University of Liverpool, Department of Physics, Liverpool, United Kingdom}}
\newcommand{\INSTFI}{\affiliation{Louisiana State University, Department of Physics and Astronomy, Baton Rouge, Louisiana, U.S.A.}}
\newcommand{\INSTIH}{\affiliation{Joint Institute for Nuclear Research, Dubna, Moscow Region, Russia}}
\newcommand{\INSTHB}{\affiliation{Michigan State University, Department of Physics and Astronomy,  East Lansing, Michigan, U.S.A.}}
\newcommand{\INSTCE}{\affiliation{Miyagi University of Education, Department of Physics, Sendai, Japan}}
\newcommand{\INSTDF}{\affiliation{National Centre for Nuclear Research, Warsaw, Poland}}
\newcommand{\INSTFJ}{\affiliation{State University of New York at Stony Brook, Department of Physics and Astronomy, Stony Brook, New York, U.S.A.}}
\newcommand{\INSTEH}{\affiliation{STFC, Rutherford Appleton Laboratory, Harwell Oxford,  and  Daresbury Laboratory, Warrington, United Kingdom}}
\newcommand{\INSTGJ}{\affiliation{Okayama University, Department of Physics, Okayama, Japan}}
\newcommand{\INSTCF}{\affiliation{Osaka Metropolitan University, Department of Physics, Osaka, Japan}}
\newcommand{\INSTGG}{\affiliation{Oxford University, Department of Physics, Oxford, United Kingdom}}
\newcommand{\INSTIC}{\affiliation{University of Pennsylvania, Department of Physics and Astronomy,  Philadelphia, Pennsylvania, U.S.A.}}
\newcommand{\INSTGC}{\affiliation{University of Pittsburgh, Department of Physics and Astronomy, Pittsburgh, Pennsylvania, U.S.A.}}
\newcommand{\INSTGD}{\affiliation{University of Rochester, Department of Physics and Astronomy, Rochester, New York, U.S.A.}}
\newcommand{\INSTHC}{\affiliation{Royal Holloway University of London, Department of Physics, Egham, Surrey, United Kingdom}}
\newcommand{\INSTBC}{\affiliation{RWTH Aachen University, III. Physikalisches Institut, Aachen, Germany}}
\newcommand{\INSTJF}{\affiliation{School of Physics and Astronomy, University of Minnesota, Minneapolis, Minnesota, U.S.A.}}
\newcommand{\INSTFB}{\affiliation{University of Sheffield, School of Mathematical and Physical Sciences, Sheffield, United Kingdom}}
\newcommand{\INSTDI}{\affiliation{University of Silesia, Institute of Physics, Katowice, Poland}}
\newcommand{\INSTIA}{\affiliation{SLAC National Accelerator Laboratory, Stanford University, Menlo Park, California, U.S.A.}}
\newcommand{\INSTBB}{\affiliation{Sorbonne Universit\'e, CNRS/IN2P3, Laboratoire de Physique Nucl\'eaire et de Hautes Energies (LPNHE), Paris, France}}
\newcommand{\INSTJE}{\affiliation{South Dakota School of Mines and Technology, 501 East Saint Joseph Street, Rapid City, SD 57701, United States}}
\newcommand{\INSTCH}{\affiliation{University of Tokyo, Department of Physics, Tokyo, Japan}}
\newcommand{\INSTBJ}{\affiliation{University of Tokyo, Institute for Cosmic Ray Research, Kamioka Observatory, Kamioka, Japan}}
\newcommand{\INSTCG}{\affiliation{University of Tokyo, Institute for Cosmic Ray Research, Research Center for Cosmic Neutrinos, Kashiwa, Japan}}
\newcommand{\INSTHF}{\affiliation{Institute of Science Tokyo, Department of Physics, Tokyo}}
\newcommand{\INSTGI}{\affiliation{Tokyo Metropolitan University, Department of Physics, Tokyo, Japan}}
\newcommand{\INSTHG}{\affiliation{Tokyo University of Science, Faculty of Science and Technology, Department of Physics, Noda, Chiba, Japan}}
\newcommand{\INSTB}{\affiliation{TRIUMF, Vancouver, British Columbia, Canada}}
\newcommand{\INSTJH}{\affiliation{University of Toyama, Department of Physics, Toyama, Japan}}
\newcommand{\INSTDJ}{\affiliation{University of Warsaw, Faculty of Physics, Warsaw, Poland}}
\newcommand{\INSTDH}{\affiliation{Warsaw University of Technology, Institute of Radioelectronics and Multimedia Technology, Warsaw, Poland}}
\newcommand{\INSTIJ}{\affiliation{Tohoku University, Faculty of Science, Department of Physics, Miyagi, Japan}}
\newcommand{\INSTFD}{\affiliation{University of Warwick, Department of Physics, Coventry, United Kingdom}}
\newcommand{\INSTEA}{\affiliation{Wroclaw University, Faculty of Physics and Astronomy, Wroclaw, Poland}}
\newcommand{\INSTHE}{\affiliation{Yokohama National University, Department of Physics, Yokohama, Japan}}
\newcommand{\INSTH}{\affiliation{York University, Department of Physics and Astronomy, Toronto, Ontario, Canada}}

\INSTHD
\INSTFE
\INSTD
\INSTGA
\INSTI
\INSTGB
\INSTFH
\INSTJA
\INSTEF
\INSTIG
\INSTIE
\INSTEG
\INSTHJ
\INSTJG
\INSTDG
\INSTCB
\INSTIB
\INSTED
\INSTJC
\INSTHH
\INSTEI
\INSTGF
\INSTBE
\INSTBF
\INSTBD
\INSTEB
\INSTHI
\INSTJD
\INSTHA
\INSTID
\INSTIF
\INSTCC
\INSTCD
\INSTEJ
\INSTII
\INSTBA
\INSTFC
\INSTFI
\INSTIH
\INSTHB
\INSTCE
\INSTDF
\INSTFJ
\INSTEH
\INSTGJ
\INSTCF
\INSTGG
\INSTIC
\INSTGC
\INSTGD
\INSTHC
\INSTBC
\INSTJF
\INSTFB
\INSTDI
\INSTIA
\INSTBB
\INSTJE
\INSTCH
\INSTBJ
\INSTCG
\INSTHF
\INSTGI
\INSTHG
\INSTB
\INSTJH
\INSTDJ
\INSTDH
\INSTIJ
\INSTFD
\INSTEA
\INSTHE
\INSTH

\author{K.\,Abe}\INSTBJ
\author{S.\,Abe}\INSTCH
\author{R.\,Akutsu}\INSTCB
\author{H.\,Alarakia-Charles}\INSTEJ
\author{Y.I.\,Alj Hakim}\INSTFB
\author{S.\,Alonso Monsalve}\INSTEF
\author{L.\,Anthony}\INSTEI
\author{S.\,Aoki}\INSTCC
\author{K.A.\,Apte}\INSTEI
\author{T.\,Arai}\INSTCH
% ...

\collaboration{The T2K Collaboration}\noaffiliation


\date{\today \\ veVersion 1.0}

\begin{abstract} We report a flux-integrated \xsec  measurement of muon neutrino interactions on water and hydrocarbon via charged current reactions without charged pions in the final state with the \wb  detector which was installed in the T2K near detector hall in 2018. The detector is located 1.5$^\circ$ off-axis and is exposed to a more energetic neutrino flux than ND280, another T2K near detector, which is located at a different off-axis position. The total flux-integrated cross section is measured to be  $1.26 \pm 0.18\,(stat.+syst.) \times 10^{-39} $ $\mathrm{cm^{2}/nucleon}$ on \ch and $1.44 \pm 0.21\,(stat.+syst.) \times 10^{-39} $ $\mathrm{cm^{2}/nucleon}$ on \water. These results are compared to model predictions provided by the NEUT v5.3.2 and GENIE v2.8.0 MC generators and the measurements are compatible with these models. Differential cross sections in muon momentum and cosine of the muon scattering angle are also reported. This is the first such measurement reported with the WAGASCI-BabyMIND detector and utilizes the 2020 and 2021 datasets. 

\end{abstract}

\maketitle

\section{\label{Sect:Intro}INTRODUCTION}

\vskip 0.5cm

In order to improve our understanding of the phenomenon of neutrino oscillations, a new generation of experiments is required with increased detector mass, higher-power beam and better near detectors yielding improved statistics and  sensitivity. Within the paradigm of three Dirac neutrinos with their flavor states being linear combinations of their mass eigenstates as determined by the PMNS mixing matrix~\cite{PMNS}, the remaining major unknowns are: the ordering of the mass eigenstates and the value of the CP symmetry violating parameter. Together with the hypothetical existence of sterile neutrinos these are the main questions to be addressed by several long- and short-baseline neutrino oscillation experiments. The precision of measurements performed in experiments like T2K~\cite{T2K:2011qtm}, NOvA~\cite{Nova}, Hyper-Kamiokande (HK)~\cite{HyperKamiokande} and DUNE~\cite{DUNE} relies strongly on reducing the systematic uncertainties originating from our limited knowledge of neutrino-nucleus cross sections~\cite{Avanzini:2021qlx}. 
T2K uses a neutrino beam peaked in the sub-GeV energy range hence the main neutrino interaction channel is Charged Current Quasi-Elastic (CCQE) scattering with a nucleon. This interaction mode is described with the standard Llewellyn Smith formalism~\cite{Llewellyn-Smith} but it is far from being fully understood due to the axial structure of the weak current and complications introduced through nuclear effects, as neutrino oscillation experiments use complex nuclear targets such as carbon, water, or argon. Other interaction modes bring their own unknowns and this motivates experimental efforts to measure neutrino \xsecs with different fluxes, targets and definitions of the experimental signal. 

The T2K experiment uses Super-Kamiokande, a water Cherenkov detector~\cite{SuperKamiokande}, as the far detector to measure neutrinos after oscillation and ND280~\cite{Kudenko:2008ia} as the near detector, measuring the neutrino flux before oscillation. The primary target material in ND280 is hydrocarbon, therefore it is highly desirable to also measure neutrino interactions on a water target at the near detector site. 
\wb with its mixed water and hydrocarbon composition allows the simultaneous measurement of the neutrino cross section on both of these target materials, motivating its construction and inclusion within the T2K experiment.

\begin{figure}[htbp]
\begin{center}
\includegraphics[width=.5\textwidth]{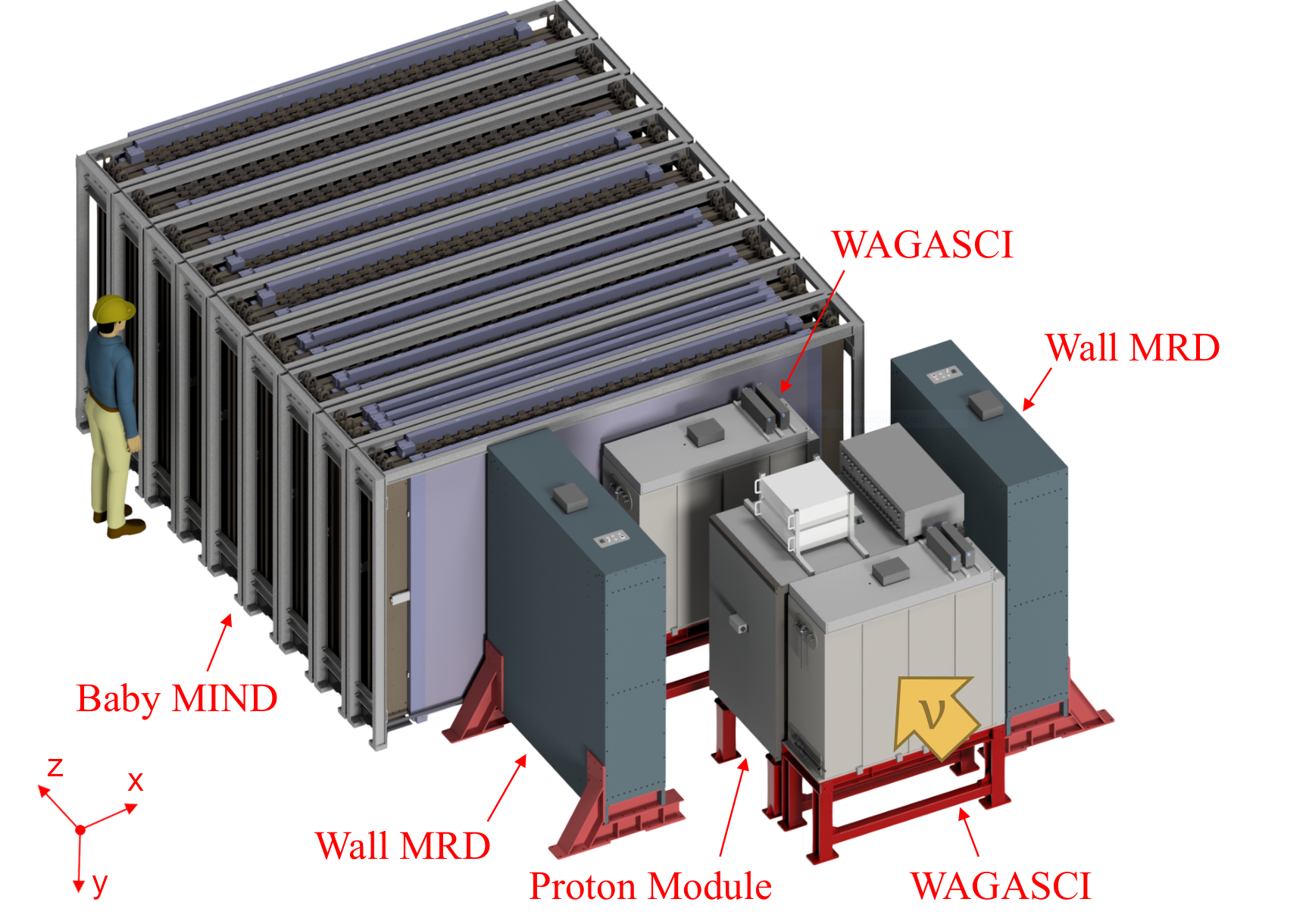}
\end{center}
\caption{Configuration of the \wb. A large orange arrow represents the neutrino beam direction.}
\label{fig:wagasci_detectors}
\end{figure}

The \wb is the collective name of the detectors shown in \RefFig{fig:wagasci_detectors}. It was installed in 2018 at the main ND280~\cite{T2K:2023qjb} complex in a  position  1.5 degrees off-axis on the B2 floor of the Neutrino Monitoring (NM) building of the  Japan-Proton Accelerator Research Complex (J-PARC). Together, the WAter Grid And SCIntillator modules (\dwgs), Proton Module (\dpm), Wall Muon Range Detectors (\dwms) and Baby Magnetized Iron Neutrino Detector (\dbm) form an integrated system of detectors. In this paper, we present new \xsec measurements made by the \wb on both \water and \ch targets.

\begin{figure}[htbp]
\begin{center}
 \includegraphics[width=.5\textwidth]{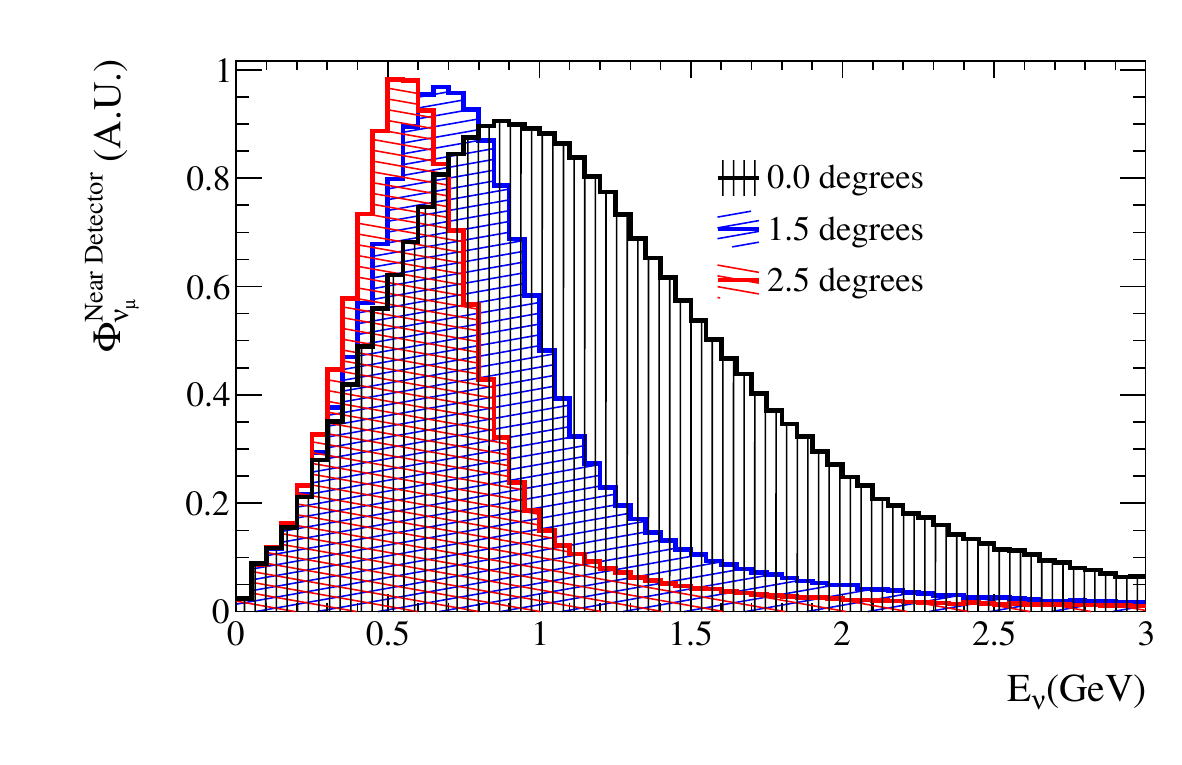}
\end{center}
 \caption{Neutrino flux produced by the J-PARC accelerator at different off-axis angles. The \wb is located 1.5$^\circ$ off-axis, where the neutrino flux is peaked at $0.7$~GeV.}
\label{fig:flux}
\end{figure}

For the neutrinos provided by the J-PARC accelerator, at the \wb position the energy spectrum has a peak around 0.7~GeV, compared to 0.6~GeV at the 2.5 degrees off-axis position where the ND280 detector is located, as shown in \RefFig{fig:flux}. 

At these higher neutrino energies interactions other than CCQE, such as pion production and multinucleon knock-out, become more significant compared to lower neutrino energy events seen by ND280. They can contribute to the experimental signal measured in this paper defined as a muon and no charged pions in the final state. We work with final states as a charged pion may be produced and then absorbed, or it may be below tracking threshold, resulting in no reconstructed track. 

We measured both flux-integrated cross sections and also the differential cross section in the resultant lepton's momentum and the cosine of it's scattering angle. 

The current statistics available for physics analysis are $2.96 \times 10^{20}$ Protons on Target (POT) in the neutrino mode accumulated since 2018. This is compared to the ND280 statistics of $9.78 \times 10^{20}$ accumulated since 2010. With this dataset it was possible to extract the one-dimensional differential cross sections. 

The results presented in this paper follow previous \xsec measurements  on a standalone \dwg module~\cite{10.1093/ptep/ptab014}. Here we report the first measurement with the full \wb set up. The results reported in the paper are the first obtained using a new near detector in T2K with a water component and measures the Carbon/Oxygen cross section ratio at a higher energy than ND280.
 
 Our paper is organized as follows. 
 In \RefSec{Sect:Method} the T2K experimental setup is described including a detailed description of the \wb in Subsect.~\ref{wb}.  The simulation and selection of events is explained in \RefSec{Sect:sampledef}.  In \RefSec{Sect:xsec_extraction}, the methods and calculation formulae are described.  The sources of systematic uncertainty are reported in \RefSec{Sect:systematics}.  We validated our statistical approach and selected samples with simulated data studies and the results of the studies are briefly summarized in \RefSec{Sect:fds}.  In \RefSec{Sect:results}, we present the integrated and two differential cross section results on \water and \ch targets.  The paper is concluded in \RefSec{Sect:Conclusions}.
 \vskip 0.5cm

\section{\label{Sect:Method}The T2K experiment}

The T2K experiment has achieved several major milestones, including the first observation of electron neutrino appearance in a muon neutrino beam~\cite{T2K:2011ypd} and precision measurements of $\theta_{23}$—a key parameter in the PMNS mixing matrix~\cite{T2K:2014ghj}. Moreover, T2K recently demonstrated sensitivity to measuring $\delta_{CP}$, a parameter of the PMNS matrix responsible for the violation of the CP symmetry in neutrino interactions~\cite{Abe2020-Nature, PhysRevD.91.072010}. These observations are crucial for understanding the fundamental properties of neutrinos and their role in the universe~\cite{Canetti_2012}.

\subsection{\label{Sect:Data}The muon neutrino beam}

The T2K experiment utilizes a muon neutrino beam produced at the J-PARC facility. Protons are accelerated to 30 GeV in stages by three accelerators~\cite{10.1093/ptep/pts020}. These high-energy protons collide with a graphite target producing secondary particles, predominantly pions. The pions are focused in the forward direction by three magnetic horns, which also defocus particles with the wrong charge, thereby enhancing the purity of the neutrino beam.  As these pions travel through a 96-meter-long decay pipe, they decay predominantly producing muon neutrinos or muon anti-neutrinos in either Forward Horn Current or Reverse Horn Current mode. 

\subsection{\label{wb}The \wb}

Near detectors, such as the \wb~\cite{10.1093/ptep/ptab014} shown in \RefFig{fig:wagasci_detectors}
and ND280~\cite{T2K:2023qjb}, measure the unoscillated neutrino flux. This allows for detailed 
measurements of neutrino interactions and cross sections before oscillation, helping to constrain the 
oscillation measurements at the far detector and deepening our understanding of neutrino interactions 
in the energy range of interest. 

In the remainder of this paper, the \dwgs and \dpm detectors are occasionally referred to as `vertex' detectors, as the neutrino interactions in these detectors were considered signals for \xsec extraction. Meanwhile, the \dbm and \dwm detectors are referred to as `muon range detectors' (MRDs), as their primary function is to track muons resulting from neutrino interactions. The \dbm and \dwm detectors were never used as target detectors in this paper.

All sub-detectors of the \wb use the same principle to track the particles from neutrino interactions. They are composed of scintillator bars of various lengths and shapes inside which are inserted wavelength-shifting fibers that gather and transport scintillation light to a Multi-Pixel Photon Counter (MPPC).

The \dwgs serve as the primary water target for neutrinos and consists of a three-dimensional grid of scintillator immersed in water, allowing for high signal purity. The internal structure and dimensions are explained in detail in Ref.\,~\cite{10.1093/ptep/ptab014}.

The \dpm is a scintillator tracking module which complements the \dwgs by providing a purely hydrocarbon target. It is used for estimating background events that originate in the WAGASCI scintillators.

The fiducial mass of the \dpm is 313~kg of \ch and that of the \dwgs is 229~kg of \water and 62~kg of \ch in total. By combining events from both the \dwgs and the \dpm in the cross section fit, we can decorrelate the cross sections on water and hydro-carbon. A similar strategy was employed by the ND280 detector as described in Ref.~\cite{PhysRevD.101.112004}. The geometry of the \dpm is briefly discussed in Ref.~\cite{T2K:2011qtm}, while the performance of the scintillator bars is described in Ref.~\cite{NITTA2004147}.

\begin{figure}[htbp]
\begin{center}
\includegraphics[width=.5\textwidth]{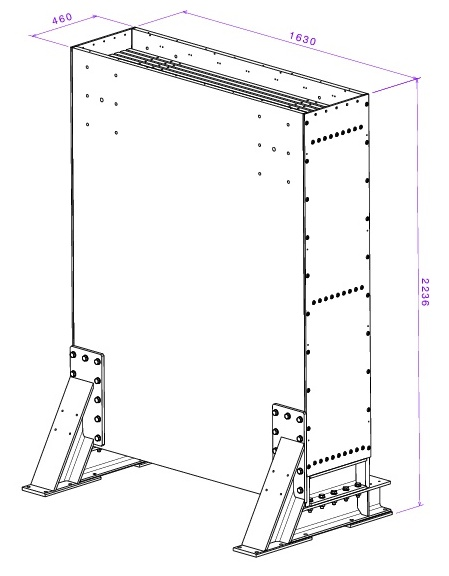}
\end{center}
\caption{External dimensions of one of the \dwm detectors with a size in mm. It consists of 11 layers of iron plate (each 1.8~meters high, 1.6~meters long and 3~cm thick) and 10 layers of scintillator. Each layer of scintillator is made up of 8 plastic scintillator plates, each 1.8 meters high, 0.2 meters long and 7 mm thick.}
\label{fig:wallmrd_structure}
\end{figure}

\begin{figure}[htbp]
\begin{center}
\includegraphics[width=.5\textwidth]{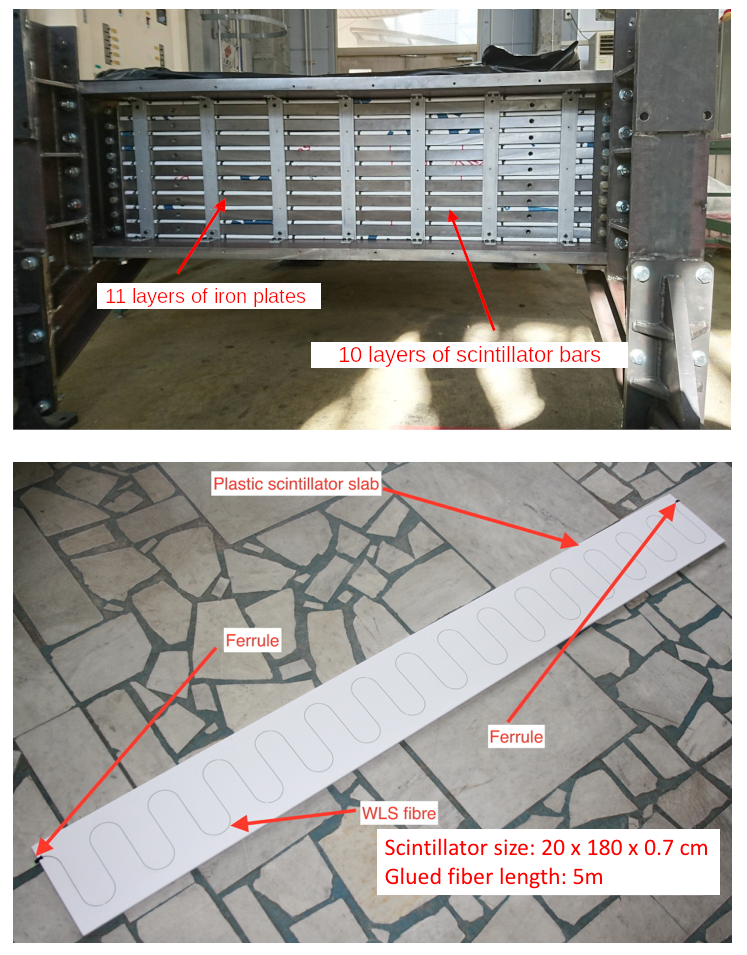}
\end{center}
\caption{Internal structure of one of the \dwm detectors. The structure and size of both detectors is identical. Top view: view of the \dwm from the bottom with the lid opened and before the readout electronics were installed. Bottom view: a single scintillator bar where the S-shape pattern of the wavelength-shifting fiber is clearly visible.}
\label{fig:wallmrd_scintillators}
\end{figure}

A preliminary \xsec measurement~\cite{10.1093/ptep/ptab014} was performed between 2017 and 2018 using only one \dwg module and the \dpm as target detectors and an INGRID module~\cite{ABE2012211} as a downstream muon range detector. In the measurement presented in this paper, we have improved the detectable phase space of the downrange muons by replacing the INGRID module with two new detectors:
\begin{enumerate}
    \item One \dbmfull, installed downstream of the \dwgs and made of 33 magnetized iron planes and 18 scintillator planes. It provides the capability to identify the charge of muons by the curvature of their tracks and precisely measure their momentum by measuring their penetration distance. A more in-depth description of the structure and performance can be found in~\cite{BabyMIND:2017mys, BabyMIND:2020lxx}. The \dbm and \dwgs are not aligned along the flux axis because of the narrow available space in the B2 pit.
    \item Two \dwm detectors were installed on either side of the \dwgs. They consist of ten scintillator tracking planes interleaved with iron planes and can detect muon tracks at high-angle relative to the beam direction, enhancing the detection efficiency for such events and increasing the observable phase-space of the neutrino interaction measurements. The mechanical structure is illustrated in \RefFig{fig:wallmrd_structure} and \RefFig{fig:wallmrd_scintillators}. Since these scintillators have a large width relative to their thickness, wavelength-shifting fibers (Kuraray Y-11) with a diameter of 1.0 mm were embedded in a wavy pattern (S-shape), as shown in \RefFig{fig:wallmrd_scintillators}; otherwise, the light yield from a hit would be decreased due to the scintillator attenuation. As a result, the wavelength-shifting fibers become longer and both sides were read out to maximize the light yield and provide timing information by comparing the relative readout time. The skewed position of the \dwms, seen in \RefFig{fig:wagasci_detectors}, was chosen to maximize the angular coverage.
\end{enumerate}
Thus, for the first time, a measurement was done on the complete and final set of detectors. In this analysis, these three MRDs were not used as target detectors.

\section{\label{Sect:sampledef}Signal definitions and selection strategy}

\subsection{\label{SubSect:signal and sideband definitions}Signal and sideband definitions}

The signal considered in this analysis was a muon neutrino charged current interaction on \ch or \water targets without charged pions in the final state. Unlike other T2K CC$0\pi$ cross sections, events with a neutral pion in the final state were accepted within the \signal sample since \wb is not capable of identifying these kind of shower events.

Charged pions, however, can be identified in the detector (see Sec. \ref{subsect:selection_criteria} for details). The \water and \ch mass ratio in \dwg is 4:1 hence samples of events with a vertex inside a \dwg module include both \water and \ch interactions, whereas samples of events with a vertex inside the \dpm include only \ch interactions. We performed a simultaneous fit of \water and \ch when calculating the measured cross sections to decouple these contributions. The fit results reflect the difference between the cross sections of CH and $\mathrm{H}_{2}$O, which was one of the goals of this analysis.

The dominant background to the \signal signal comes from \sideband events when a charged  pion was misidentified as a muon or was not reconstructed. We defined the sideband as \sideband to give data-based constraints on this interaction process in the cross section fitting (see \RefSec{Sect:xsec_extraction} for details). Furthermore, there may be Out-Of-Fiducial-Volume (OOFV) background events whose secondary particles penetrate the fiducial volume and contaminate our signal events. The main components of this background were cosmic muons and the interaction of beam neutrinos with the concrete walls of the experimental hall, which can produce secondary particles that contaminate signal events. A ``sand muon'' sample is mainly produced in the sand or concrete wall surrounding the detector hall and can be selected as single track events (usually a muon) originating outside the fiducial volume, going through the first layer of \dwg and reaching one of the downstream muon detectors. They were usually discarded as background, but can be used as a tool for reconstruction studies, especially in cases like ours where the statistics of the signal sample were relatively low.

\subsection{\label{subsect:MC_simulation}Monte-Carlo simulation}

The selection criteria were studied using Monte Carlo (MC) simulations. A MC framework was developed to generate simulated datasets for this analysis with the approach consistent with that adopted in the previous analysis~\cite{10.1093/ptep/ptab014}. Simulated datasets for each interaction target were produced and included the initial neutrino flux properties, particle information, hit/track information, etc. The T2K neutrino flux at 1.5 degrees off-axis was simulated using the T2K  package  JNUBEAM (version 2021.2)~\cite{PhysRevD.87.012001}. The output of the simulated flux information was fed into the T2K neutrino interaction generator, NEUT (version 5.3.2)~\cite{Hayato:2002sd}. 
Using the kinematic variables of the generated final state particles of the simulated neutrino-nucleus interaction as input, the detector response was then simulated by Geant4 (version 4.10.6.2)~\cite{AGOSTINELLI2003250} with the \lq QGSP BERT\rq\ model which also simulates the subsequent secondary or tertiary interactions within the detector material. 

\subsection{\label{subsect:Reconstruction}Reconstruction}
The reconstruction techniques were based on those implemented for the previous analysis~\cite{10.1093/ptep/ptab014} but extended so that the methods could be applied to all the detectors including the magnetized detector, \dbm. The modifications were done by tuning the reconstruction parameters on a detector-by-detector basis. The track reconstruction process was composed of three steps: track seeding, track matching and reconstruction of track properties. The track seeding tries to find track candidates using hit information. After track candidates were found in each detector, tracks were connected between multiple detectors. Once a matched track was reconstructed, the algorithm reconstructs a vertex by identifying the most upstream hit of the matched track. Finally, we extracted track parameters such as momentum, angle, charge and particle type based on the reconstructed tracks.

\subsection{\label{subsect:signal_purity_efficiency} Signal purity and efficiency}
We set cut values as part of the event selection to maximize the figure of merit defined by $N_{\mathrm{sig}} / \sqrt{N_{\mathrm{sig}} + N_{\mathrm{bg}}}$, where $N_{\mathrm{sig(bg)}}$ was the number of signal (background) events. As the figure of merit depends on the signal purity and efficiency, increasing purity whilst maintaining the signal efficiency as much as possible was the key to optimizing the cut values. Signal purity ($\rho$) and efficiency ($\epsilon$) are defined by 

\begin{equation}
\rho = \frac{N_{\mathrm{sig, allsel}}}{N_{\mathrm{allsel}}}, \epsilon = \frac{N_{\mathrm{sig, allsel}}}{N_{\mathrm{sig}}},
\end{equation}
where $N_{\mathrm{sig, allsel}}$ means the number of signal events satisfying all selection criteria and $N_{\mathrm{allsel}}$ is the total number of events passing all the selection criteria.

\subsection{\label{subsect:PS_restriction}Phase-space restrictions}
Phase-space restrictions on the cosine of the muon angle ($\cos\theta_{\mu}$) and the muon momentum ($p_{\mu}$) were imposed in this analysis to maintain high signal efficiency in the selected sample. These restrictions were:  $\cos\theta_{\mu}>0.34$ (i.e. $\theta_\mu < 70^o$) and $p_{\mu}>300\ \mathrm{MeV/c}$.

\subsection{\label{subsect:selection_criteria}Selection criteria}
\begin{figure*}[htbp]
\begin{center}
\includegraphics[width=1.0\textwidth]{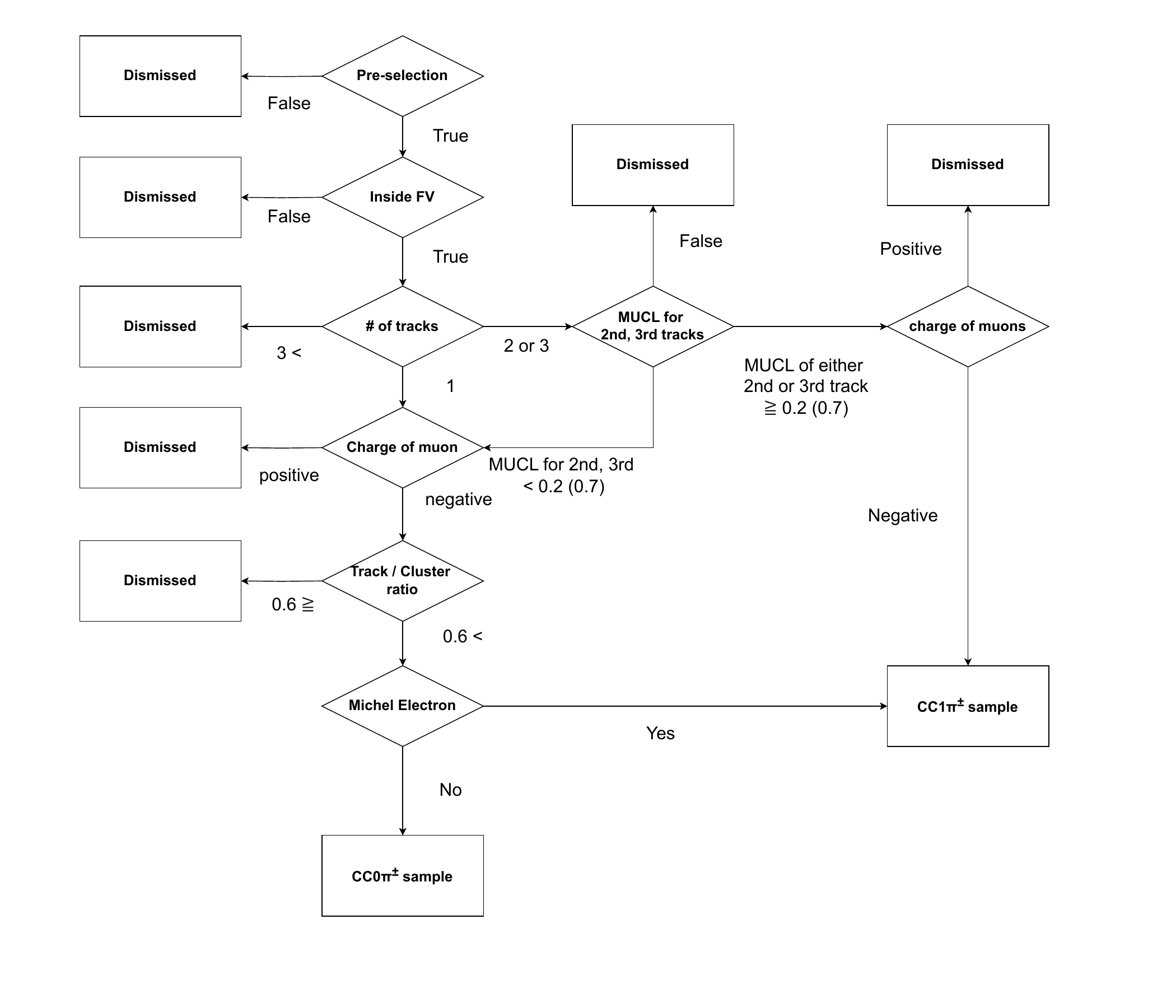}
\end{center}
\caption{Overview of the selection criteria for the \signal and \sideband samples. }
\label{fig:cutflow}
\end{figure*}

The flow of cuts used in this analysis is shown in \RefFig{fig:cutflow}. In this analysis, we used only events where a muon candidate produced in a vertex detector was matched with either \dbm or one of the \dwms. Unmatched events were rejected with these criteria defining the `Pre-selection'. The subsequent cut criteria are described as follows.

\begin{figure*}[htbp]

 \begin{tabular}{c}
  \begin{minipage}{0.5\linewidth}
    \centering
    \includegraphics[width=1.0\textwidth]{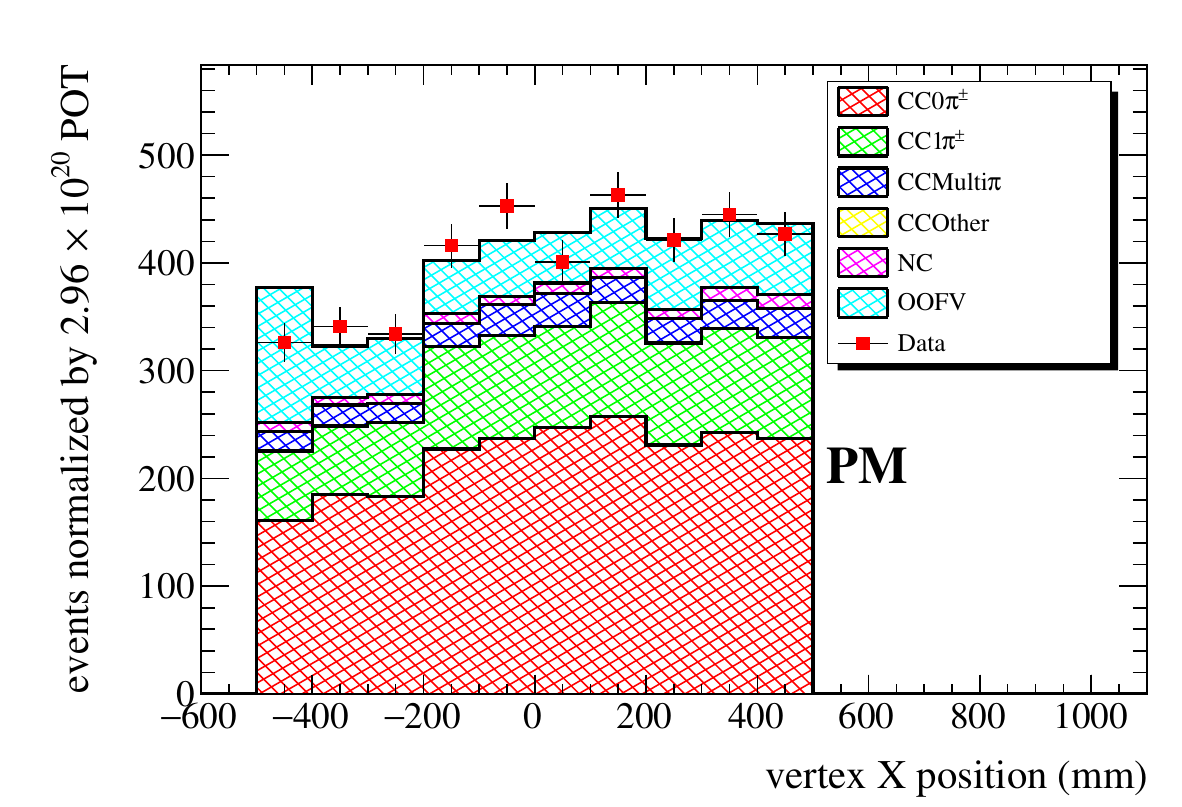} 
  \end{minipage}
  \begin{minipage}{0.5\linewidth}
    \centering
    \includegraphics[width=1.0\textwidth]
    {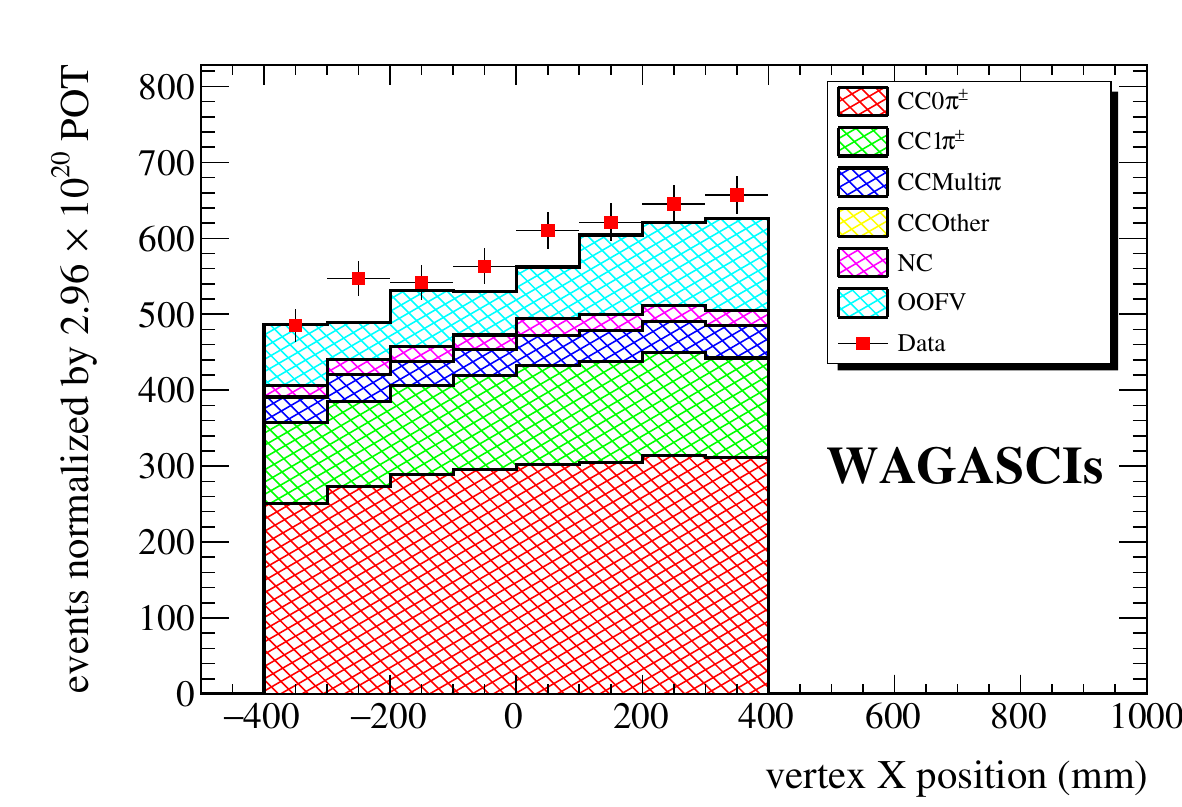} 
  \end{minipage}

\\

  \begin{minipage}{0.5\linewidth}
    \centering
    \includegraphics[width=1.0\textwidth]
    {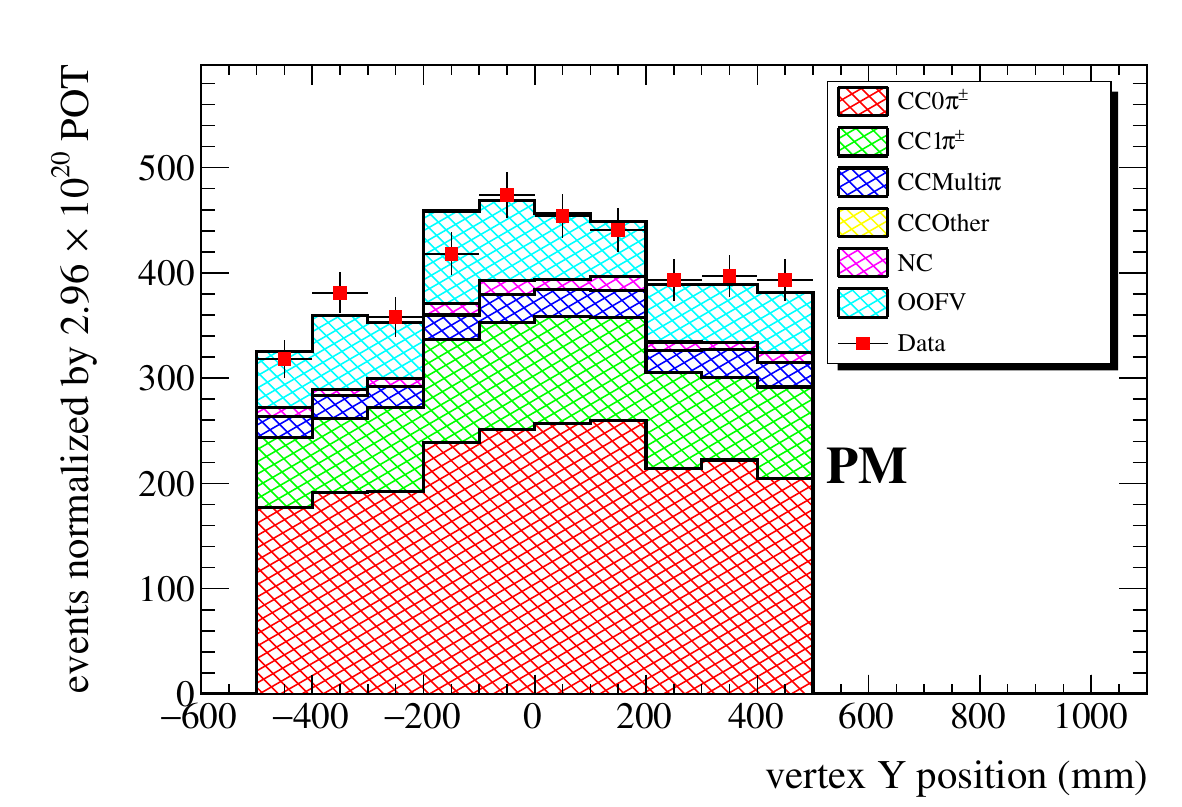} 
  \end{minipage}
  \begin{minipage}{0.5\linewidth}
    \centering
    \includegraphics[width=1.0\textwidth]
    {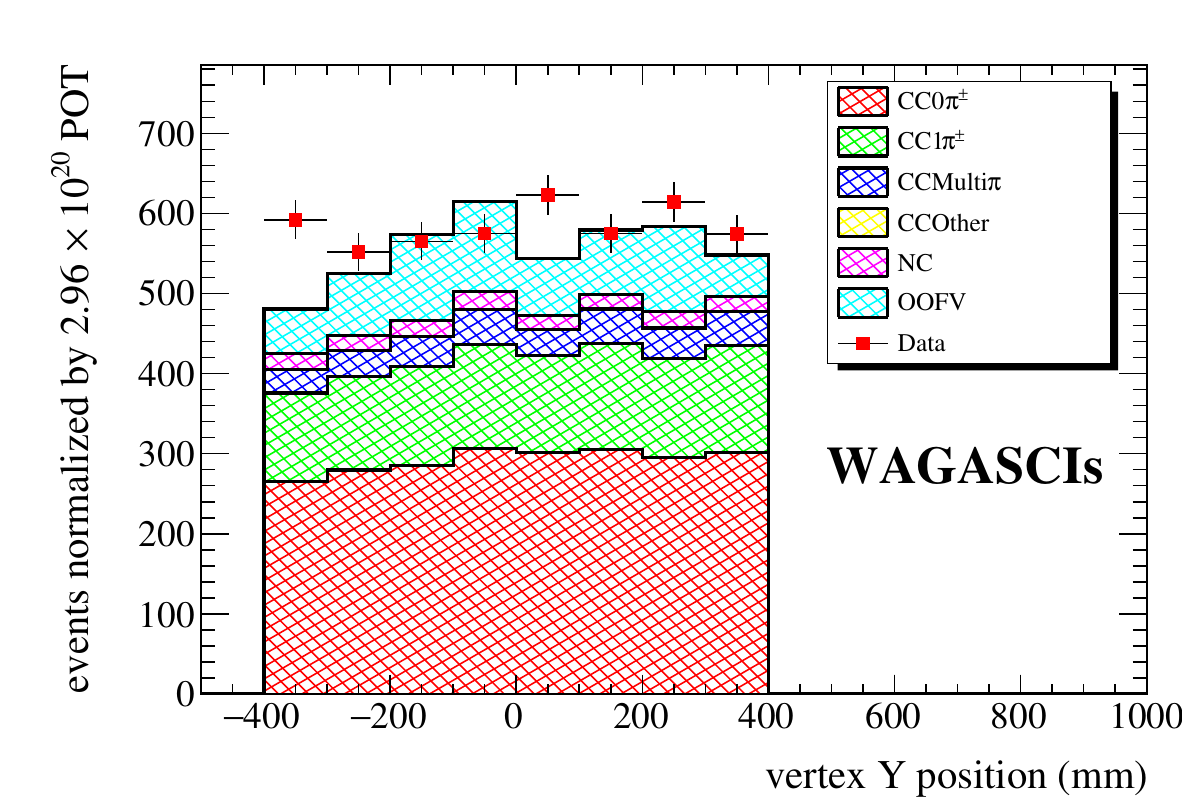} 
  \end{minipage}

\\

  \begin{minipage}{0.5\linewidth}
    \centering
    \includegraphics[width=1.0\textwidth]
    {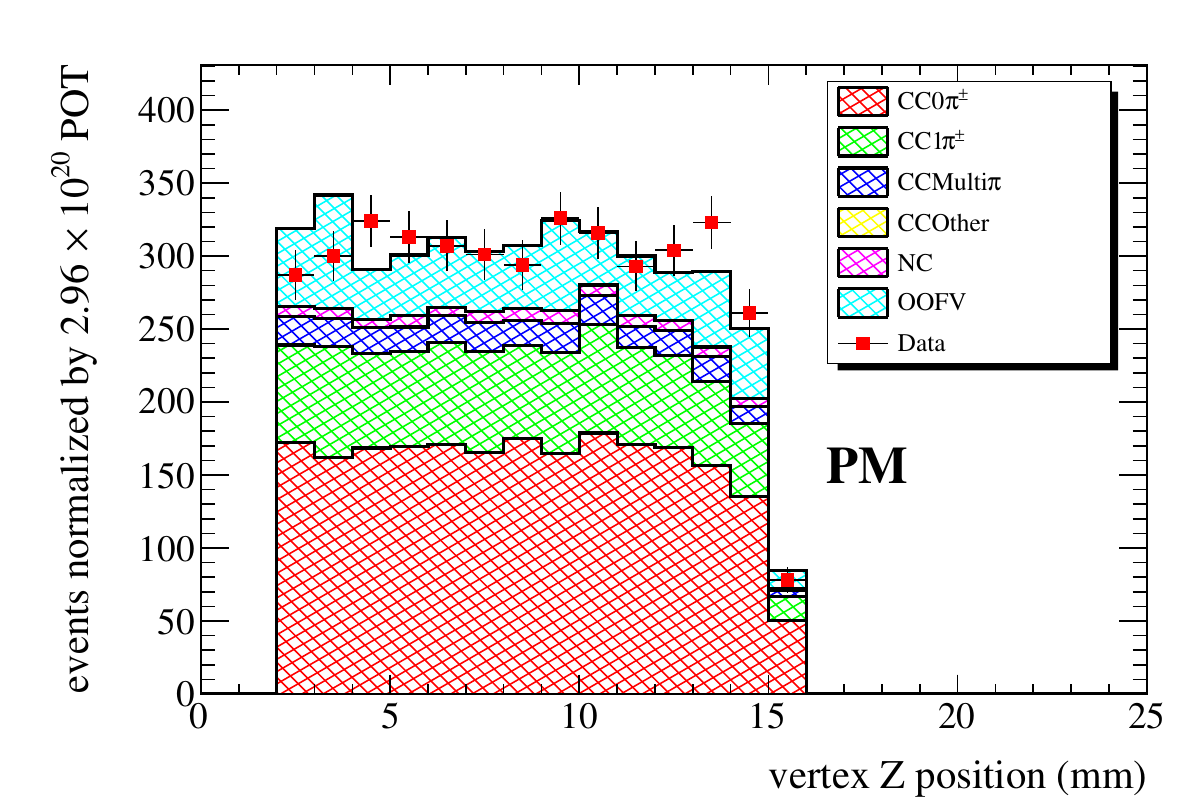} 
  \end{minipage}
  \begin{minipage}{0.5\linewidth}
    \centering
    \includegraphics[width=1.0\textwidth]{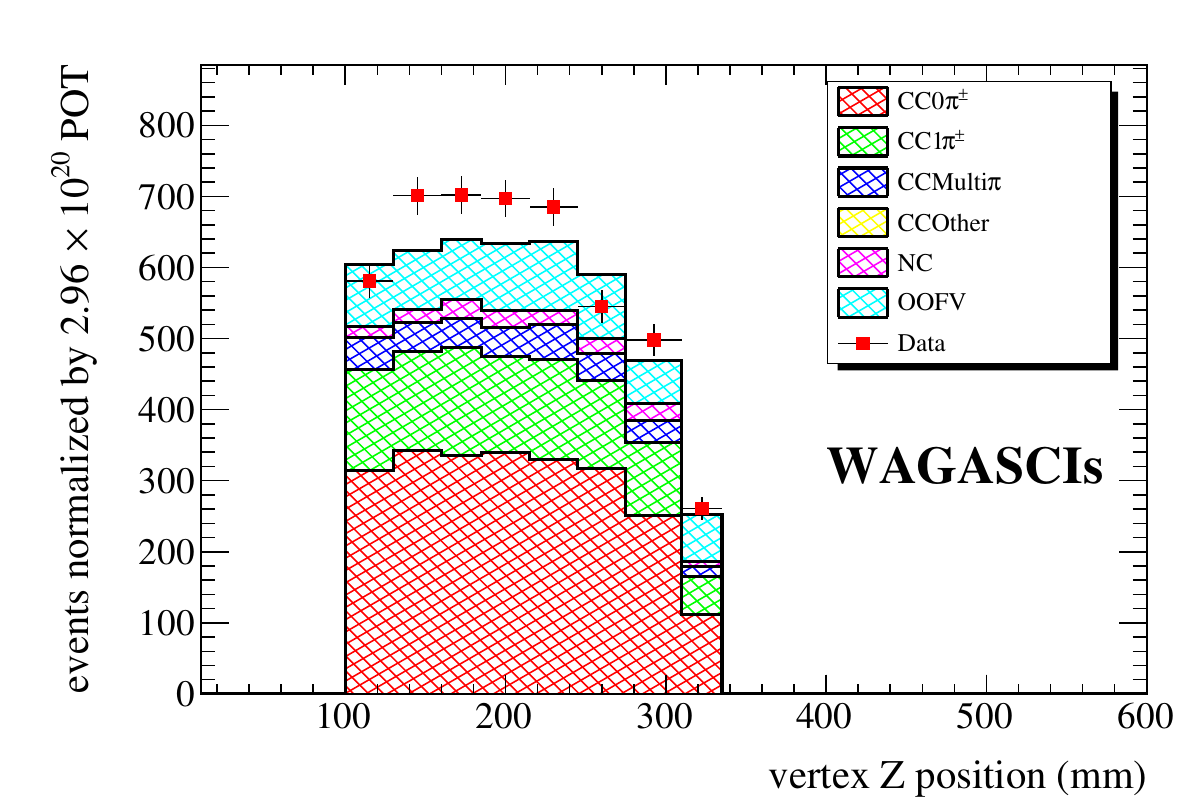} 
  \end{minipage}

  \end{tabular}
\caption{Data-MC comparison of the vertex distributions in the X (top), Y (middle) and Z (bottom) directions for the control samples for the \dpm (left) and the \dwgs (right). The MC predictions are normalized by accumulated POT.}
   \label{fig:CompVTX}
\end{figure*}

\begin{figure*}[htbp]

 \begin{tabular}{c}
  \begin{minipage}{0.5\linewidth}
    \centering
    \includegraphics[width=1.0\textwidth]{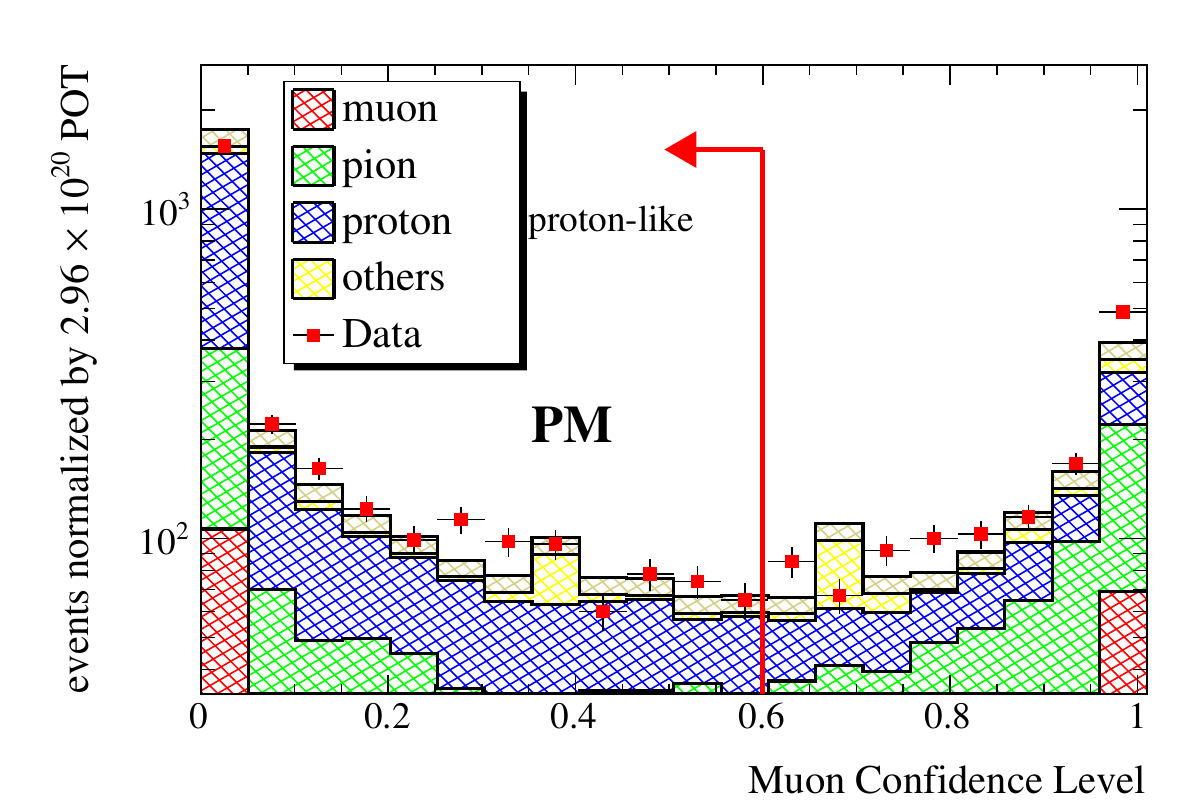} 
  \end{minipage}
  \begin{minipage}{0.5\linewidth}
    \centering
    \includegraphics[width=1.0\textwidth]{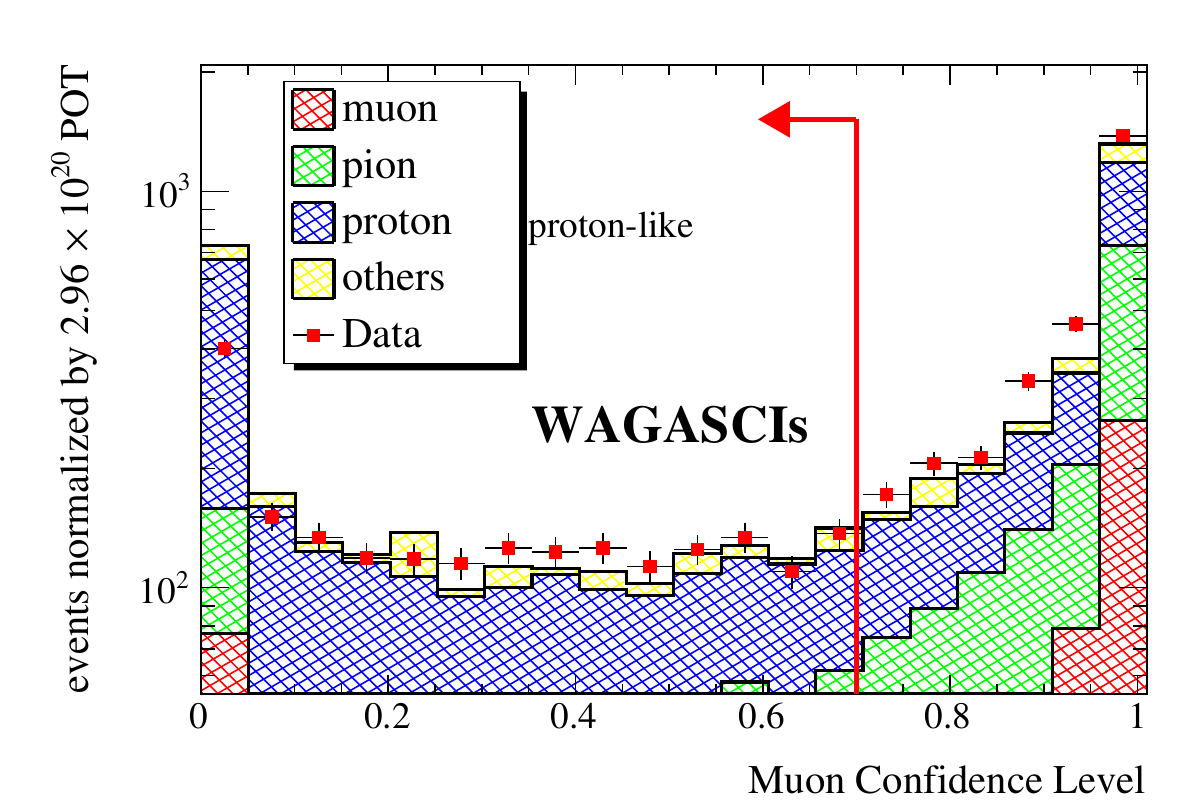} 
  \end{minipage}

  \end{tabular}
   \caption{Data-MC comparison of the muon confidence level for the \dpm (left) and the \dwgs (right) for the CC-inclusive samples}
   \label{fig:CompPID}
\end{figure*}

\begin{figure}[htbp]
\begin{center}
\includegraphics[width=.5\textwidth]
{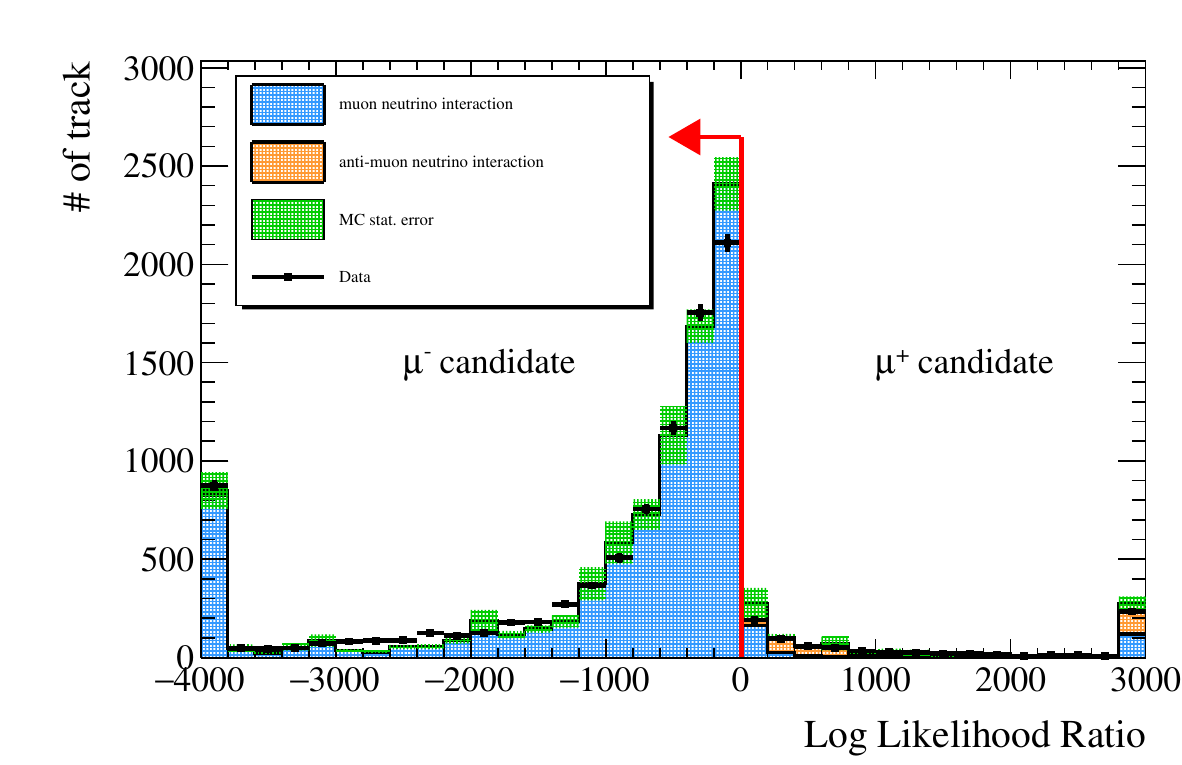}
\end{center}
\caption{Data-MC comparison of the Log-likelihood ratio of track charge for the stopping muon sample.}
\label{fig:CompLLR}
\end{figure}

\begin{description} 

\item[Fiducial volume] \hfill \\
We defined a fiducial volume cut for each of the vertex detectors. The fiducial volume cut played an important role in suppressing the beam-induced muon background. \RefFig{fig:CompVTX} shows the comparisons of the vertex distributions after pre-selection and the fiducial volume cut, the events selected by these cuts are defined as the control sample. The coordinate system used in  \RefFig{fig:CompVTX} is the same as defined in \RefFig{fig:wagasci_detectors}. For each detector the origins of the X and Y are the center of the scintillator and that of Z is the most downstream scintillator location.

The data and MC differed by 10\% for the WAGASCI sample. A possible reason for the smaller number of MC events compared to the data was that some tracks were misreconstructed due to less efficient MPPC channels in the WAGASCIs. We found the data MC agreement, especially for the vertex Z position, was improved by lowering the hit detection efficiency of several MPPC channels in the WAGASCIs in the simulation. The difference in the number of selected events between the nominal and modified detection efficiency of MPPC channels was accounted as one of the detector systematic uncertainties for this analysis.

\item[Number of tracks] \hfill \\
In the \signal sample, one muon and up to two protons were expected in the final state. When the number of tracks exceeds three, the purity of the \signal events in the selected sample was low. In this analysis, only events with at most three tracks were selected. This cut reduc edthe background due to $\nu_\mu$ CC interactions other than CC0$\pi^{\pm}$ and CC1$\pi^{\pm}$ which were dominated by Deep Inelastic Scattering (DIS).

\item[Particle identification] \hfill \\
When there was only one track associated with an event, that track was taken as the muon candidate.
The \signal events sometimes have multiple tracks if a proton or kaon undergoes secondary interactions.
In that case, we used particle identification (PID) on all tracks except the longest matching track, which was always assumed to be a muon. 
We introduced a discriminator, `MUon Confidence Level (MUCL)', for the particle identification by  energy deposition~\cite{PhysRevD.91.112002}. MUCL is defined by 
\begin{equation}
\mathrm{MUCL} = P \times \sum_{i=0}^{n=1}\frac{{(-\ln{P})}^{i}}{i!}, P = \prod_{i=1}^{n}, \mathrm{CL}_{i}
\end{equation}
where $n$ refers to the number of planes having a hit and $\mathrm{CL}_{i}$ is each confidence level for a hit on the $i$-th plane, which is defined by a cumulative function of $dE/dx$ distribution for each detector.

The MUCL distributions are shown in \RefFig{fig:CompPID}. When the MUCL exceeds 0.2 (0.7) for a \dpm (\dwg) event the track was identified as a muon-like particle and when it was under 0.2 (0.7) it was identified as a proton-like track. The \dpm has more light yield inside the scintillator than the \dwgs and tended to have higher MUCL values for muon-like  tracks. The different values come from the figure of merit optimization. When there were two or three tracks from a reconstructed vertex, it was required that the second and third tracks were proton-like tracks. This selection step increases the sample purity by 10\%.

\item[Charge of muon] \hfill \\

As we measured the cross section of muon neutrino the muon candidate should be negatively charged. The charge identification was applied if the muon candidate was contained within \dbm. This detector has an iron-core magnet in each steel plane, producing a magnetic field where charged particles are bent upwards if positively charged, or downwards if negatively charged. The discriminator for the charge identification was defined by a log-likelihood ratio, which is a ratio of the likelihood of $\mu^{-}$ or $\mu^{+}$ calculated using the track curvature. The distribution of the log-likelihood ratio is shown in \RefFig{fig:CompLLR}. When the log-likelihood ratio was less than 4, the track was selected as a negatively charged particle. This selection helps to reduce backgrounds from opposite-sign neutrino interactions. This background was due to a contamination of muon antineutrinos in the predominantly muon neutrino flux. 

\item[Track-associated hit ratio] \hfill \\
The track-associated hit ratio is defined as: 

\begin{equation}
R_{\mathrm{track-associated\ hit}} = \frac{N_{\mathrm{rechits}}}{N_{\mathrm{allhits}}}
\end{equation}
where  $N_{\mathrm{rechits}}$ was the number of hits in the vertex detector from all reconstructed tracks associated with the same vertex. $N_{\mathrm{allhits}}$ was the number of hits in all tracks obtained in the vertex detector including tracks not associated with the vertex. The hit ratio represented the fraction of hits in an event associated with a given track. This was applied only to the \dwg events because the wall background constituted around 20\% of the total number of \dwg events without this cut. This selection reduced the background from OOFV events, in particular from interactions in the wall. Particles produced within the wall sometimes produced tracks in the vertex detectors within the neutrino bunch timing, such an event has a lower track-associated hit ratio. When all hits in a vertex detector end up in the reconstructed tracks for the event the ratio was 1.0. When the track-associated hit ratio of an event was larger than 0.6, it was selected as a signal event. After this selection was applied, the contamination of OOFV background was reduced to around 10\%, a similar level as in the \dpm sample.

\item[Michel electron tagging] \hfill \\
A Michel electron is the electron from a decay of a muon. Particles are produced by the J-PARC accelerator in bunches with each bunch around 500\,ns in length, whereas the typical lifetime of a muon is 2.2\,$\mu$s. We searched for Michel electron hits originating from a pion track in each event in order to identify non-reconstructed pions. The presence of a Michel electron was a sign that a pion was produced in the neutrino interaction. The tagging proceeded in the following way.

\begin{itemize}
\item{Identify hit clusters outside the beam bunch timing. Here a hit cluster was a group of hits within a 100\,ns time window.}
\item{If the distance between the earliest hit in the cluster and the vertex of the neutrino event was less than 150~mm in the $X$, $Y$ \& $Z$ directions, select it for the next step.}
\item{Count the number of hits in the cluster that satisfy the above criteria. When the number of hits exceeds two, the event was tagged as having a Michel electron.}
\end{itemize}

As our signal did not contain any charged pion in the final state, events with a Michel electron must be rejected. The Michel electron tagging increases the signal purity by about 5\%. 

\item[Stop inside the MRD detectors] \hfill \\
When measuring the differential \xsec against momentum, an additional requirement was added that the track stops inside the muon range detectors so that the track momentum by range could be evaluated. 

\end{description}

The uncertainties on the selection cut parameters were considered by alternating the cut values, which is described in Sec.\ref{Sect:systematics}

\subsection{Kinematic distributions}

\begin{figure*}[htbp]
 \begin{tabular}{c}
  \begin{minipage}{0.5\linewidth}
    \centering
    \includegraphics[width=1.0\textwidth]{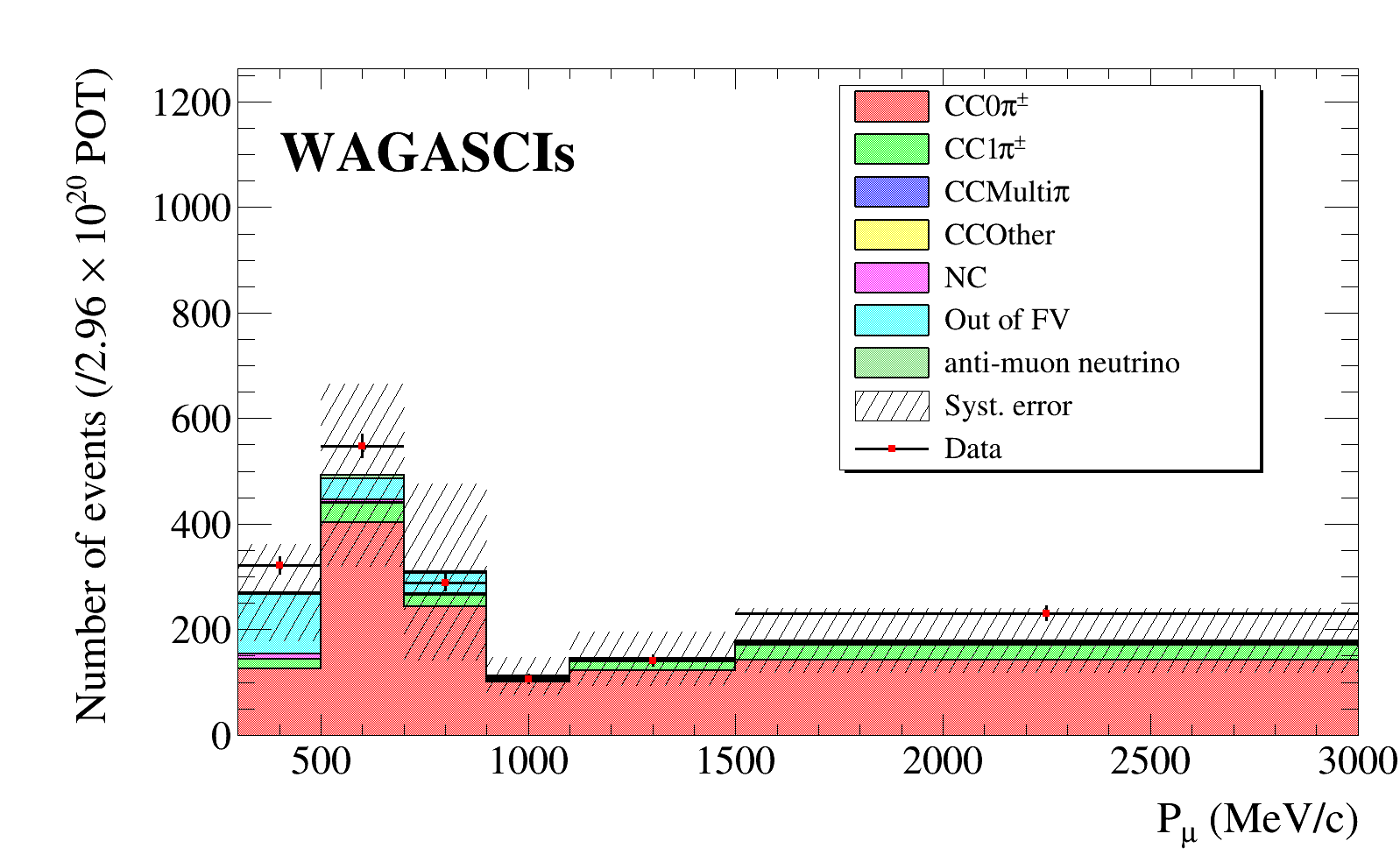} 
  \end{minipage}
  \begin{minipage}{0.5\linewidth}
    \centering
    \includegraphics[width=1.0\textwidth]{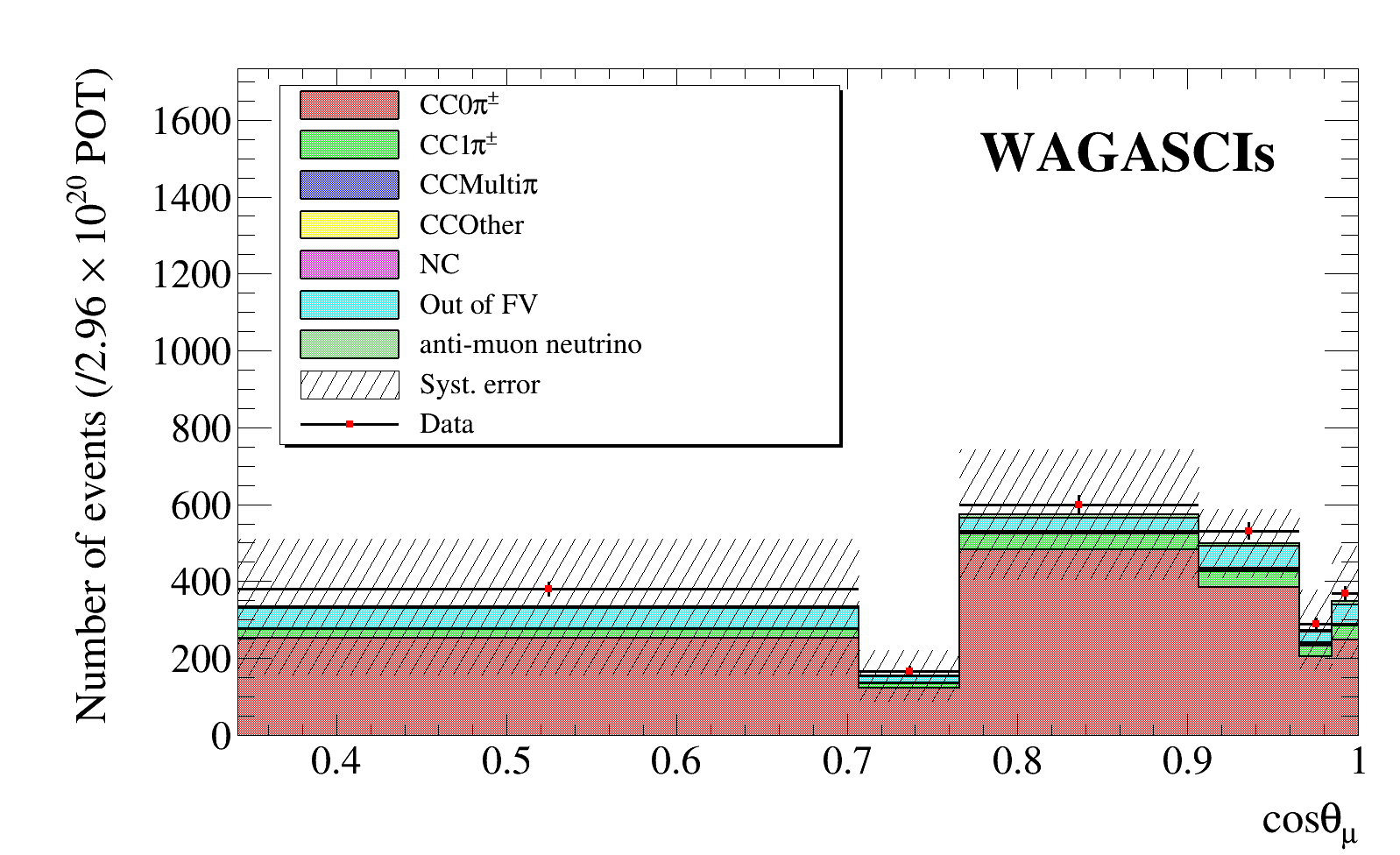} 
  \end{minipage}

\\

  \begin{minipage}{0.5\linewidth}
    \centering
    \includegraphics[width=1.0\textwidth]{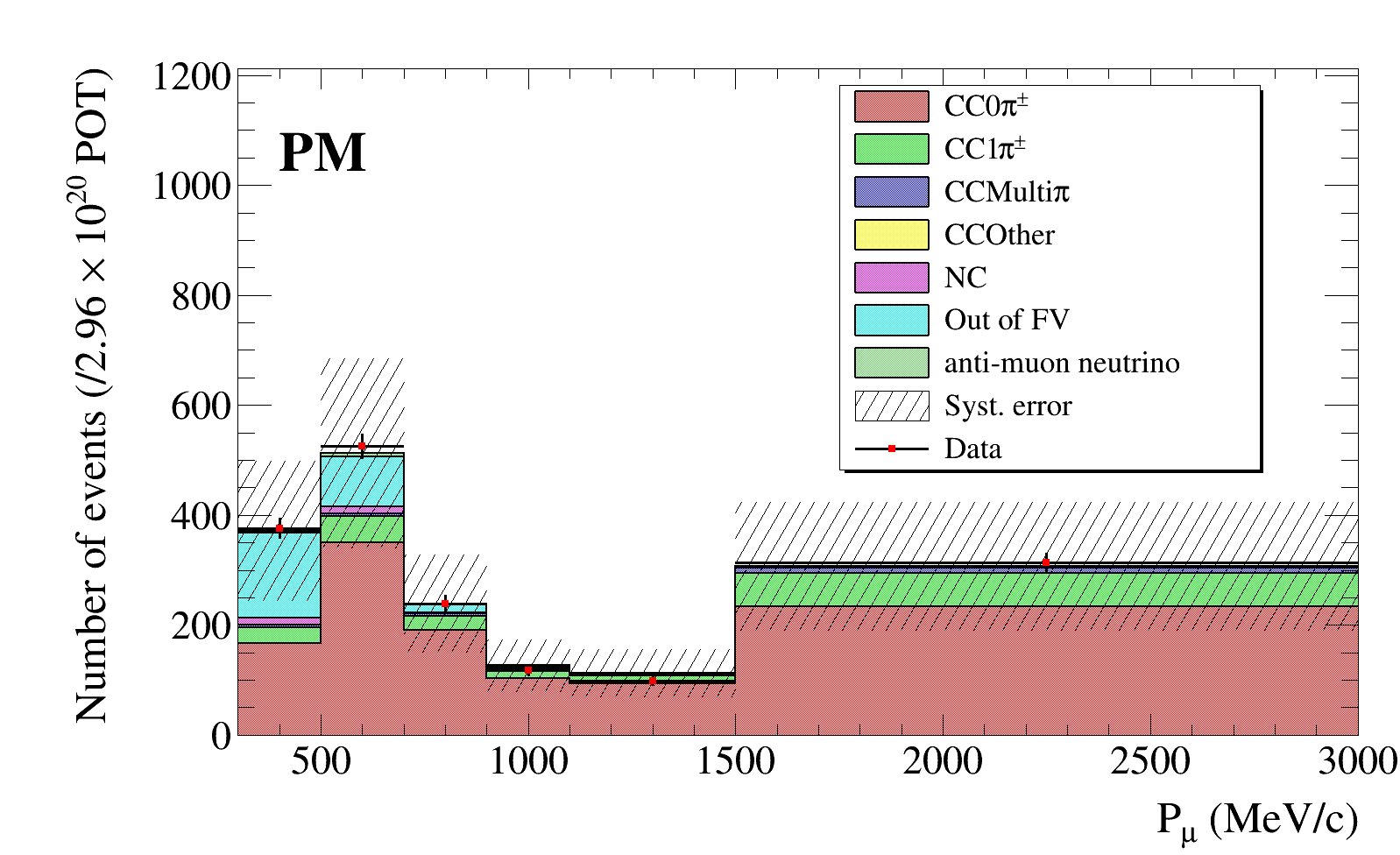} 
  \end{minipage}
  \begin{minipage}{0.5\linewidth}
    \centering
    \includegraphics[width=1.0\textwidth]{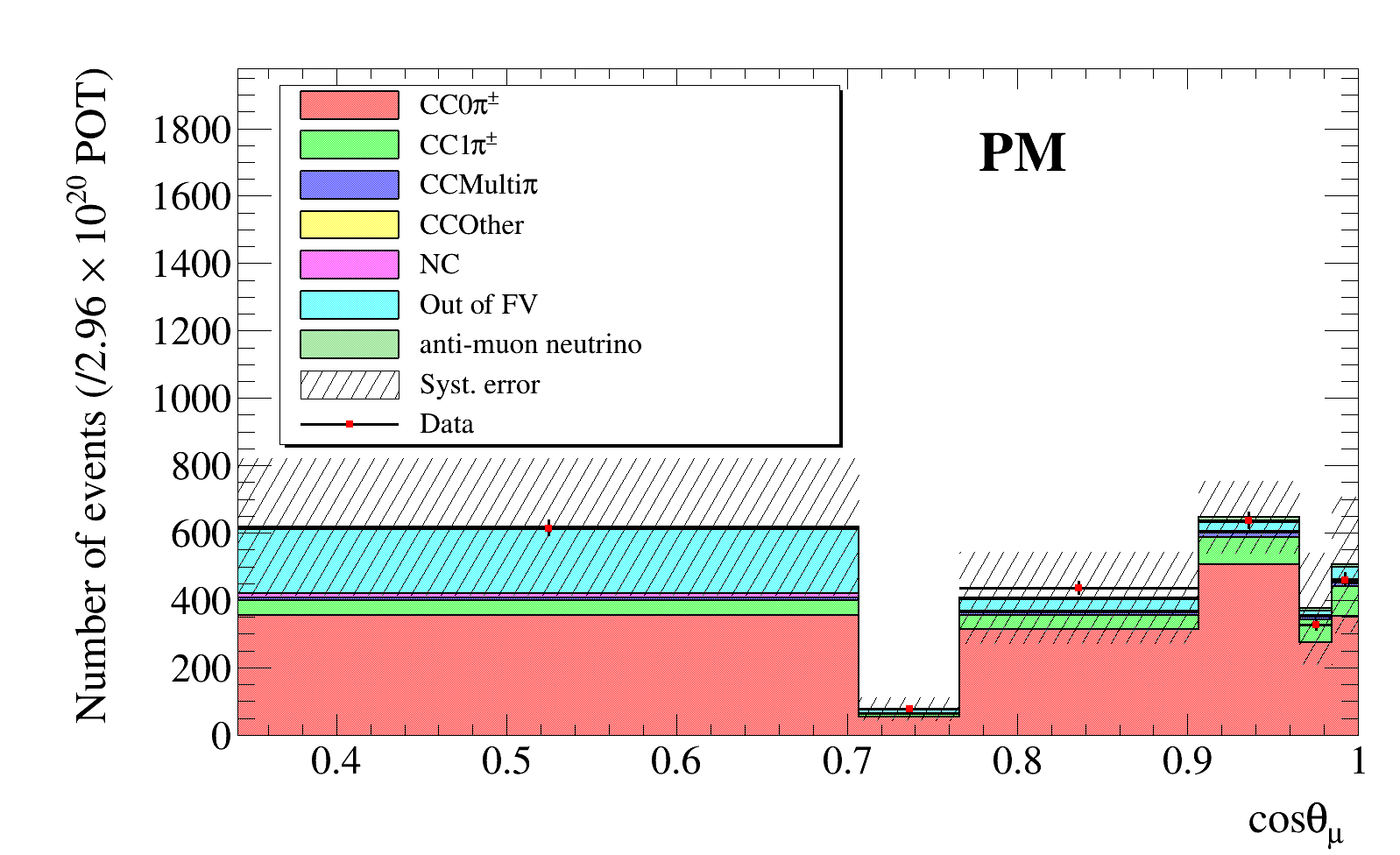} 
  \end{minipage}

  \end{tabular}
   \caption{Kinematic variables (momentum and angle) of the outgoing lepton for the selected events in  the \dwgs (top) and \dpm (bottom) samples.} 
   \label{fig:KineDist}
\end{figure*}

\RefFig{fig:KineDist} shows the kinematic distributions of the outgoing leptons for selected events in the \dwgs and the \dpm. The purity of the CCQE sample on the \ch and \water target is 49.7\% and 53.0\% respectively. Focusing on the data-MC comparison, the \dpm sample shows good consistency between data and MC, however, the number of events in the data was about 10\% larger than in the MC for the \dwg sample. 

\subsection{Sideband selection \label{4_sideband}}

\RefFig{fig:cutflow} also shows the selection scheme for the muon neutrino \sideband, which was the sideband sample of this measurement. Each cut has been already described in the \signal selection (Sec.~\ref{subsect:selection_criteria}) and to select the sideband only the Michel electron cut was reversed. 

The selection was divided into two branches depending on whether a pion was reconstructed as a track or not. The first branch of \sideband selection includes pion tracks below the tracking threshold. The whole selection criteria in this path had the same scheme as in the selection of the \signal sample, except for reversing the final cut for Michel electron tagging. For this \sideband selection, the Michel electron must be detected. The second branch includes pion tracks that were above the tracking threshold. A pion track is considered to be more like a muon-like track than a proton-like track in terms of its MUCL value, hence in this scenario the MUCL cut was reversed. The charge cut was applied in order to reduce the opposite-sign background and select muon neutrino \sideband interactions with the appropriate sign.

\subsection{Number of target nucleons}

The cross sections were normalized per nucleon (either proton or neutron) in the target material.
\RefTab{tab:NumTarget} summarizes the number of target nucleons in the \ch and \water targets.

 \begin{table}[htb]
  \caption{The number of target nucleons inside the fiducial volume for each vertex detector. Each number refers to the number of nucleons (both protons and neutrons) in the CH or \water.}
  \label{tab:NumTarget}
  \vspace{3mm}
  
  \begin{ruledtabular}
  \begin{tabular}{cc|cc}

     \multicolumn{2}{c|}{ \multirow{2}{*}{Target material} }			& \multicolumn{2}{c}{Number of target nucleons}\\			
    						& 							& Measured ($10^{28}$) 		&  Uncertainty ($10^{28}$)        \\    
    \hline	
     
     \multirow{1}{*}{ PM }      & CH                        & 18.66                 & 0.045 (0.24\%)                           \\

      \hline	
     \multirow{2}{*}{ WAGASCI }& CH						   & 3.64 				   & 0.011 (0.30\%)						\\
					        & \water					& 13.77				    & 0.048 (0.35\%) 					\\												
           
  \end{tabular}
  \end{ruledtabular}
 \end{table}

\section{\label{Sect:xsec_extraction}cross section extraction}

The selected samples of data contain \signal events as a signal and backgrounds such as events with pions produced in neutrino interactions, neutral current interactions, etc. In order to extract the signal components in the selected samples, we applied a binned likelihood approach performed with the MINUIT2~\cite{MINUIT2} algorithm. The fitting method was similar to that previously used in T2K \xsec analyses \cite{PhysRevD.108.112009}. Here, a brief summary of the method is provided. 

The binned likelihood function is schematically defined as

\begin{equation}
\label{eq:wagasci_babymind/cross_section_fitter/fitter_negative_loglikelihood_total}
-2\ln{\mathcal{L}(\overrightarrow{y}; \overrightarrow{\theta})} = -2\ln{\mathcal{L}_{\mathrm{stat}}(\overrightarrow{y}; \overrightarrow{\theta})} -2\ln{\mathcal{L}_{\mathrm{syst}}(\overrightarrow{y}; \overrightarrow{\theta})},
\end{equation}
where $\mathcal{L}$ is the likelihood function, $\overrightarrow{y}$ is a vector of data and $\overrightarrow{\theta}$ is a set of parameter values.

The statistical part of the likelihood function is given by

\begin{equation}
\label{eq:wagasci_babymind/cross_section_analysis/fitter_stat_negative_loglikelihood}
\begin{split}
-2\ln{\mathcal{L}_{\mathrm{stat}}(\overrightarrow{y}; \overrightarrow{\theta})}
&=\sum_{j}^{\mathrm{reco\ bins}}2 \left( \beta_{j}N^{\mathrm{exp}}_{j} - N^{\mathrm{obs}}_{j} \right. \\
&+\quad \left. N^{\mathrm{obs}}_{j}\ln{\frac{N_{j}^{\mathrm{obs}}}{\beta_{j}N^{\mathrm{exp}}_{j}}} + \frac{(\beta_{j}-1)^{2}}{2\sigma^{2}_{j}} \right),
\end{split}
\end{equation}
where $j$ runs over each reconstructed bin for all samples and $N^{\mathrm{exp}} (N^{\mathrm{obs}})$ is the number of selected events predicted by MC (obtained in the data), $\beta_j$ are the Barlow-Beeston scaling parameters which account for the uncertainty of finite MC simulation (for the definition see \cite{PhysRevD.108.112009}) and $\sigma$ is the MC statistical uncertainty.

The binned expected number of selected events, $N^{\mathrm{exp}}_{j}$, is given for each value of $\overrightarrow{\theta}$ as a sum of two contributions one from the signal, $N^{\mathrm{exp,\, sig}}_{j}$, and another from the background, $N^{\mathrm{exp,\, bg}}_{j}$. The signal contribution was calculated with the true signal events multiplied with a detector smearing matrix to provide the number of true signal events in the reconstructed bin. 

We then introduce multiplicative parameters to weight events. Template parameters denoted by $c_{i}$ are given for each true bin $i$, to weight the number of selected events in MC, $N^{\mathrm{MC,\, sig}}_{i}$.  The template parameters scale the number of signal events of MC in each true bin to reproduce the data and  are free parameters with no prior uncertainty (or penalty term).
Besides the template parameters, there were three types of systematic parameters from neutrino flux, neutrino interaction and detector systematics. The expected number of events was weighted by these parameters too. They were mainly sensitive to the shape of the input distributions.

The complete expression for the expected number of events is
\begin{equation}
\label{eq:wagasci_babymind/cross_section_analysis/fitter_method_reconstructed_events}
\begin{split}
N^{\mathrm{exp}}_{j} 
&= \sum_{i}^{\mathrm{true}} \left [ c_{i} \left( N_{i}^{\mathrm{MC,\, sig}} \prod_{a}^{\mathrm{int}} w(a)_{i}^{\mathrm{sig}} \right) \right.\\
&+ \quad \left. \sum_{k}^{\mathrm{\mathrm{bg}}}N_{ik}^{\mathrm{MC,\, bg}} \prod_{a}^{\mathrm{int}}w(a)_{i}^{\mathrm{k}}  \right] t_{ij}r_{j} \sum_{n}^{E_{\nu}}v_{in}f_{n}.
\end{split}
\end{equation}
where $a$ runs over the interaction parameters, $w(a)$ refers to each weight and $k$ is an index running over interaction modes. They affect the number of selected events in the MC changing its shape or normalization. $t_{ij}$ is the transfer matrix. $n$ runs over the neutrino energy corresponding to the true bin $i$. The $v_{in}$ is the fractional contribution in the i-th bin by neutrinos having energy $n$. $f_{n}$ terms are the weights from flux systematics. The reconstructed signal events were multiplied by the smearing matrix to calculate the true number of signal events.

With the assumption that the systematics uncertainties follow a Gaussian probability distribution the systematic part of the likelihood function is given in

\begin{equation}
\label{eq:wagasci_babymind/cross_section_analysis/fitter_syst_negative_loglikelihood}
-2\ln{\mathcal{L}_{\mathrm{syst}}} = \sum_{\mathrm{syst}}\left( (\overrightarrow{p}-\overrightarrow{p}_{\mathrm{prior}})(V_{\mathrm{cov}}^{\mathrm{syst}})^{-1}(\overrightarrow{p}-\overrightarrow{p}_{\mathrm{prior}})^T\right),
\end{equation}
where $p$ runs over all systematic parameters and $V_{\mathrm{cov}}^{\mathrm{syst}}$ are covariance matrices for each prior parameter. This acts as a penalty term for moving the systematic parameters away from their prior values. We have a sufficient number of events in each bin for this assumption.

The \xsec fitter provides the best-fit values for template parameters and systematic parameters. The expected signal events were then calculated by

\begin{equation}
\label{eq:wagasci_babymind/cross_section_analysis/recon_to_true_conversion}
\hat{N}^{\mathrm{exp,\, sig}}_{i} =  \sum_{j}^{\mathrm{recon}} \left(\hat{c}_{i}N_{i}^{\mathrm{MC,\, sig}} \prod_{a}^{\mathrm{int}} \hat{w}(a)_{i}^{\mathrm{sig}} \right) (t_{ij})^{-1} \hat{r}_{j} \sum_{n}^{E_{\nu}}v_{in}\hat{f}_{n},
\end{equation}
where the `hat' means the best fit value. 
Then, the differential cross section is calculated by

\begin{equation}
\label{eq:wagasci_babymind/cross_section_analysis/flux_integrated_cross_section}
\frac{d \sigma}{dx_{i}} = \frac{\hat{N}^{\mathrm{exp, sig}}_{i}}{\epsilon_{i}\Phi N^{\mathrm{FV}}_{\mathrm{nucleons}}} \times \frac{1}{\Delta x_{i}},
\end{equation}
where $\sigma$ is the flux-integrated \xsec and $x$ is a muon kinematic variable, $\epsilon, \Phi, N^{\mathrm{FV}}_{\mathrm{nucleons}}, \Delta x_{i}$ are the detection efficiency, integrated flux, number of target nucleons in the fiducial volume and the bin width of the true bin $i$ respectively. Since we report a single differential \xsec as a function of muon kinematics, $x$ refers to either \Pmu or $\cos\theta_\mu$.

When we calculate a \xsec there were multiple ways of treating the neutrino flux. In this approach a flux-integrated \xsec rather than a flux-unfolded one was calculated to avoid model dependence (or at least to reduce it) in terms of the shape of the neutrino energy spectrum.

\section{\label{Sect:systematics} Systematics uncertainties}

This section is organized into three sub-sections: neutrino flux, neutrino interaction and detector systematics. For neutrino flux and detector systematics, systematic parameters were summarized in covariance matrices which were used when fitting the data. For neutrino interactions, a different approach based on MC event reweighting was used.

\subsection{Neutrino Flux Systematics}
\begin{figure}[htbp]
\begin{center}
\includegraphics[width=.48\textwidth]{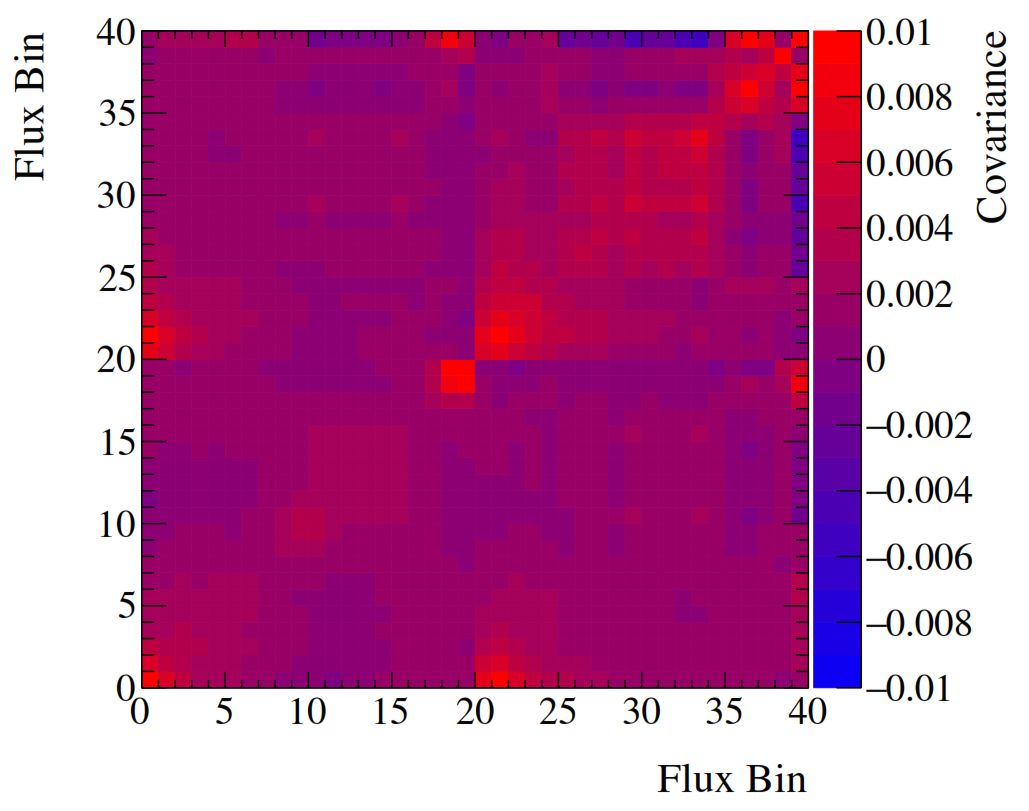}
\end{center}
\caption{Covariance matrix for the neutrino flux uncertainty at the $1.5^\circ$ off-axis angle where the \wb is located. The bins 0-19 correspond to the \numu component and bins 20-39 correspond to the \numubar component of the beam.}
\label{fig:neutrino_flux_fhc_covariance_matrix}
\end{figure}

Neutrino flux uncertainties were evaluated using a method employed previously in other T2K studies~\cite{PhysRevD.87.012001}. The covariance matrix  binning has been chosen to ensure approximately equal statistics in each bin as can be seen in \RefTab{tab:neutrino_flux_fhc_binning} and the resultant matrix is shown in \RefFig{fig:neutrino_flux_fhc_covariance_matrix}.

\begin{table}[htb]
\begin{ruledtabular}
\begin{tabular}{c|c|c}	
Energy range & Number of bins & Width in energy per bin \\    
\hline
0.0 - 3.0 GeV      & 15                       & 0.2 GeV    \\
3.0 - 4.0 GeV      & 1                        & 1.0 GeV    \\
4.0 - 10.0 GeV     & 3                        & 2.0 GeV    \\
10.0 - 30.0 GeV    & 1                        & 20.0 GeV   \\
\end{tabular}
\end{ruledtabular}
\caption{Energy binning for flux uncertainty distribution. Bins in a given energy range have the same width.}
\label{tab:neutrino_flux_fhc_binning}
\end{table}

\begin{figure}[htbp]
\begin{center}
\includegraphics[width=.5\textwidth]{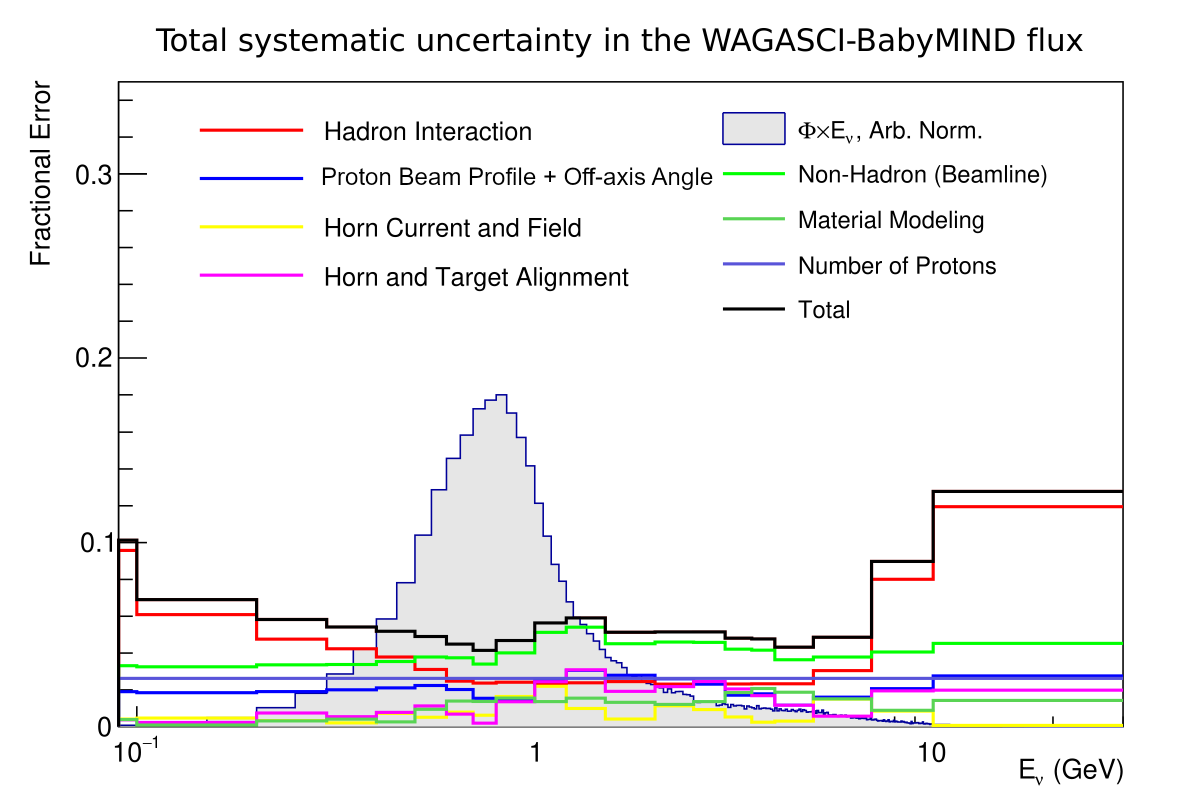}
\end{center}
\caption{Flux uncertainties for \numu in FHC mode at the \wb position.}
\label{fig:flux_error_t2k_nd7_numode_numu}
\end{figure}

In \RefFig{fig:flux_error_t2k_nd7_numode_numu} the fractional error on the muon neutrino flux is shown by the colored spectrum. In the background, the curve filled with gray shows the neutrino flux as a function of neutrino energy at the 1.5$^\circ$ off-axis angle. The uncertainties stem from several factors, including the hadron production model, the profile of the proton beam, the off-axis angle, the horn current, horn alignment, wrong-sign ($\nu/ \bar{\nu} $) neutrino contamination and others. The T2K replica target hadron production measurements in NA61/SHINE \cite{Abgrall:2019EPJC.79.2, Abgrall:2016EPJC.76.11, Abgrall:2016EPJC.76.2} were used to define the hadron uncertainties reported here. How each source of error impacts the flux prediction is detailed in~\cite{PhysRevD.87.012001}. 

\subsection{Neutrino Interaction Systematics}

Our MC simulation generated particles from interaction vertices based on theoretical models that incorporate our current understanding of neutrino interactions. However, these models rely on approximations due to limited experimental data and/or the complexity of underlying theoretical models. We used the NEUT software~\cite{Hayato:2002sd} to simulate neutrino interactions.

Neutrino interaction models include a set of parameters called dials. Dials can be scalar quantities, such as a dimensional parameter affecting a physical process, a non-dimensional quantity tuning interaction strength or a boolean variable controlling a model feature.

Our method for estimating uncertainties due to neutrino interaction models assumed that varying all dials within their stated uncertainty fully accounts for the overall uncertainty of the \xsec measurement. A limitation of this method was that the models themselves might be incomplete and some sources of uncertainty might not be accounted for. To cover such cases, we performed the validations described in \RefSec{Sect:fds}.

Prior knowledge about each dial comes from theoretical calculations and previous measurements. 
While a complete account of each and every dial is out of the scope of this paper, in \RefTab{tab:dials} we list all the dials used in our \xsec measurement. How the variation of each dial was reflected into the \xsec measurement is summarized in \RefSec{sec:reweighting-of-mc-events}. In \RefSec{sec:axial-mass}, \RefSec{sec:multi-nucleon-interactions-2p2h} and \RefSec{sec:single-pion-production}, we discuss the dials expected to have the greatest impact.

It is important to stress that the systematic uncertainties listed below do not directly impact the fitting results, but affect them through changes in detection efficiencies, as each bin's signal count was fitted with the corresponding template parameter to reproduce the data.
\begin{table*}
\footnotesize 
\begin{tabular}{lllll}
\cline{1-5}
\textbf{Dial Name} & \textbf{Prior} & \textbf{Uncertainty} & \textbf{Target} & \textbf{Description} \\
\cline{1-5}
\multicolumn{5}{c}{\textbf{CCQE}} \\
\cline{1-5}
MaCCQE & 1.21 GeV & -0.3, +0.3 & Both & CCQE axial mass \\
QETwk\_High\_Q2\_Weight\_1 & 1 & -0.11, +0.11 & Both & Reweight CCQE events with $0.25 < Q^2 < 0.5$ \\
QETwk\_High\_Q2\_Weight\_2 & 1 & -0.18, +0.18 & Both & Reweight CCQE events with $0.5 < Q^2 < 1$ \\
QETwk\_High\_Q2\_Weight\_3 & 1 & -0.4, +0.4 & Both & Reweight CCQE events with $Q^2 > 1$ \\
SF\_OptPotTwkDial\_O16 & 0 & -0, +0.49 & Water & SF Optical Potential Correction, O \\
SF\_OptPotTwkDial\_C12 & 0 & -0, +0.49 & CH & SF Optical Potential Correction, C \\
SF\_P1\_2Shell\_MeanF\_Norm\_O & 0 & -0.2, +0.2 & Water & Norm of Gaussian positioned at some $E_m$ for restricted $p_m$ \\
SF\_P1\_2Shell\_MeanF\_PMissShape\_O & 0 & -1, +1 & Water & Template normalisation in $p_m$ in some range of $p_m$, $E_m$ \\
SF\_P3\_2Shell\_MeanF\_Norm\_O & 0 & -0.4, +0.4 & Water & Norm of Gaussian positioned at some $E_m$ for restricted $p_m$ \\
SF\_P3\_2Shell\_MeanF\_PMissShape\_O & 0 & -1, +1 & Water & Template normalisation in $p_m$ in some range of $p_m$, $E_m$ \\
SF\_PBTwkDial\_Hybrid\_C12\_nu & 0 & -2, +0.6 & CH & Hybrid nucl. mom cut off and template normalisation in $q_3$, $q_0$ \\
SF\_PBTwkDial\_Hybrid\_C12\_nubar & 0 & -2, +0.6 & CH & Hybrid nucl. mom cut off and template normalisation in $q_3$, $q_0$ \\
SF\_PBTwkDial\_Hybrid\_O16\_nu & 0 & -2, +0.6 & Water & Hybrid nucl. mom cut off and template normalisation in $q_3$, $q_0$ \\
SF\_PBTwkDial\_Hybrid\_O16\_nubar & 0 & -2, +0.6 & Water & Hybrid nucl. mom cut off and template normalisation in $q_3$, $q_0$ \\
SF\_PShell\_MeanF\_Norm\_C & 0 & -0.2, +0.2 & CH & Norm of Gaussian positioned at some $E_m$ for restricted $p_m$ \\
SF\_PShell\_MeanF\_PMissShape\_C & 0 & -1, +1 & CH & Template normalisation in $p_m$ in some range of $p_m$, $E_m$ \\
SF\_SRC\_Norm\_C & 1 & -1, +1 & CH & Overall normalisation in some range of $p_m$, $E_m$ \\
SF\_SRC\_Norm\_O & 1 & -1, +1 & Water & Overall normalisation in some range of $p_m$, $E_m$ \\
SF\_SShell\_MeanF\_Norm\_C & 0 & -0.4, +0.4 & CH & Norm of Gaussian positioned at some $E_m$ for restricted $p_m$ \\
SF\_SShell\_MeanF\_Norm\_O & 0 & -0.2, +0.2 & Water & Norm of Gaussian positioned at some $E_m$ for restricted $p_m$ \\
SF\_SShell\_MeanF\_PMissShape\_C & 0 & -1, +1 & CH & Template normalisation in $p_m$ in some range of $p_m$, $E_m$ \\
SF\_SShell\_MeanF\_PMissShape\_O & 0 & -1, +1 & Water & Template normalisation in $p_m$ in some range of $p_m$, $E_m$ \\
\cline{1-5}
\multicolumn{5}{c}{\textbf{2p2h}} \\
\cline{1-5}
MECTwkDial\_Norm\_C12 & 1 & -0.99, +0.99 & CH & 2p2h Normalization Carbon \\
MECTwkDial\_Norm\_O16 & 1 & -0.99, +0.99 & Water & 2p2h Normalization Oxygen \\
MECTwkDial\_Norm\_Other & 1 & -0.49, +0.49 & Both & 2p2h Normalization Others \\
MECTwkDial\_PDDWeight\_C12\_NN & 0 & +0.49, -0.49 & CH & 2p2h shape on Carbon: contributions change for NN pairs \\
MECTwkDial\_PDDWeight\_C12\_np & 0 & -0.49, +0.49 & CH & 2p2h shape on Carbon:  contributions change for NP pairs \\
MECTwkDial\_PDDWeight\_O16\_NN & 0 & -0.49, +0.49 & Water & 2p2h shape on Oxygen:  contributions change for NN pairs \\
MECTwkDial\_PDDWeight\_O16\_np & 0 & -0.49, +0.49 & Water & 2p2h shape on Oxygen:  contributions change for NP pairs \\
MECTwkDial\_PNNN\_Shape & 0 & -0.33, +0.33 & Both & 2p2h shape: NN or Np nucleon pair \\
\cline{1-5}
\multicolumn{5}{c}{\textbf{SPP}} \\
\cline{1-5}
RES\_E\_b\_C\_numu & 25 MeV & -24.9, +24.9 & CH & Resonant SPP $E_b$ value (MeV) for muon neutrinos on Carbon \\
RES\_E\_b\_O\_numu & 25 MeV & -24.9, +24.9 & Water & Resonant SPP $E_b$ value (MeV) for muon neutrinos on Oxygen \\
RES\_E\_b\_C\_numubar & 25 MeV & -24.9, +24.9 & CH & Resonant SPP $E_b$ value (MeV) for \numubar on Carbon \\
RES\_E\_b\_O\_numubar & 25 MeV & -24.9, +24.9 & Water & Resonant SPP $E_b$ value (MeV) for \numubar on Oxygen \\
BgSclRES & 1.3 & -0.15, +0.15 & Both & I½ non-resonant background \\
CA5RES & 1.01 & -0.15, +0.15 & Both & CA5(0): value at $Q^2=0$ of the axial form factor \\
MaRES & 0.95 GeV & -0.15, +0.15 & Both & RES axial mass \\
\hline
\multicolumn{5}{c}{\textbf{FSI}} \\
\cline{1-5}
PionFSI\_AbsProb & 1.404 & -0.432, +0.432 & Both & Pion FSI absorption (FEFABS) \\
PionFSI\_CExHighMomProb & 1.8 & -0.288, +0.288 & Both & Pion FSI Single charge exchange (high energy) (FEFCXH) \\
PionFSI\_CExLowMomProb & 0.697 & -0.305, +0.305 & Both & Pion FSI Single charge exchange (low energy) (FEFCX) \\
PionFSI\_InelProb & 1.002 & -1.101, +1.101 & Both & Pion FSI Hadron (N+n pi) production (FEFINEL) \\
PionFSI\_QEHighMomProb & 1.824 & -0.859, +0.859 & Both & Pion FSI QE scattering (high energy) (FEFQEH) \\
PionFSI\_QELowMomProb & 1.069 & -0.313, +0.313 & Both & Pion FSI QE scattering (low energy) (FEFQE) \\
TwkDial\_FateNucleonFSI & 0 & -0.3, +0.3 & Both & Fate of final state nucleons after the FSI interactions \\
\cline{1-5}
\multicolumn{5}{c}{\textbf{Other}} \\
\cline{1-5}
CC\_DIS\_norm\_nu & 1 & -0.15, +0.15 & Both & Normalisation of CC DIS for neutrinos \\
CC\_DIS\_norm\_nubar & 1 & -0.15, +0.15 & Both & Normalisation of CC DIS for anti-neutrinos \\
CC\_MultiPi\_norm\_nu & 1 & -0.2, +0.2 & Both & Multi-pion production normalisation (nu) \\
CC\_MultiPi\_norm\_nubar & 1 & -0.2, +0.2 & Both & Multi-pion production normalisation (nubar) \\
\cline{1-5}
\end{tabular}
\caption{List of all systematic parameters from neutrino interaction models that were included in the analysis. The column `Dial Name' is the name of the parameter as it appears in the source code. The column `Prior' is the parameter's prior value before variation. The column `Uncertainty' represents the $\pm 1 \sigma$ prior uncertainty. The column `Target' indicates which type of target material is affected, "CH" is a shorthand for hydrocarbon, while "Both" refers to both water and hydrocarbon. The column `Description' contains a short description of the meaning of the parameter. No single parameter dominates the systematic uncertainty from the neutrino interaction models, in the sense that no parameter contributes more than 50\% to the total systematic error.}
\label{tab:dials}
\end{table*}

\subsubsection{Reweighting of MC events}

\label{sec:reweighting-of-mc-events}

In the course of a \xsec analysis, it is customary to use MC simulations to model changes to the event selection under changes of a generic \xsec dial ($\vec{x}$). We calculated a series of correction factors to the MC, i.e., for a change to dial $\vec{x} \rightarrow \vec{x'}$ for each event $i$, then we calculated a weight:

\begin{equation}
w_i = \frac{\sigma(\vec{x'})}{\sigma(\vec{x})}
\end{equation}
which is the ratio of the nominal \xsec $\sigma(\vec{x})$ to an updated \xsec $\sigma(\vec{x'})$. Applying the individual weights to the MC sample will function in much the same way as a regenerated MC sample with the modified dial and this procedure is called “reweighting”. 

The weights were then organized into a set of splines, allowing for smooth interpolation between the node in a certain analysis bin, when the value of the dial was varied with respect to the nominal value. These splines allow for smooth interpolation between dial values, enabling continuous variation of model parameters. They constitute one of the inputs passed to the \xsec fitter. This approach is the same as the one adopted in many other T2K \xsec analyses ~\cite{10.1093/ptep/ptab014}.

\subsubsection{CCQE and the axial mass}

\label{sec:axial-mass}

CCQE interactions are dominant at T2K energies. 
The CCQE \xsec on a single nucleon depends on vector and axial form factors~\cite{Llewellyn-Smith}. Vector form factors are known precisely from electron scattering data~\cite{PhysRevD.102.074012}, due to the conserved vector current hypothesis. However, for the axial form factor the situation is different and it was a dominant source of uncertainty. A dipole expression is assumed with an axial mass parameter $M_A^{QE} = 1.21$ GeV, which is constrained by past neutrino-deuterium scattering measurements~\cite{PhysRevD.104.093001}. 

In CCQE interactions on a nucleon, within a nucleus such as Oxygen or Carbon, the effects of nuclear binding were modeled using the Benhar Spectral Function (SF)  model 
\cite{Benhar:1999bg}, which provides a realistic shell-model-based description of the nuclear ground state and is strongly favored over relativistic Fermi gas models~\cite{1975547} by electron scattering data~\cite{Sobczyk:2017mts}.

In our study, we treated $M_A^{QE}$ like an effective parameter to account for uncovered effects such as a non-dipole expression for the axial form factor; hence, a large uncertainty was assumed to let the data determine its best value.

\subsubsection{Multi-Nucleon Interactions (2p2h)}

\label{sec:multi-nucleon-interactions-2p2h}

The term `2p2h process' refers to neutrino interactions on pairs of correlated nucleons. They give rise to an important fraction of the events observed in T2K's energy range. In NEUT simulations 2p2h includes contributions from Meson Exchange Currents, nucleon-nucleon correlations (NN) and their interference~\cite{RuizSimo:2016rtu, PASCHALIS2020135110} as described by the Valencia model~\cite{Nieves:2011pp}. It was assumed that the 2p2h contribution populates a kinematic region of momentum transfer $|\vec{q}|\leq 1.2$\,GeV/c. Since precise 2p2h \xsec measurements are unavailable, a conservative approach was taken in setting priors for its dials. A set of normalization dials scaling the whole 2p2h contribution (depending on the neutrino flavour and target nucleus) were considered in this analysis.

\subsubsection{Single Pion Production}

\label{sec:single-pion-production}

Single-pion final states arise from resonance excitation, coherent pion production and multi-pion processes. As already mentioned in \RefSec{Sect:Intro}, the neutrino flux observed at the off-axis angle of the \wb is shifted towards higher energies with respect to ND280 resulting in an increased probability of pion production from the neutrino interactions, thus making their correct characterization important for our analysis. The resonance excitation mechanism is modeled by NEUT with the Rein-Sehgal model~\cite{REIN198179, Rein:1987cb}.
 
The most important dials were $M_A^{RES}$ (RESonance axial-vector mass), $C_5^A(0)$ (normalization of the leading axial form factor) and $I_{\nicefrac{1}{2}}$, (the amount of non-resonance contribution with isospin $\nicefrac{1}{2}$ in the Rein-Sehgal model). 
The nominal values were taken from theoretical models tuned to external data~\cite{Graczyk:2009qm, Kuzmin:2006dh}.

\subsection{Detector systematics}

\label{sec:detector-systematics}

In this analysis, detector systematic uncertainties were separated into three categories: detector-related parameters (\RefSec{sec:detector-related-parameters}), track-reconstruction-related parameters (\RefSec{sec:track-reconstruction-related-parameters}) and sample-selection-related parameters (\RefSec{sec:sample-selection-related-parameters}). The covariance matrix, $V_{ij}$, for each systematic uncertainty was calculated from the change in the predicted number of events by:

\begin{equation}
\label{eq:detector-covariance-matrix}
\begin{split}
V_{ij} = \frac{1}{2} \frac{(\phi_{\text{nom}}^{i} - \phi_{+}^{i})(\phi_{\text{nom}}^{j} - \phi_{+}^{j})}{\phi_{\text{nom}}^{i} \phi_{\text{nom}}^{j}} + \\ + \frac{1}{2} \frac{(\phi_{\text{nom}}^{i} - \phi_{-}^{i})(\phi_{\text{nom}}^{j} - \phi_{-}^{j})}{\phi_{\text{nom}}^{i} \phi_{\text{nom}}^{j}}
\end{split}
\end{equation}

where $i$ and $j$ cover the binning of \RefTab{tab:wagasci_babymind/differential_cross_section_momentum} and \RefTab{tab:wagasci_babymind/differential_cross_section_angle}. $\phi_{nom}^{i}$ is the number of selected events in the nominal setting in the $i$-th bin while $\phi_{+}^{i}$ and $\phi_{-}^{i}$ refer to the number of selected events in the $i$-th bin when the parameter was varied from the right boundary (+) to the left boundary (-) of its range. \RefEq{eq:detector-covariance-matrix} is applicable when the parameter in question affects the event rate in a monotonic manner, which holds true for all parameters considered in this analysis.

\begin{figure*}[htbp]

 \begin{tabular}{c}
  \begin{minipage}{0.5\linewidth}
    \centering
    \includegraphics[width=1.0\textwidth]{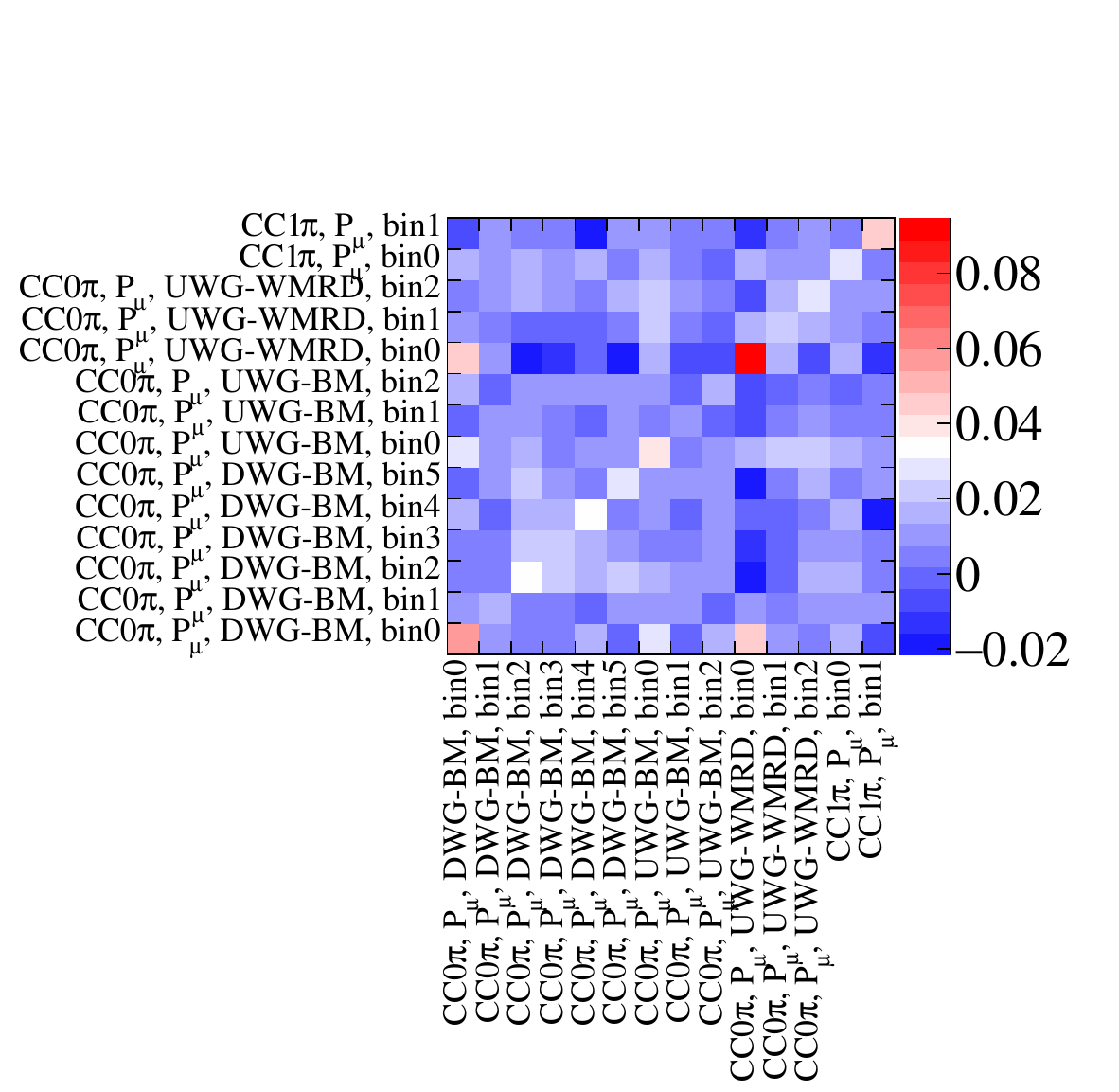} 
    \\ (a) \dwgs: muon momentum
  \end{minipage}
  \begin{minipage}{0.5\linewidth}
    \centering
    \includegraphics[width=1.0\textwidth]{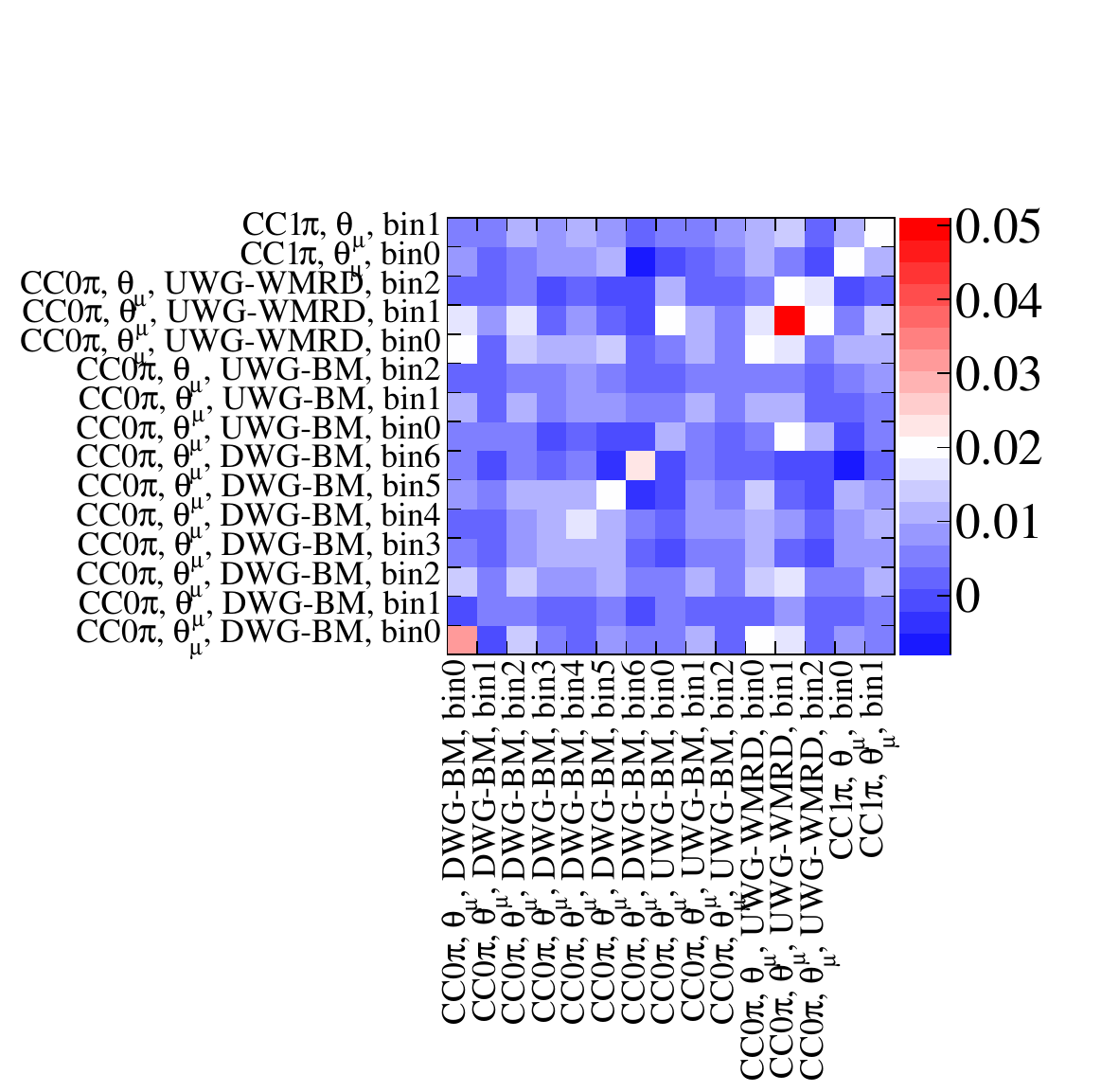} 
    \\ (b) \dwgs: muon angle
  \end{minipage}

\\

  \begin{minipage}{0.5\linewidth}
    \centering
    \includegraphics[width=1.0\textwidth]{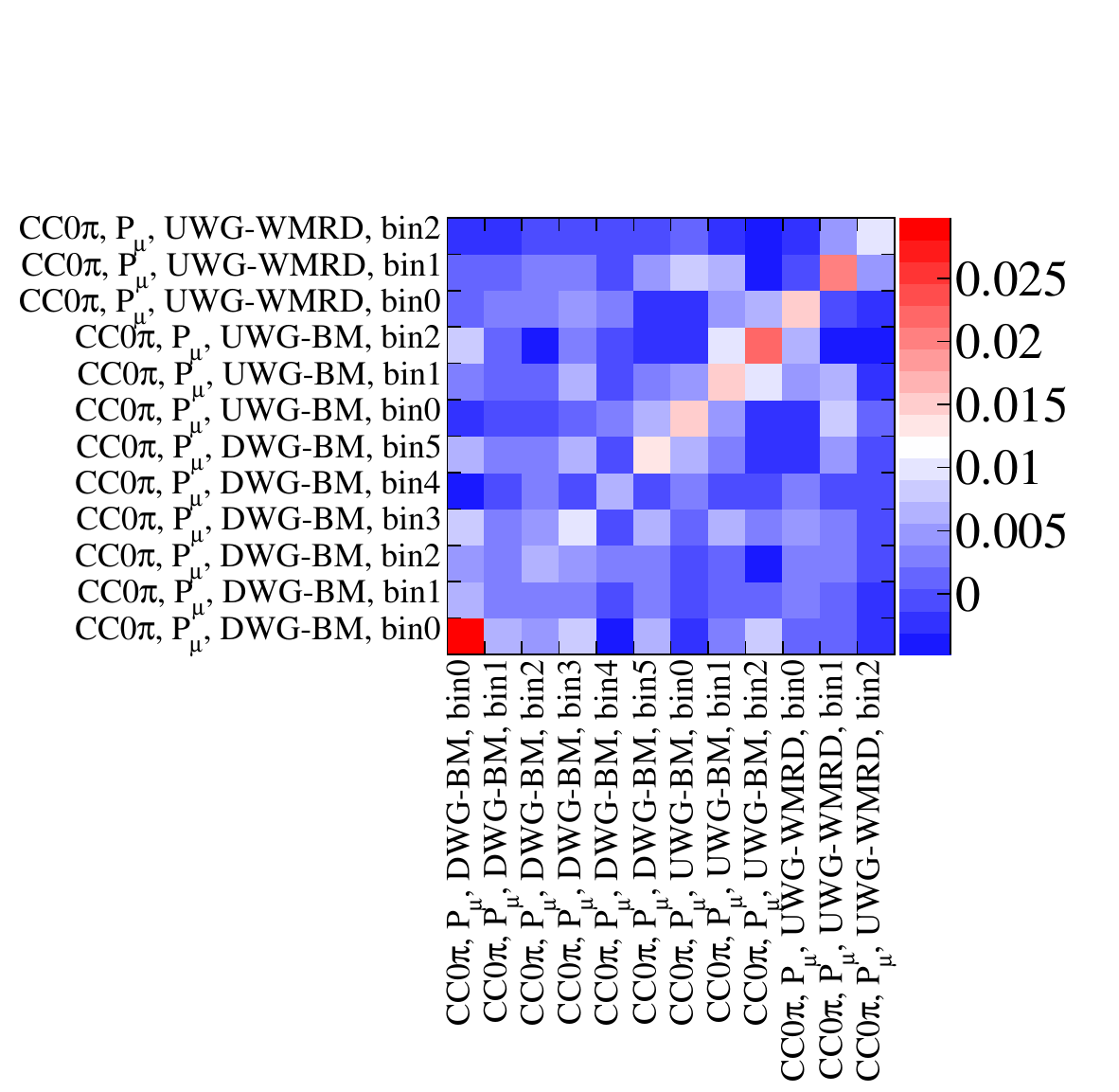} 
    \\ (c) \dpm: muon momentum
  \end{minipage}
  \begin{minipage}{0.5\linewidth}
    \centering
    \includegraphics[width=1.0\textwidth]{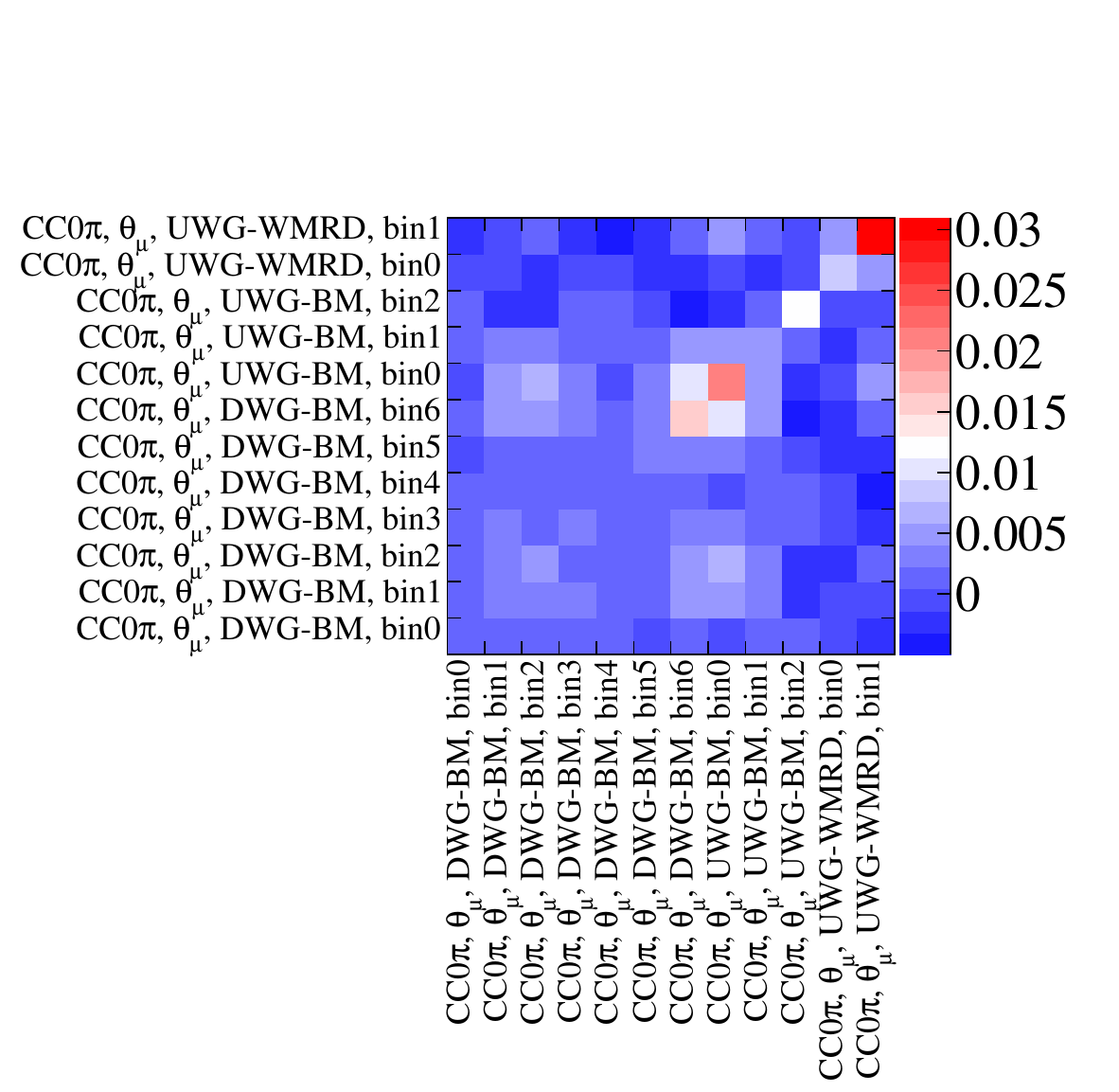} 
    \\ (d) \dpm: muon angle
  \end{minipage}

  \end{tabular}
   \caption{Covariance matrix for the detector systematic uncertainties on the number of selected events given as a fractional change for the \dwgs ((a) and (b)) and \dpm ((c) and (d)).} 
   \label{fig:DetectorCovariance}
\end{figure*}

\begin{figure*}[htbp]

 \begin{tabular}{c}
  \begin{minipage}{0.5\linewidth}
    \centering
    \includegraphics[width=\linewidth]{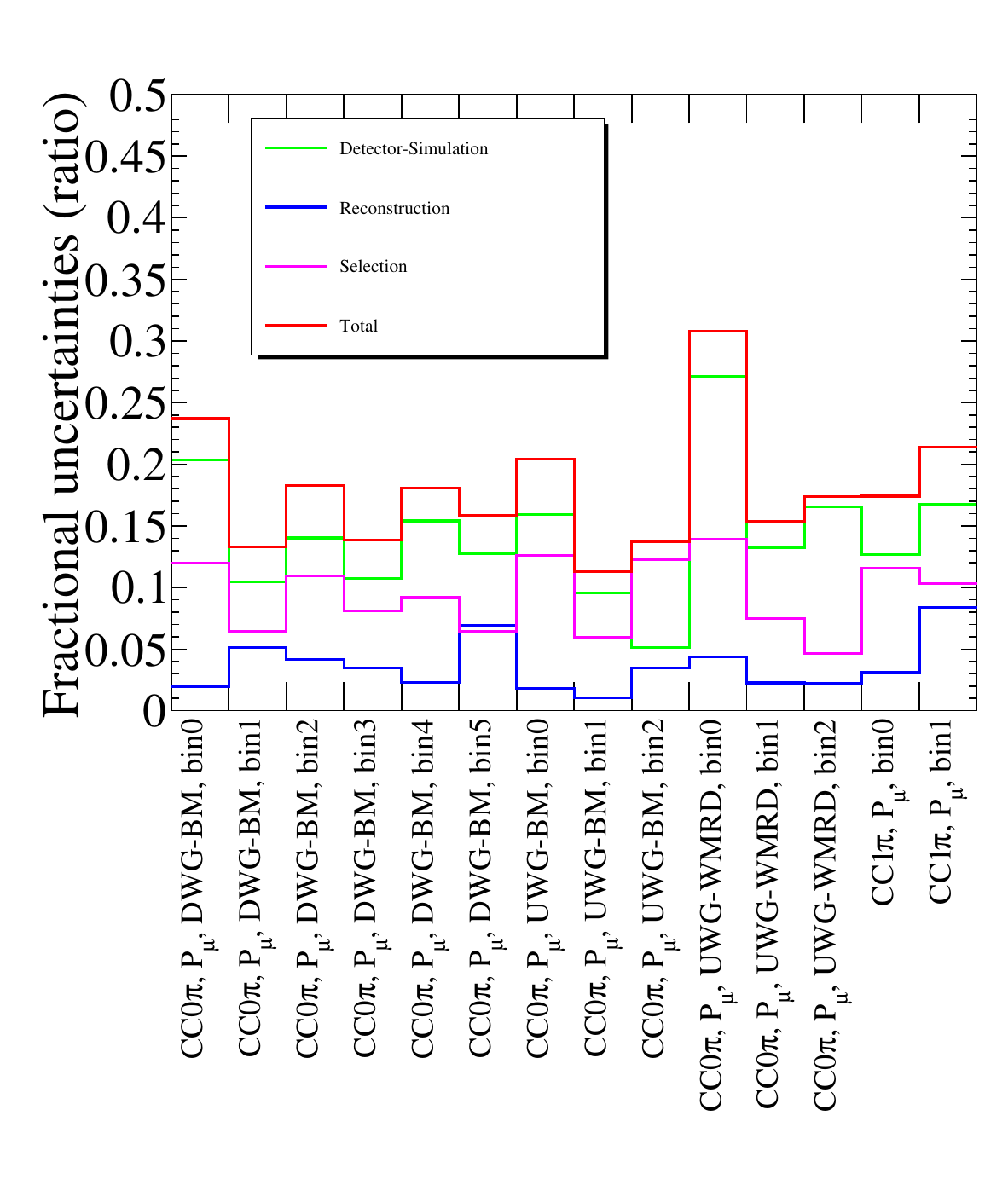} \\
      (a) \dwgs: muon momentum
  \end{minipage}
  \begin{minipage}{0.5\linewidth}
    \centering
     \includegraphics[width=\linewidth]{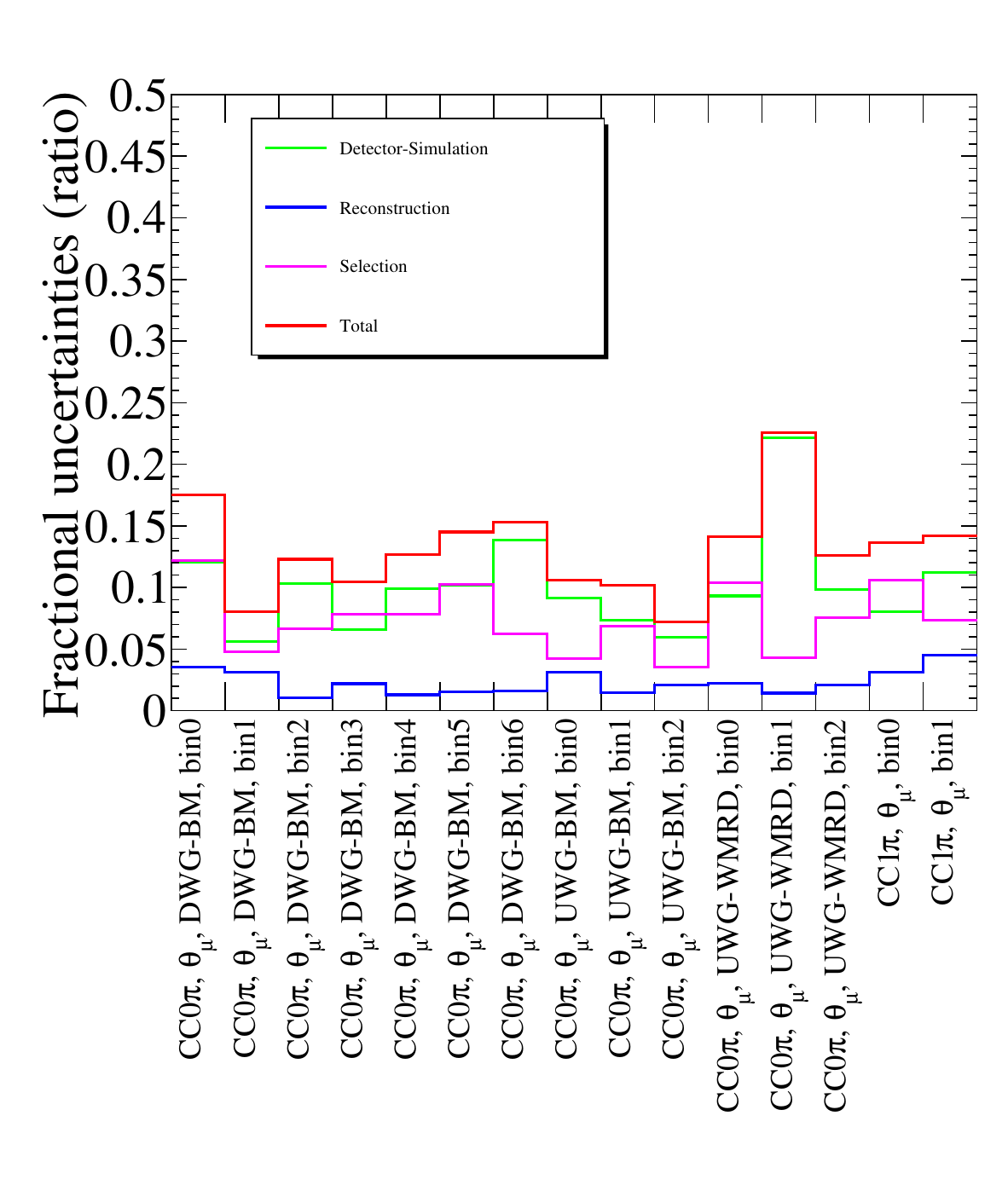} \\
      (b) \dwgs: muon angle
  \end{minipage}

\\

  \begin{minipage}{0.5\linewidth}
    \centering
    \includegraphics[width=\linewidth]{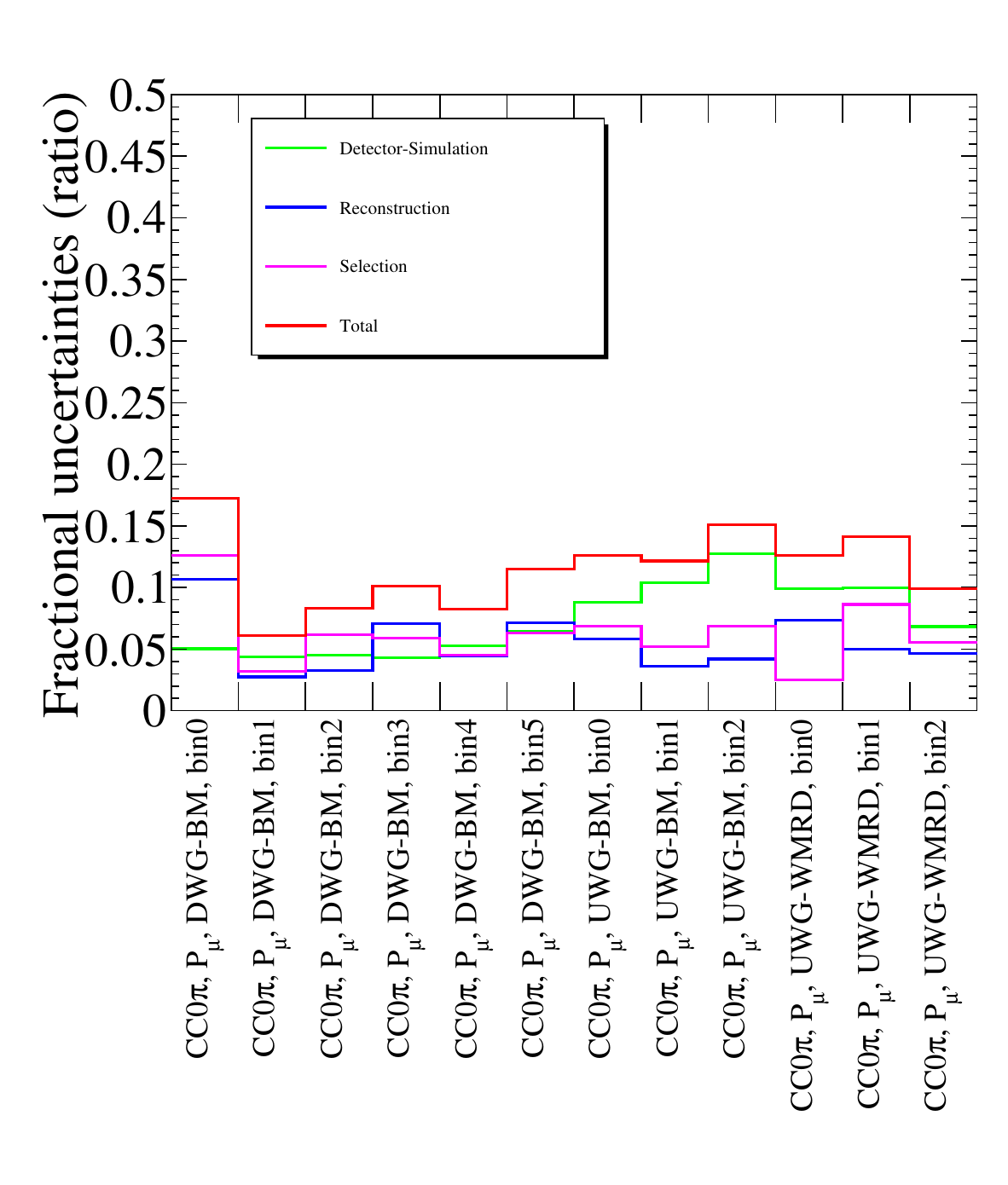} \\
      (c) \dpm: muon momentum
  \end{minipage}
  \begin{minipage}{0.5\linewidth}
    \centering
    \includegraphics[width=\linewidth]{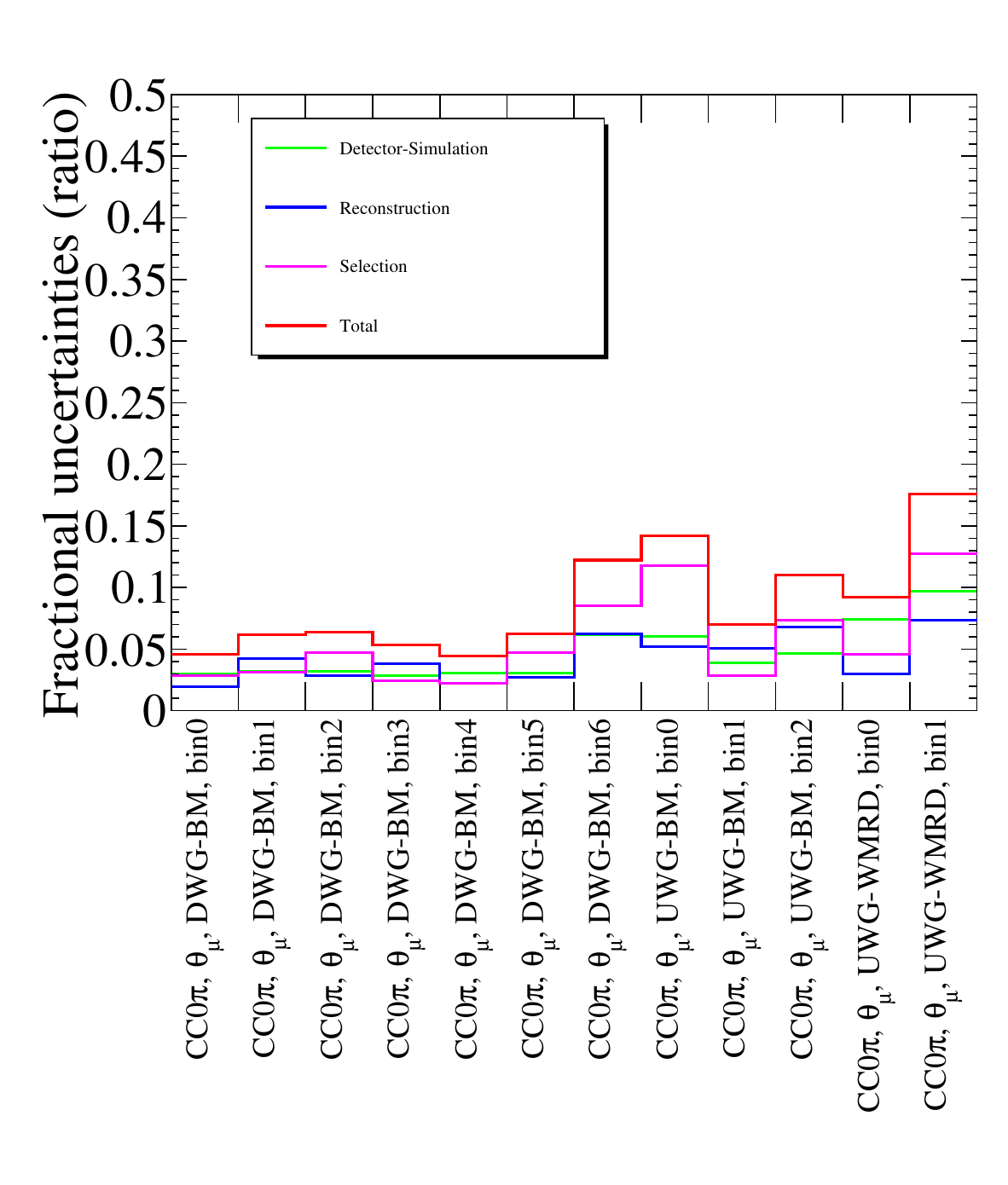} \\
      (d) \dpm: muon angle
  \end{minipage}

  \end{tabular}
   \caption{Fractional errors due to the detector systematic uncertainties on the number of selected events. Subfigures (a) and (b) correspond to the \dwgs sample, while (c) and (d) correspond to the \dpm sample. The muon momentum binning is the same as in \RefTab{tab:wagasci_babymind/differential_cross_section_momentum} and the muon angle binning is the same as in \RefTab{tab:wagasci_babymind/differential_cross_section_angle}. Systematic uncertainties arising from the neutrino interaction model are not included in this plot, only detector and reconstruction-related systematic uncertainties are included.}
  \label{fig:detector_systematics_summary_combined}
\end{figure*}

The sum of the covariance matrices of all of the parameters are shown in \RefFig{fig:DetectorCovariance}. The fractional uncertainty on the number of selected events for each category of detector systematic uncertainties are shown in  \RefFig{fig:detector_systematics_summary_combined}. The obtained  covariance matrix was used to extract the \xsec as described in \RefEq{eq:wagasci_babymind/cross_section_analysis/fitter_syst_negative_loglikelihood} of \RefSec{Sect:xsec_extraction}. \RefTab{tab:DetectorParameter} shows the detector parameters considered in this analysis. We discuss these parameters in the Section \ref{sec:detector-related-parameters}.

\begin{table*}
 \caption{List of the detector systematic parameters}

 \label{tab:DetectorParameter}
 \vspace{3mm}
 \centering
 
 \begin{ruledtabular}
 \begin{tabular}{cccc}

     Parameter                                          & Category         \\			   
     \hline     	

     Target mass                                        & Detector performance      \\
     Magnetic field                                     & Detector performance      \\
     Detector alignment                                 & Detector performance      \\
     MPPC noise                                         & Detector performance      \\
     Light yield                                        & Detector performance      \\
     Hit threshold                                      & Detector performance      \\
     Cross talk                                         & Detector performance      \\
     Scintillator inefficiency                          & Detector performance      \\
     Pion Secondary interaction                         & Detector performance      \\ 

     Two-dimensional tracking efficiency                & Track reconstruction      \\
     Threshold for track connection between detectors   & Track reconstruction      \\
     Threshold for determination of vertex              & Track reconstruction      \\

     Bunch timing selection                             & Event selection           \\
     Fiducial volume                                    & Event selection           \\
     Contained volume                                   & Event selection           \\
     Particle identification                            & Event selection           \\
     Charge identification                              & Event selection           \\
     Track-associated hit ratio                         & Event selection           \\
     Michel electron tagging                            & Event selection           \\
     Event pile up                                      & Event selection           \\     
    
 \end{tabular}
 \end{ruledtabular}
 \end{table*}

\subsubsection{Detector-Related Parameters}

\label{sec:detector-related-parameters}

Detector-related parameters encompass various factors intrinsic to the detector’s physical properties and operational conditions. 
They were tuned with actual measurements of the detector performance. The variations around the nominal value were chosen so that they cover the discrepancy between MC and the measurements.

\subsubsection{Track-Reconstruction-Related Parameters}

\label{sec:track-reconstruction-related-parameters}

The change of the detection efficiency was evaluated by varying the tolerances of each of the track-reconstruction steps and calculating a covariance matrix in a similar way as described in \RefSec{sec:detector-related-parameters}.

\subsubsection{Sample-Selection-Related Parameters}

\label{sec:sample-selection-related-parameters}

These parameters refer to criteria and thresholds used to select specific events of interest from the overall data collected by the detector. 
Below, we highlight those that contribute most significantly to the overall uncertainty.

\begin{enumerate}
    \item \textbf{Contained Volume Cut:} To ensure accurate momentum measurements, we analyzed only muons which stopped inside a defined volume of the muon range detectors. We defined variations in the contained volume for different detector modules to estimate the impact on event selection.
    \item \textbf{Fiducial Volume Cut:} We adjusted the boundaries of the fiducial volume, which was defined within the target detector mass, to account for uncertainties in selecting events which happen near the edge of the fiducial volume.
\end{enumerate}

\subsubsection{Out of Fiducial Volume background}

\label{sec:out-of-fiducial-volume-background}

We used a sand muon sample to estimate the normalization of wall-generated background events. A 13\% difference (data exceeding MC) was observed in the number of sand muon events estimated by the MC and data with a partial data set of $1.5 \times 10^{20}$ POT. This difference might arise from disparities in the density and composition of the pit wall materials \textit{in situ} and as modelled in the MC simulation. A 13\% normalization error was assigned to cover the difference between MC and data in this sample.

\section{\label{Sect:fds}Simulated data studies}

We performed simulated data studies using the selected samples and the evaluated systematic uncertainties. These studies aimed to confirm that no significant bias was introduced into the analysis and validates the fitting method and prepared inputs. Several kinds of alternative data sets were prepared and used in turn as input to the fitter. The pseudo data sets were made using the alternative interaction models or using the nominal interaction model where some parameters were tweaked from their nominal value. The psuedo data sets used are listed below.

\begin{description} 

\item[Tweaked $M_{A}^{\mathrm{RES}}$] \hfill \\
The parameter value of $M_{A}^{\mathrm{RES}}$ was changed by 20\% leaving all other parameters unchanged. 

\item[Alternative RPA model (BeRPA)] \hfill \\
The pseudo data were made with an alternate Random Phase Approximation model (BeRPA) for CCQE interactions instead of the default RPA Nieves model~\cite{RPA-Default, Nieves:2011pp}.

\item[Alternative CCRES model (low-Q$^{2}$ suppression)] \hfill \\
This alternative model accounts for the difference between data and MC reported by the MINERvA experiment~\cite{MINERVA_pion_production} and the MINOS experiment~\cite{MINOS_ND_CCQE}. The difference emerges in the low-energy transfer region (sub-GeV), where the data were smaller than the MC prediction.

\item[Alternative NEUT model] \hfill \\
One of the significant changes in the alternative NEUT model is that the $M_{A}^{\mathrm{QE}}$ value was set to 1.05 \massGeV while the nominal value was 1.21 \massGeV. Another change was the use of the Nieves 1p1h model without spectral function as a CCQE model.

\item[Signal from GENIE] \hfill \\
The pseudo data were produced by the GENIE neutrino interaction generator (version:v3, tune: $\mathrm{G18\_02b}$~\cite{GENIE}). Only the signal events were replaced by the GENIE prediction (\signal events), while the background events remained the same as in the nominal NEUT prediction.

\end{description}

We validated both inputs and fitter by confirming that the calculated differential cross sections using each psuedo data set were consistent with the nominal simulation within the systematics uncertainty.

\section{\label{Sect:results}Results}

\subsection{Integrated cross section result}

We report as our main result the flux integrated total \xsec within the phase space, $\cos\theta_\mu>0.34$ and $p_\mu>300$~MeV/c, on both targets using the angular binning samples (without the contained cut in the MRDs). The integrated \xsec values are shown in \RefTab{tab:IntegratedXSEC}. 

\begin{table*}
 \caption{Results for the integrated \xsec obtained from \Tmu and \Pmu distributions on \water and \ch targets. Different event selections were used for the angular differential \xsec and the momentum differential \xsec, which causes the $\sim1$\% difference in the \xsecs for the \water target.}

 \label{tab:IntegratedXSEC}
 \vspace{3mm}
 \centering
 
 \begin{ruledtabular}
 \begin{tabular}{cccc}

     \multirow{2}{*}{Kinematics} & \multirow{2}{*}{Target}  & \multicolumn{2}{c}{Integrated cross section(\xsecunit)}         \\			
    		                   & 				          & Data 		                                 &  MC               \\    
     \hline     	

     \multirow{2}{*}{\Tmu}          & \water            & $1.44 \pm 0.21\ (\mathrm{stat. + syst.})$         &   1.197           \\
		                          & \ch               & $1.26  \pm 0.18\ (\mathrm{stat. + syst.})$		  &   1.165           \\

     \hline	
     \multirow{2}{*}{\Pmu}          & \water	           & $1.40 \pm 0.28\ (\mathrm{stat. + syst.})$ 	       &   1.215	       \\
			                         & \ch	           & $1.24 \pm 0.17\ (\mathrm{stat. + syst.})$	       &   1.165           \\			
    
 \end{tabular}
 \end{ruledtabular}
 \end{table*}

\subsection{Differential cross section result}

\begin{figure*}[htbp]

 \begin{tabular}{c}
  \begin{minipage}{0.5\linewidth}
    \centering
    \includegraphics[width=1.0\textwidth]{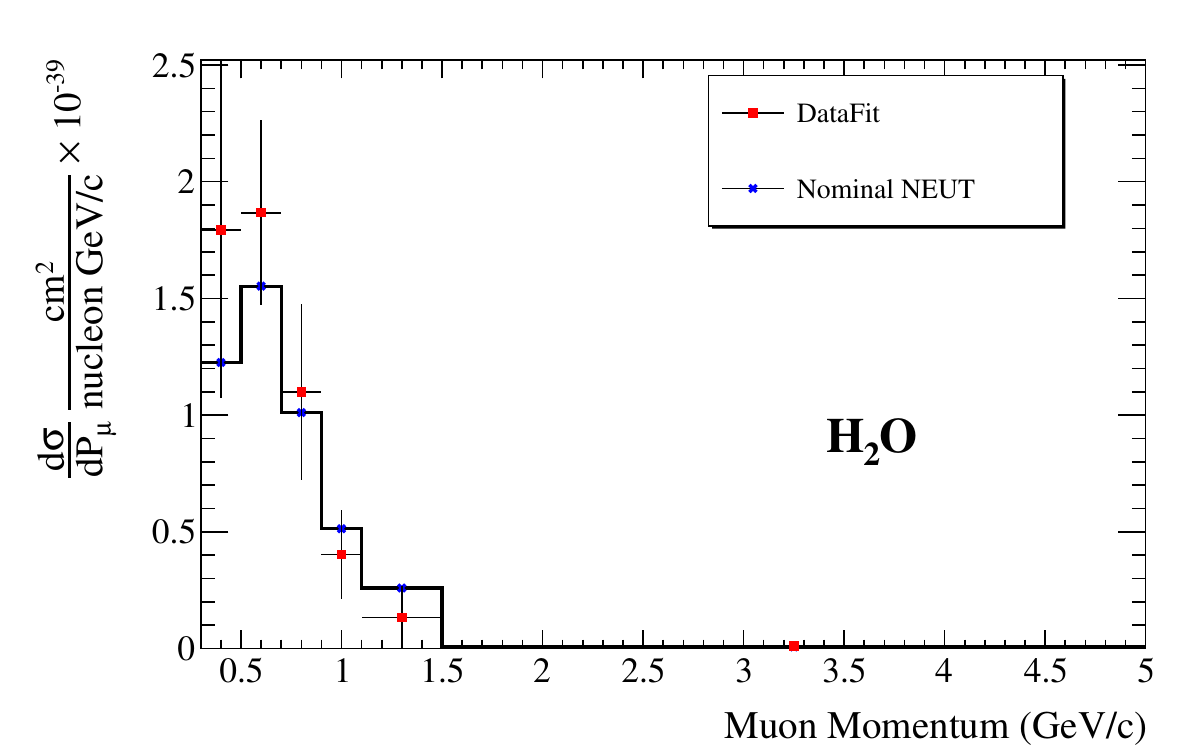} 
  \end{minipage}

  \begin{minipage}{0.5\linewidth}
    \centering
    \includegraphics[width=1.0\textwidth]{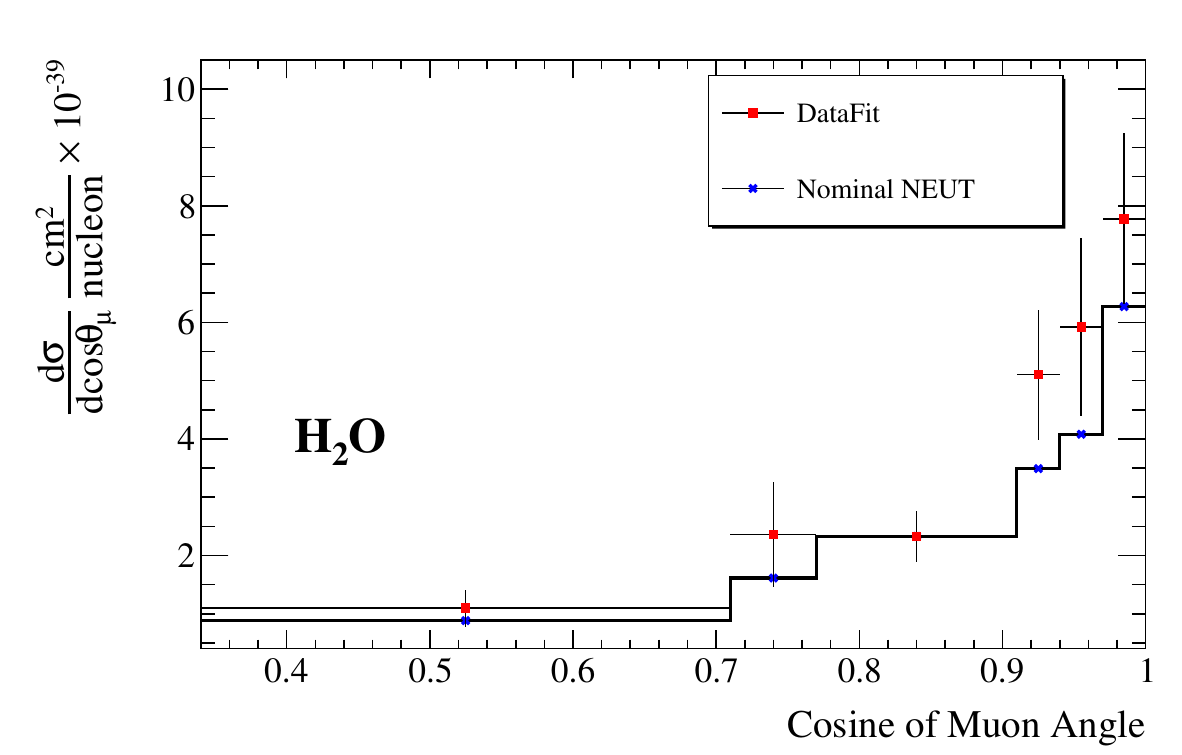} 
  \end{minipage}
  
\\

  \begin{minipage}{0.5\linewidth}
    \centering
    \includegraphics[width=1.0\textwidth]{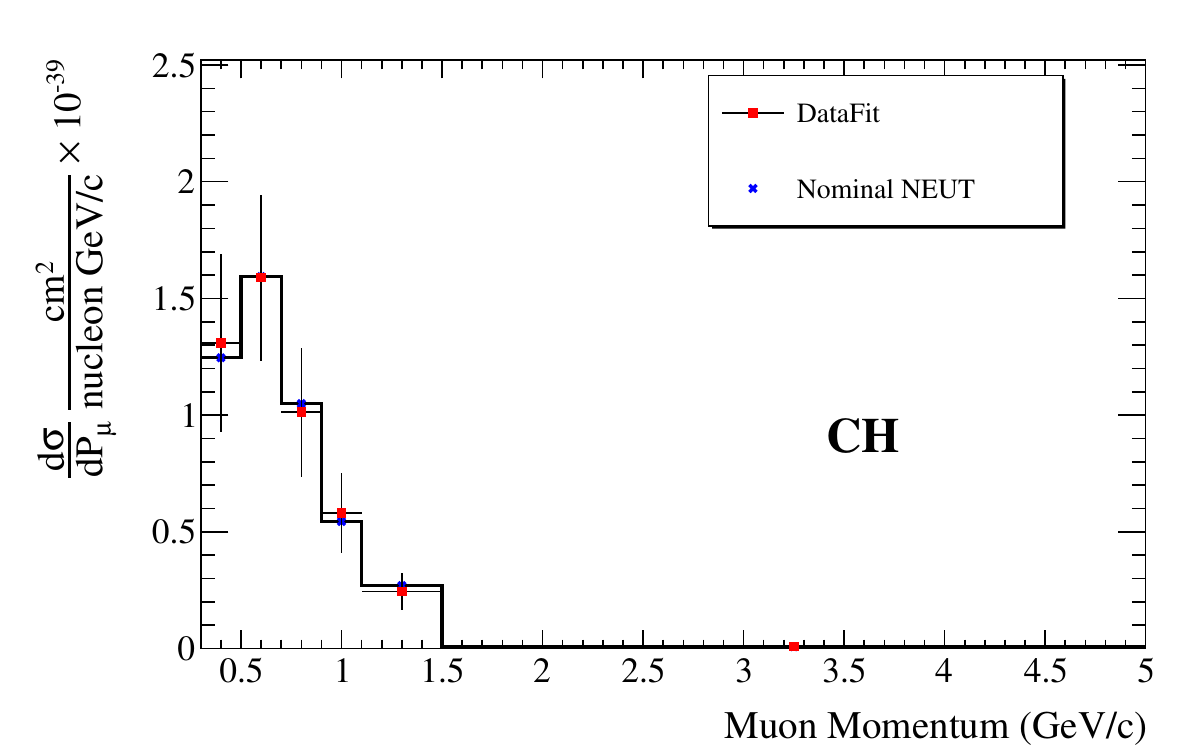} 
  \end{minipage}

  \begin{minipage}{0.5\linewidth}
    \centering
    \includegraphics[width=1.0\textwidth]{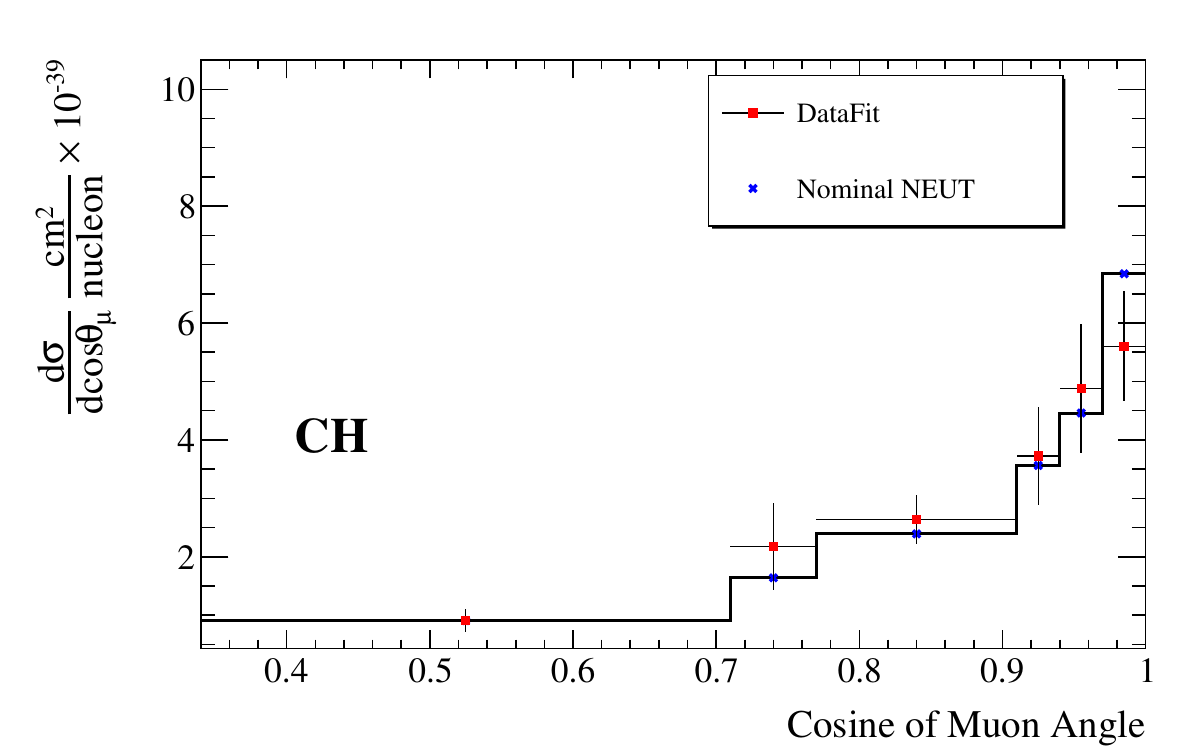} 
  \end{minipage}

  \end{tabular}
   \caption{Differential \xsec result. The plots on the top refer to \water target and the ones on the bottom to the \ch target}
   \label{fig:DifXSEC}
\end{figure*}

The results for the differential \xsec in momentum (angle) bins are shown in \RefTab{tab:wagasci_babymind/differential_cross_section_momentum} (\RefTab{tab:wagasci_babymind/differential_cross_section_angle}). The errors include both statistical and systematic contributions. The corresponding plots are shown in \RefFig{fig:DifXSEC}.
The differential \xsec{s} on the \water and \ch targets are consistent within errors with the MC prediction in most of the bins. 
The \water and \ch measurements are shown in \RefFig{fig:DifXSEC}.

\begin{table*} 
 \centering
  \caption{The results of the differential \xsec measurement as a function of the muon momentum in units of $\mathrm{(cm^{2})} / (\mathrm{nucleon} \cdot \mathrm{GeV/c})$.}
  \label{tab:wagasci_babymind/differential_cross_section_momentum}
   \vspace{3mm}

 \begin{ruledtabular}
 \begin{tabular}{c|c|c|c|c|c|c}
    
    Target Material	& Range (GeV/c) &  \xsec ($10^{-39}$)   & error ($10^{-39}$)    &Rel. error (\%)   & MC ($10^{-39}$)    & Data/MC				\\
    \hline

	\multirow{7}{*}{\water}
    				& 0.3 - 0.5					& 1.794		& 0.707		& 39.4    & 1.225 & 1.464						\\
    				& 0.5 - 0.7					& 1.867		& 0.388		& 20.8    & 1.553 & 1.202						\\
    				& 0.7 - 0.9					& 1.099		& 0.365		& 33.2    & 1.010 & 1.089						\\
    				& 0.9 - 1.1					& 0.404		& 0.184		& 45.5    & 0.513 & 0.787						\\
    				& 1.1 - 1.5					& 0.133		& 0.138		& 103.5   & 0.259 & 0.514						\\
    				& 1.5 - 30.0					& 0.011		& 0.003		& 25.5    & 0.007 & 1.623					    \\

    \hline
    \hline      
  
 	\multirow{7}{*}{\ch}
    				& 0.3 - 0.5					& 1.309		& 0.373		& 28.5    & 1.227 & 1.067						\\
    				& 0.5 - 0.7					& 1.589		& 0.359		& 22.6    & 1.570 & 1.012 						\\
    				& 0.7 - 0.9					& 1.013	    & 0.283		& 28.0    & 1.034 & 0.980						\\
    				& 0.9 - 1.1					& 0.581		& 0.168		& 28.9    & 0.536 & 1.084						\\
    				& 1.1 - 1.5					& 0.244		& 0.077		& 31.5    & 0.266 & 0.919						\\
    				& 1.5 - 30.0					& 0.009		& 0.002		& 21.1    & 0.008 & 1.089						\\

  \end{tabular}
  \end{ruledtabular}

 \end{table*}
 
 \begin{table*}
 \centering
  \caption{The results of the differential \xsec measurement as a function of the cosine of the muon scattering angle in units of $\mathrm{cm^{2}} / \mathrm{nucleon}$.}

  \label{tab:wagasci_babymind/differential_cross_section_angle}
  \vspace{3mm}

  \begin{ruledtabular}
  \begin{tabular}{c|c|c|c|c|c|c}

    Target Material	& Range in cosine angle/angular bins	&  \xsec ($10^{-39}$)		& error ($10^{-39}$)	& Rel. error (\%)	& MC ($10^{-39}$) & Data/MC	\\
    \hline
    
	\multirow{7}{*}{\water}
    				& 0.34 - 0.71			& 1.097				& 0.311 		& 28.4 			& 0.884  & 1.241 		\\
    				& 0.71 - 0.77			& 2.358				& 0.906  		& 38.4			& 1.613  & 1.462			\\
    				& 0.77 - 0.91			& 2.326 			& 0.440 		& 18.9			& 2.329  & 0.998      			\\
    				& 0.91 - 0.94			& 5.103 			& 1.095			& 21.5			& 3.491  & 1.462  			\\
    				& 0.94 - 0.97			& 5.922 			& 1.519 		& 25.7			& 4.080  & 1.451  			\\
    				& 0.97 - 1.00		    & 7.776			    & 1.496			& 19.2			& 6.269  & 1.240 			\\
  						   
    \hline
    \hline      
 
	\multirow{7}{*}{\ch}
    				& 0.34 - 0.71			& 0.911 			& 0.184 		& 20.2		   & 0.895 & 1.018 			\\
    				& 0.71 - 0.77			& 2.176 			& 0.718 		& 33.0		   & 1.615 & 1.347 			\\
    				& 0.77 - 0.91			& 2.639				& 0.402 		& 15.2		   & 2.357 & 1.133		 	\\
    				& 0.91 - 0.94		    & 3.724 			& 0.819 		& 22.0		   & 3.509 & 1.067			\\
    				& 0.94 - 0.97			& 4.882				& 1.087 		& 22.3		   & 4.392 & 1.196 			\\
    				& 0.97 - 1.00			& 5.599 			& 0.947 		& 16.9		   & 6.739 & 0.893 			\\

  \end{tabular}
  \end{ruledtabular}
  
 \end{table*}

\subsection{Comparison with different interaction models}

The reported measurement results were compared to predictions from MC generators GENIE and NEUT to see how well they agree with our measurements. The agreement between the measurement and models was quantified by the value of $\chi^{2}$ defined by 

\begin{equation*}
\label{eq:chi_square_model_comparison}
\begin{split}
\chi^{2} = \sum_{ij}^{N} \left( \left(\frac{d\sigma}{dx}\right)_{i,\mathrm{data}} - \left(\frac{d\sigma}{dx}\right)_{i, \mathrm{model}} \right) {\bf{V}_{ij}}^{-1} \\
\left( \left(\frac{d\sigma}{dx}\right)_{j, \mathrm{data}} - \left(\frac{d\sigma}{dx}\right)_{j, \mathrm{model}} \right)
\end{split}
\end{equation*}
where $N$ is the number of \xsec bins, $i, j$ are the $i$th and $j$th kinematic bin respectively and $\bf{V}_{ij}$ is the covariance matrix. The MC generators used in this study were: 

\begin{description}

\item[NEUT alternative version] \hfill \\ 
We used NEUT with a different choice of interaction models. One difference was in the treatments of the nuclear ground state. The alternative model described the ground state with a Global Fermi Gas (GFG) model~\cite{NEUT_update}. The CCQE model was the Nieves 1p1h model without the RPA correction, with $M_{A}^{\mathrm{QE}}$ set to 1.03~GeV, compared to 1.21~GeV in the original NEUT model. The expected number of CCQE events decreased by about 15\% in this model.

\item[GENIE] \hfill \\
GENIE allows for  various combinations of interaction models to be simulated and tested. In this study the CCQE model was the same as NEUT but with a different $M_{A}^{\mathrm{QE}}$ (0.99~GeV). Final State Interactions were modeled with the hN model~\cite{GENIE_FSI} and the treatment of the ground state of a nucleus was also different. The NEUT model utilized the Benhar SF model, whereas GENIE adopts a Relativistic Fermi Gas (RFG) model. 

\end{description}

\begin{figure*}  
 \begin{tabular}{c}
  \begin{minipage}{0.5\linewidth}
    \centering
    \includegraphics[width=1.0\textwidth]{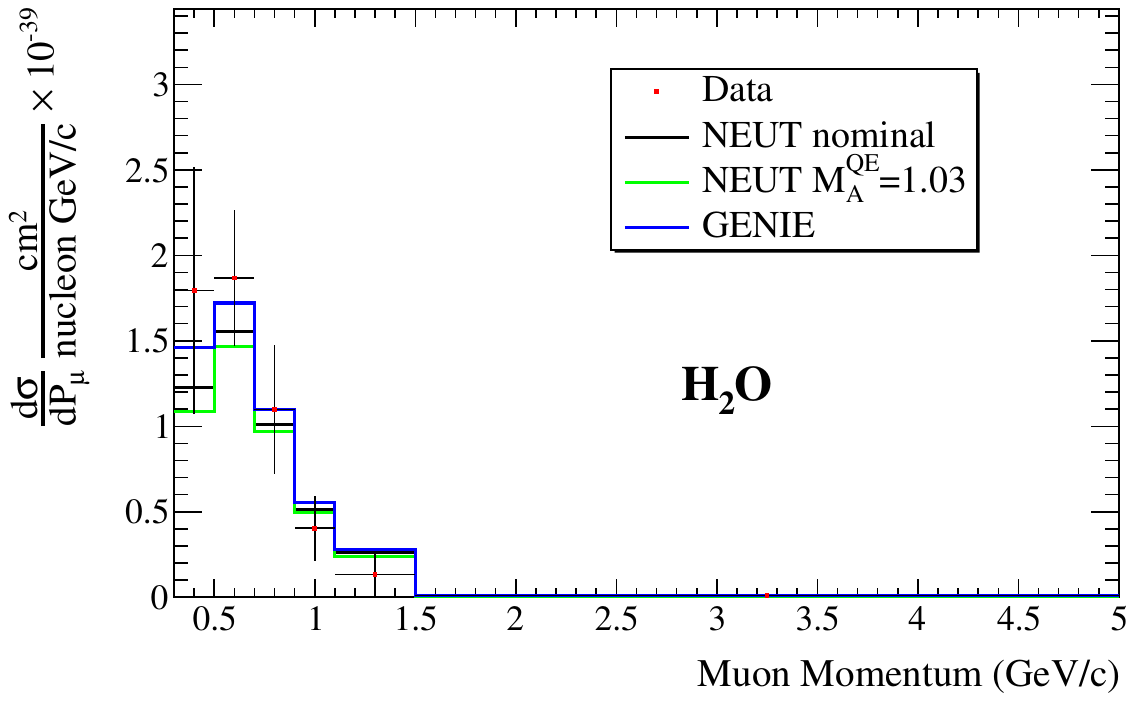}
  \end{minipage}
  \begin{minipage}{0.5\linewidth}
    \centering
    \includegraphics[width=1.0\textwidth]{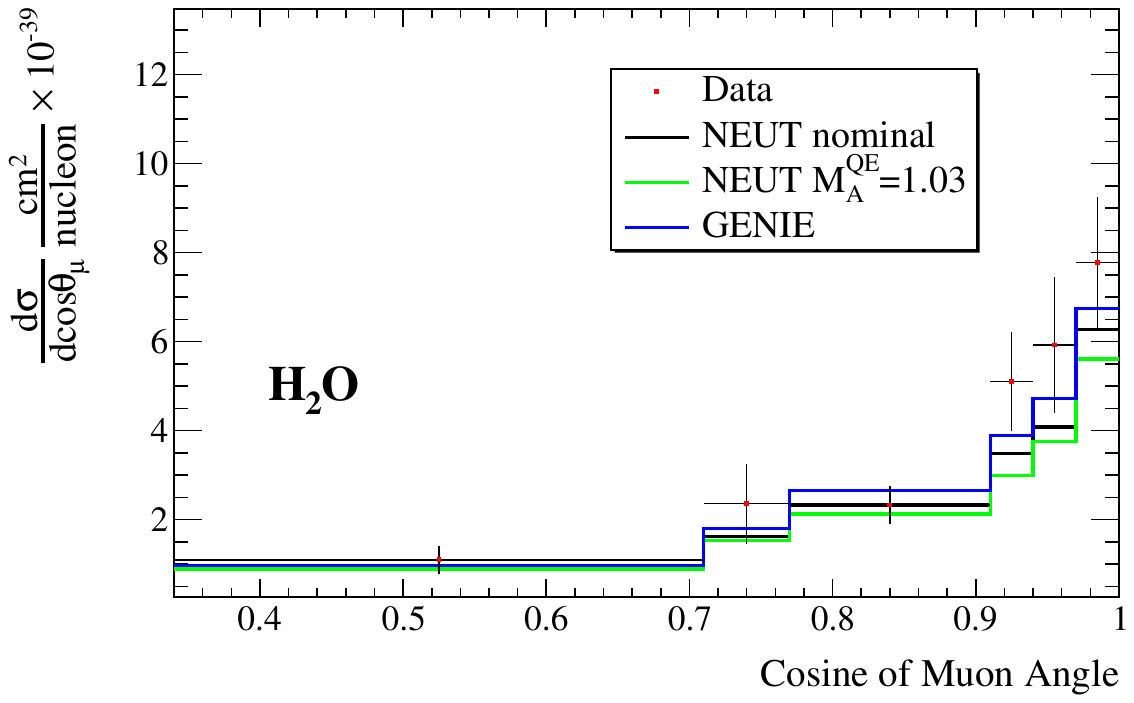}
  \end{minipage}  
  
\\
  
  \begin{minipage}{0.5\linewidth}
    \centering
    \includegraphics[width=1.0\textwidth]{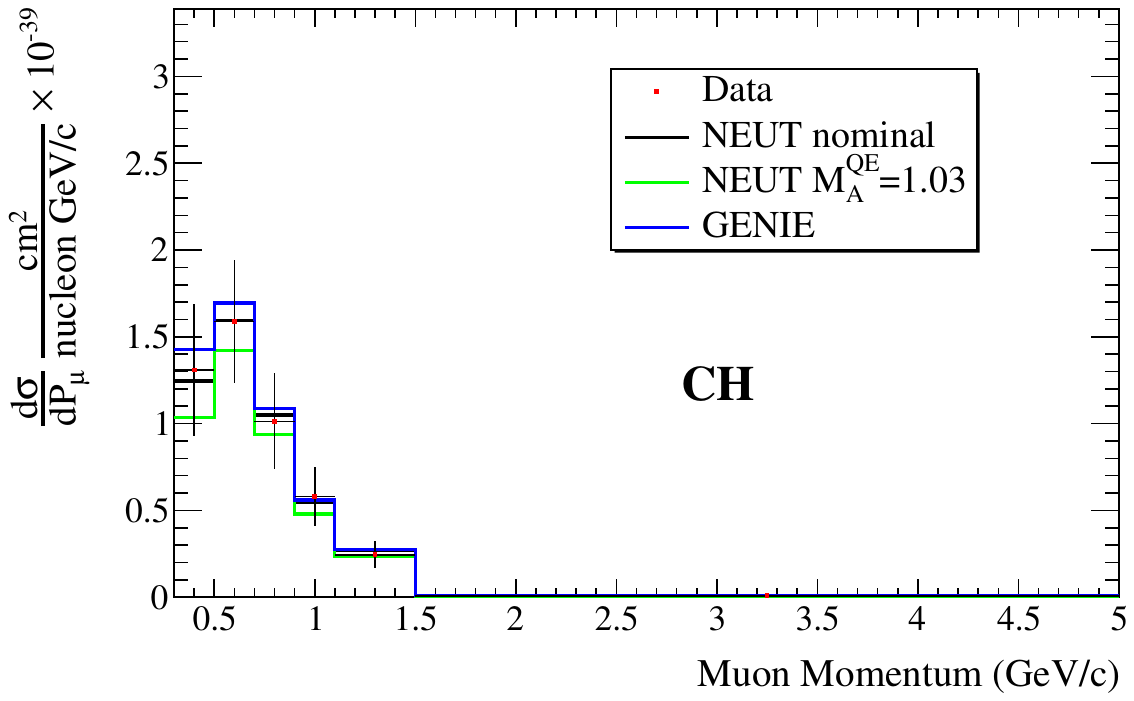}
  \end{minipage}
  \begin{minipage}{0.5\linewidth}
    \centering
    \includegraphics[width=1.0\textwidth]{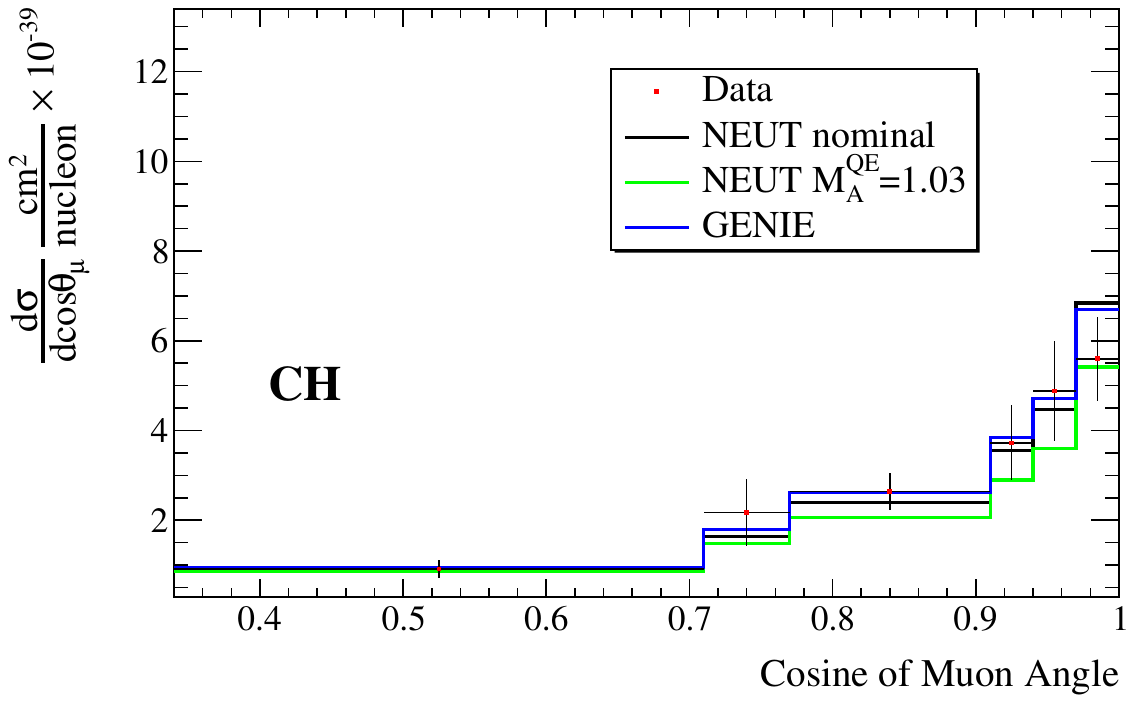}
  \end{minipage}
\end{tabular}
    \caption[Extracted differential \xsec compared to nominal MC and alternative MC models as a function of $P_{\mu}$ and $\cos{\theta_{\mu}}$]
    {Extracted differential \xsec compared to nominal MC and alternative MC models (see the text) as a function of momentum (left) and cosine of muon angle (right). The \water results are shown on the top and \ch results on the bottom. The color corresponds to each model; black: the nominal NEUT version, green: the alternative version of NEUT with $\mathrm{M}_{A}^{\mathrm{QE}}$ set to 1.03~\massGeV, blue: GENIE, red: post-fit ND tune. }
  \label{fig:wagasci_babymind/differential_cross_section_model_comparison}
\end{figure*}

The calculated $\chi^{2}$ values are shown in \RefTab{tab:agreement_model_comparisons}. Model prediction for the differential cross section are shown in \RefFig{fig:wagasci_babymind/differential_cross_section_model_comparison} together with the data points. The number of degrees of freedom is 12 for when binning the differential cross section by both momentum and cosine of the angle. No particular model is rejected or favored based on the results and all are within the stated experimental uncertainty. This result is consistent with the simulated data studies, where simulated data were generated by alternative NEUT and GENIE models, showing their agreement with the nominal NEUT model.

\begin{table*}
  \centering
  \caption{Comparison between the results obtained in this paper and those simulated with various models as measured by the $\chi^{2}$, see Eq. \ref{eq:chi_square_model_comparison}. The number of degrees of freedom is 12.}
  \label{tab:agreement_model_comparisons}
  \begin{ruledtabular}
  \begin{tabular}{c|ccc|ccc}
    Model								& \multicolumn{3}{c}{$\chi^{2}$/NDF in momentum binning} 			& \multicolumn{3}{c}{$\chi^{2}$/NDF in angle binning	}		\\
   									& \ch			& \water		& total						&  \ch			& \water 			& total  		\\
    
    \hline
    NEUT nominal						& 0.493/12		& 5.619/12	& 6.673/12						& 4.040/12			& 5.082/12			& 9.005/12		\\   
    \hline
    NEUT alternative version				& 1.827/12		& 5.550/12		& 7.611/12						& 4.279/12			& 6.615/12			& 11.29/12		\\
    \hline			
    GENIE			 					& 1.086/12		& 6.030/12		& 7.667/12						& 2.199/12			& 4.783/12			& 6.955/12		
  \end{tabular}
 \end{ruledtabular}
\end{table*}

\section{\label{Sect:Conclusions}CONCLUSIONS}

In this paper we presented the measurement of the $\nu_{\mu}$ \xsec on \water and \ch with the \wb using $2.96\times10^{20}$ POT of data collected during J-PARC Run 10 (November 2019 to February 2020) and Run 11 (March to April 2021). This is the first $\nu_{\mu}$ \xsec measurement of neutrino interactions at 1.5 degrees off-axis with T2K neutrino mode data. The results are consistent with the neutrino interaction model used for the T2K neutrino oscillation measurements. The data-MC agreement of the differential cross section is better in the momentum binning than in the angle binning with the current data set, although the difference is not statistically significant. This study has laid the foundation for the next-generation of analyses with \wb. Future planned operations will further increase the datasets of \wb and we therefore expect future results to report measurements of neutrino interaction cross sections in water and carbon with the \wb with even greater precision, including double differential measurements on muon kinematics for events with and without charged pions. The measurement presented in this article is the first step towards using the data of the \wb to benefit future T2K neutrino oscillation analyses by validating the required neutrino interaction models.

\FloatBarrier

\appendix*
\section{\label{Sect:Appendix}}

\begin{acknowledgments}

The T2K collaboration would like to thank the J-PARC staff for superb accelerator performance. We thank the CERN NA61/SHINE Collaboration for providing valuable particle production data. We acknowledge the support of MEXT,   JSPS KAKENHI (JP16H06288, JP18K03682, JP18H03701, JP18H05537, JP19J01119, JP19J22440, JP19J22258, JP20H00162, JP20H00149, JP20J20304, JP24K17065) and bilateral programs (JPJSBP120204806, JPJSBP120209601),  Japan; NSERC, the NRC, and CFI, Canada; the CEA and CNRS/IN2P3, France; the Deutsche Forschungsgemeinschaft (DFG, German Research Foundation) 397763730, 517206441, Germany; the NKFIH (NKFIH 137812 and TKP2021-NKTA-64), Hungary; the INFN, Italy; the Ministry of Science and Higher Education (2023/WK/04) and the National Science Centre (UMO-2018/30/E/ST2/00441, UMO-2022/46/E/ST2/00336 and UMO-2021/43/D/ST2/01504), Poland;  the RSF (RSF 24-12-00271) and the Ministry of Science and Higher Education, Russia;  MICINN  (PID2022-136297NB-I00 /AEI/10.13039/501100011033/ FEDER, UE, PID2024-157541NB-I00 (UAM) and PID2023-146401NB-I00 (US), Severo Ochoa Centres of Excellence Programme 2025-2029 (CEX2024001442-S),  Government of Andalucia (FQM160) and the University of Tokyo ICRR's Inter-University Research Program FY2025 Ref. J1, and ERDF and European Union (UAM: H2020-MSCA-RISE-GA872549- SK2HK) and NextGenerationEU funds (PRTR-C17.I1) and  Generalitat de Catalunya (AGAUR 2021-SGR-01506, CERCA program) University of Seville grant (RYC2022-035203-I funded by MICIU/AEI/10.13039/501100011033, "ERDF a way of making Europe" and FSE+, Ayudas "Atracción de Investigadores con Alto Potencial". Ref. VIIPPIT-2025, and Secretariat for Universities and Research of the Ministry of Business and Knowledge of the Government of Catalonia and the European Social Fund (2022FI\_B 00336), Spain; the SNSF and SERI (200021\_185012, 200020\_188533, 20FL21\_186178I), Switzerland; the STFC and UKRI, UK; the DOE, USA; and NAFOSTED (103.99-2023.144,IZVSZ2.203433), Vietnam. We also thank CERN for the UA1/NOMAD magnet, DESY for the HERA-B magnet mover system, the BC DRI Group, Prairie DRI Group, ACENET, SciNet, and CalculQuebec consortia in the Digital Research Alliance of Canada, and GridPP in the United Kingdom, and the CNRS/IN2P3 Computing Center in France and NERSC (HEP-ERCAP0028625). In addition, the participation of individual researchers and institutions has been further supported by funds from the ERC (FP7), “la Caixa” Foundation  (ID 100010434, fellowship code LCF/BQ/IN17/11620050), the European Union’s Horizon 2020 Research and Innovation Programme under the Marie Sklodowska-Curie grant agreement numbers 713673 and 754496, and H2020 grant numbers  RISE-GA822070-JENNIFER2 2020 and RISE-GA872549-SK2HK, the Horizon Europe Marie Sklodowska-Curie Staff Exchange project JENNIFER3 grant 101183137; the JSPS, Japan; the Royal Society, UK; French ANR grant number ANR-19-CE31-0001 and ANR-21-CE31-0008; and  Sorbonne Université Emergences programmes; the SNF Eccellenza grant number PCEFP2\_203261;  the VAST-JSPS (No. QTJP01.02/20-22);  and the DOE Early Career programme, USA. For the purposes of open access, the authors have applied a Creative Commons Attribution license to any Author Accepted Manuscript version arising.

Definitions

ANR:  French National Research Agency 
CEA: Commissariat à l’Energie Atomique (France) http://www.cea.fr
CERCA: Centres de Recerca de Catalunya (Spain) https://cerca.cat/en/
CERN: European Organization for Nuclear Research (derived from the name "Conseil Européen pour la Recherche Nucléaire”)
CFI: Canada Foundation for Innovation (Canada)
CNRS/IN2P3: Centre National de la Recherche Scientifique—Institut National de Physique Nucléaire et de Physique des Particules (France) http://www.in2p3.fr
DFG: Deutsche Forschungsgemeinschaft [German Research Foundation] (Germany) http://www.dfg.de/en/
DESY: Deutsches Elektronen-Synchrotron, (Germany)
DOE: Department of Energy (United States) https://energy.gov
ERC: European Research Council (European Union) https://erc.europa.eu
ERDF: European Regional Development Fund (European Union) http://ec.europa.eu/regional\_policy/en/funding/erdf/
GridPP: Grid for Particle Physics https://www.gridpp.ac.uk
HERA-B: Hadron-Elektron-Ring-Anlage-B, "Hadron Electron Ring Facility for Bs“, http://www-hera-b.desy.de
INFN: Istituto Nazionale di Fisica Nucleare [National Institute for Nuclear Physics] (Italy) http://home.infn.it/en/
J-PARC: Japan Proton Accelerator Research Complex (Japan)
JSPS: Japan Society for the Promotion of Science (Japan) http://www.jsps.go.jp/english/
MES: Ministry of Education and Science  http://dx.doi.org/10.13039/501100003443 (Russia) 
MEXT: Ministry of Education, Culture, Sports, Science and Technology (Japan)
MICINN: Ministerio de Ciencia, Innovacion y Universidades, (Spain) http://www.ciencia.gob.es
NA61: North Area experiment 61 (CERN)
NERSC: National Energy Research Scientific Computing Center https://www.nersc.gov/
NCN: Narodowe Centre Nauki [National Science Centre] (Poland) https://www.ncn.gov.pl
NII: National Institute of Informatics (Japan) http://www.nii.ac.jp/en/
NRC: National Research Council (Canada)
NSERC: Natural Sciences and Engineering Research Council (Canada)
NKFIH: National Research, Development and Innovation Office (Hungary)  https://nkfih.gov.hu/palyazoknak
RFBR: Russian Foundation for Basic Research, (Russia) 
Royal Society:  http://dx.doi.org/10.13039/501100000288 (United Kingdom) https://royalsociety.org
RSF: Russian Science Foundation (Russia)
SciNet: (Canada) https://www.scinethpc.ca
SINET4: Science Information Network 4 (Japan) http://w4a.sinet.ad.jp
SNSF: Swiss National Science Foundation  http://dx.doi.org/10.13039/501100001711 (Switzerland) http://www.snf.ch/en/Pages/default.aspx
SERI: State Secretariat for Education, Research and Innovation  http://dx.doi.org/10.13039/501100007352 (Switzerland) https://www.sbfi.admin.ch/sbfi/en/home.html
STFC: Science and Technology Facilities Council, (United Kingdom)
UKRI: UK Research and Innovation, (United Kingdom) https://www.ukri.org/
UA1: Underground Area experiment 1
Digital Research Alliance of Canada : https://alliancecan.ca

ADDITIONAL INFO

Alfred P. Sloan Foundation, FundRef ID http://dx.doi.org/10.13039/100000879 (United States/New York) 
Ministry of Education, Culture, Sports, Science and Technology, FundRef ID http://dx.doi.org/10.13039/501100001700 (Japan)
Natural Sciences and Engineering Research Council of Canada, FundRef ID http://dx.doi.org/10.13039/501100000038 (Canada) 
National Research Council Canada, FundRef ID http://dx.doi.org/10.13039/501100000046 (Canada/Ontario) 
Canada Foundation for Innovation, FundRef ID http://dx.doi.org/10.13039/501100000196 (Canada/Ontario) 
Centre National de la Recherche Scientifique, FundRef ID  http://dx.doi.org/10.13039/501100004794 (France)
Commissariat à l'Énergie Atomique et aux Énergies Alternatives, FundRef ID http://dx.doi.org/10.13039/501100006489 (Republic of France) 
Deutsche Forschungsgemeinschaft, FundRef ID http://dx.doi.org/10.13039/501100001659 (Federal Republic of Germany) 
Department of Energy, FundRef ID http://dx.doi.org/10.13039/100000015 (United States/District of Columbia) 
European Regional Development Fund, FundRef ID http://dx.doi.org/10.13039/501100008530 (Kingdom of Belgium) 
Istituto Nazionale di Fisica Nucleare, FundRef ID http://dx.doi.org/10.13039/501100004007 (Repubblica Italiana) 
H2020 European Research Council, FundRef ID http://dx.doi.org/10.13039/100010663 (Kingdom of Belgium)
Narodowe Centrum Nauki, FundRef ID http://dx.doi.org/10.13039/501100004281 (Republic of Poland) 
Ministerstwo Nauki i Szkolnictwa Wy?szego, FundRef ID http://dx.doi.org/10.13039/501100004569 (Republic of Poland) 
Ministry of Education and Science, FundRef ID  http://dx.doi.org/10.13039/501100003443 (Russia)
Ministerio de Economía y Competitividad, FundRef ID http://dx.doi.org/10.13039/501100003329 (Kingdom of Spain)
Russian Science Foundation, FundRef ID http://dx.doi.org/10.13039/501100006769 (Russian Federation) 
Russian Foundation for Basic Research, FundRef ID http://dx.doi.org/10.13039/501100002261 (Russian Federation) 
Royal Society, FundRef ID  http://dx.doi.org/10.13039/501100000288 (United Kingdom of Great Britain and Northern Ireland)
Science and Technology Facilities Council, FundRef ID http://dx.doi.org/10.13039/501100000271 (United Kingdom of Great Britain and Northern Ireland)
State Secretariat for Education, Research and Innovation, FundRef ID  http://dx.doi.org/10.13039/501100007352 (Switzerland)
Swiss National Science Foundation, FundRef ID  http://dx.doi.org/10.13039/501100001711 (Switzerland)
Canada Foundation for Innovation (Canada) https://www.innovation.ca
Japan Society for the Promotion of Science (Japan) http://www.jsps.go.jp/english/

\end{acknowledgments}

\bibliography{apssamp}

@preamble{
 "\providecommand{\noopsort}[1]{}" 
 # "\providecommand{\singleletter}[1]{#1}%" 
}

@article{Kudenko:2008ia,
    author = "Kudenko, Yury",
    editor = "Buzulutskov, A.",
    collaboration = "T2K",
    title = "{The Near neutrino detector for the T2K experiment}",
    eprint = "0805.0411",
    archivePrefix = "arXiv",
    primaryClass = "physics.ins-det",
    doi = "10.1016/j.nima.2008.08.029",
    journal = "Nucl. Instrum. Meth. A",
    volume = "598",
    pages = "289--295",
    year = "2009"
}

@article{T2K:2011qtm,
    author = "Abe, K. and others",
    collaboration = "T2K",
    title = "{The T2K Experiment}",
    eprint = "1106.1238",
    archivePrefix = "arXiv",
    primaryClass = "physics.ins-det",
    doi = "10.1016/j.nima.2011.06.067",
    journal = "Nucl. Instrum. Meth. A",
    volume = "659",
    pages = "106--135",
    year = "2011"
}

@article{Nieves:2011pp,
    title        = {{Inclusive Charged--Current Neutrino--Nucleus Reactions}},
    author       = {Nieves, J. and Ruiz Simo, I. and Vicente Vacas, M. J.},
    year         = 2011,
    journal      = {Physical Review C},
    volume       = 83,
    pages        = {045501},
    doi          = {10.1103/PhysRevC.83.045501},
    eprint       = {1102.2777},
    archiveprefix = {arXiv},
    primaryclass = {hep-ph}
}

@article{Sobczyk:2017mts,
    title        = {{Intercomparison of lepton-nucleus scattering models in the quasielastic region}},
    author       = {Sobczyk, Joanna Ewa},
    year         = 2017,
    journal      = {Physical Review C},
    volume       = 96,
    number       = 4,
    pages        = {045501},
    doi          = {10.1103/PhysRevC.96.045501},
    eprint       = {1706.06739},
    archiveprefix = {arXiv},
    primaryclass = {nucl-th}
}

@article{Benhar:1999bg,
    title        = {{Two nucleon spectral function in infinite nuclear matter}},
    author       = {Benhar, Omar and Fabrocini, Adelchi},
    year         = 2000,
    journal      = {Physical Review C},
    volume       = {C62},
    pages        = {034304},
    doi          = {10.1103/PhysRevC.62.034304},
    eprint       = {nucl-th/9909014},
    archiveprefix = {arXiv},
    primaryclass = {nucl-th},
    slaccitation = {%%CITATION = NUCL-TH/9909014;%%}
}

@article{Avanzini:2021qlx,
    title        = {{Comparisons and challenges of modern neutrino-scattering experiments}},
    author       = {Avanzini, M. Buizza and others},
    year         = 2022,
    journal      = {Physical Review D},
    volume       = 105,
    number       = 9,
    pages        = {092004},
    doi          = {10.1103/PhysRevD.105.092004},
    eprint       = {2112.09194},
    archiveprefix = {arXiv},
    primaryclass = {hep-ex},
    reportnumber = {FERMILAB-PUB-21-299-ND-SCD}
}

@article{T2K:2011ypd,
    title        = {{Indication of Electron Neutrino Appearance from an Accelerator-produced Off-axis Muon Neutrino Beam}},
    author       = {Abe, K. and others},
    year         = 2011,
    journal      = {Physical Review Letters},
    volume       = 107,
    pages        = {041801},
    doi          = {10.1103/PhysRevLett.107.041801},
    collaboration = {T2K},
    eprint       = {1106.2822},
    archiveprefix = {arXiv},
    primaryclass = {hep-ex}
}

@article{T2K:2014ghj,
    title        = {{Precise Measurement of the Neutrino Mixing Parameter $\theta_{23}$ from Muon Neutrino Disappearance in an Off-Axis Beam}},
    author       = {Abe, K. and others},
    year         = 2014,
    journal      = {Physical Review Letters},
    volume       = 112,
    number       = 18,
    pages        = 181801,
    doi          = {10.1103/PhysRevLett.112.181801},
    collaboration = {T2K},
    eprint       = {1403.1532},
    archiveprefix = {arXiv},
    primaryclass = {hep-ex}
}

@article{PhysRevD.87.012001,
    title        = {T2{K} neutrino flux prediction},
    author       = {Abe, K. and others},
    year         = 2013,
    month        = {Jan},
    journal      = {Physical Review D},
    publisher    = {American Physical Society},
    volume       = 87,
    pages        = {012001},
    doi          = {10.1103/PhysRevD.87.012001},
    url          = {https://link.aps.org/doi/10.1103/PhysRevD.87.012001},
    collaboration = {T2K Collaboration},
    issue        = 1,
    numpages     = 34
}

@article{PhysRevD.108.112009,
    title        = {First measurement of muon neutrino charged-current interactions on hydrocarbon without pions in the final state using multiple detectors with correlated energy spectra at {T}2{K}},
    author       = {Abe, K. and others},
    year         = 2023,
    month        = {Feb},
    journal      = {Physical Review D},
    publisher    = {American Physical Society},
    volume       = 108,
    pages        = 112009,
    doi          = {10.1103/PhysRevD.108.112009},
    url          = {https://doi.org/10.1103/PhysRevD.108.112009},
    collaboration = {T2K Collaboration},
    issue        = 11,
    numpages     = 32
}

@article{10.1093/ptep/pts020,
    title        = {{Neutrino facility and neutrino physics in J-PARC}},
    author       = {Sekiguchi, Tetsuro},
    year         = 2012,
    month        = 10,
    journal      = {Progress of Theoretical and Experimental Physics},
    volume       = 2012,
    number       = 1,
    pages        = {02B005},
    doi          = {10.1093/ptep/pts020},
    issn         = {2050-3911},
    url          = {https://doi.org/10.1093/ptep/pts020},
    eprint       = {https://academic.oup.com/ptep/article-pdf/2012/1/02B005/11589211/pts020.pdf}
}

@article{10.1093/ptep/ptab014,
    title        = {Measurements of $\bar{\nu}_{\mu}$ and $\bar{\nu}_{\mu} + \nu_{\mu}$ charged-current cross-sections without detected pions nor protons on water and hydrocarbon at mean antineutrino energy of 0.86 {G}e{V}},
    author       = {Abe, K. and others},
    year         = 2021,
    month        = {03},
    journal      = {Progress of Theoretical and Experimental Physics},
    volume       = 2021,
    pages        = {},
    doi          = {10.1093/ptep/ptab014}
}

@inproceedings{BabyMIND:2020lxx,
    title        = {{Baby MIND detector first physics run}},
    author       = {Ajmi, A. and others},
    year         = 2020,
    month        = 4,
    booktitle    = {{Prospects in Neutrino Physics}},
    collaboration = {Baby MIND},
    eprint       = {2004.05245},
    archiveprefix = {arXiv},
    primaryclass = {physics.ins-det}
}

@article{BabyMIND:2017mys,
    title        = {{Baby MIND: A Magnetized Segmented Neutrino Detector for the WAGASCI Experiment}},
    author       = {Antonova, M. and others},
    year         = 2017,
    journal      = {Journal of Instrumentation},
    volume       = 12,
    number       = {07},
    pages        = {C07028},
    doi          = {10.1088/1748-0221/12/07/C07028},
    editor       = {Shekhtman, Lev},
    collaboration = {Baby MIND},
    eprint       = {1705.10406},
    archiveprefix = {arXiv},
    primaryclass = {physics.ins-det},
    reportnumber = {FERMILAB-CONF-17-218-APC}
}

@article{MINERVA_pion_production,
    title        = {{Cross section for $\nu_\mu$ and $\nu_\mu$ induced pion production on hydrocarbon in the few-GeV region using MINERvA}},
    author       = {C. L. McGivern and others},
    year         = 2016,
    journal      = {Physical Review D},
    volume       = 94,
    number       = {052005},
    pages        = {},
    doi          = {10.1088/1748-0221/12/07/C07028},
    editor       = {},
    collaboration = {MINERvA collaboration},
    eprint       = {},
    archiveprefix = {},
    primaryclass = {},
    reportnumber = {}
}

@article{MINOS_ND_CCQE,
    title        = {Study of quasielastic scattering using charged-current $\nu_\mu$-iron interactions in the {MINOS} near detector},
    author       = {P. Adamson and others},
    year         = 2015,
    month        = {03},
    journal      = {Physical Review D},
    volume       = 91,
    number       = {012005},
    pages        = {},
    doi          = {},
    collaboration = {MINOS collaboration}
}

@article{PhysRevD.91.072010,
    title        = {Measurements of neutrino oscillation in appearance and disappearance channels by the {T}2{K} experiment with $6.6\ifmmode\times\else\texttimes\fi{}1{0}^{20}$ protons on target},
    author       = {Abe, K. and others},
    year         = 2015,
    month        = {Apr},
    journal      = {Physical Review D},
    publisher    = {American Physical Society},
    volume       = 91,
    pages        = {072010},
    doi          = {10.1103/PhysRevD.91.072010},
    url          = {https://link.aps.org/doi/10.1103/PhysRevD.91.072010},
    collaboration = {T2K Collaboration},
    issue        = 7,
    numpages     = 50
}

@article{Abe2020-Nature,
    title        = {Constraint on the matter-antimatter symmetry-violating phase in neutrino oscillations},
    author       = {Abe, K. and others},
    year         = 2020,
    month        = {Apr},
    day          = {01},
    journal      = {Nature},
    volume       = 580,
    number       = 7803,
    pages        = {339--344},
    doi          = {10.1038/s41586-020-2177-0},
    issn         = {1476-4687},
    url          = {https://doi.org/10.1038/s41586-020-2177-0},
    collaboration = {T2K Collaboration}
}

@article{T2K:2023qjb,
    title        = {First measurement of muon neutrino charged-current interactions on hydrocarbon without pions in the final state using multiple detectors with correlated energy spectra at {T}2{K}},
    author       = {Abe, K. and others},
    year         = 2023,
    journal      = {Physical Review D},
    volume       = 108,
    number       = 11,
    pages        = 112009,
    doi          = {10.1103/PhysRevD.108.112009},
    collaboration = {T2K},
    eprint       = {2303.14228},
    archiveprefix = {arXiv},
    primaryclass = {hep-ex}
}

@article{PhysRevD.104.093001,
    title        = {Resonance axial-vector mass from experiments on neutrino-hydrogen and neutrino-deuterium scattering},
    author       = {Kakorin, Igor D. and Kuzmin, Konstantin S.},
    year         = 2021,
    month        = {Nov},
    journal      = {Physical Review D},
    publisher    = {American Physical Society},
    volume       = 104,
    pages        = {093001},
    doi          = {10.1103/PhysRevD.104.093001},
    url          = {https://link.aps.org/doi/10.1103/PhysRevD.104.093001},
    issue        = 9,
    numpages     = 16
}

@article{1975547,
    title        = {Neutrino reactions on nuclear targets, Nucl. Phys. B43 (1972) 605.},
    author       = {Smith, R.A. and Moniz, E.J.},
    year         = 1975,
    journal      = {Nuclear Physics B},
    volume       = 101,
    number       = 2,
    pages        = 547,
    doi          = {https://doi.org/10.1016/0550-3213(75)90612-4},
    issn         = {0550-3213},
    url          = {https://www.sciencedirect.com/science/article/pii/0550321375906124}
}

@article{Hayato:2002sd,
    title        = {{NEUT}},
    author       = {Hayato, Y.},
    year         = 2002,
    journal      = {Nuclear Physics B - Proceedings Supplements},
    volume       = 112,
    pages        = {171--176},
    doi          = {10.1016/S0920-5632(02)01759-0},
    editor       = {Morfin, J. G. and Sakuda, M. and Suzuki, Y.}
}

@article{REIN198179,
    title        = {Neutrino-excitation of baryon resonances and single pion production},
    author       = {Dieter Rein and Lalit M Sehgal},
    year         = 1981,
    journal      = {Annals of Physics},
    volume       = 133,
    number       = 1,
    pages        = {79--153},
    doi          = {https://doi.org/10.1016/0003-4916(81)90242-6},
    issn         = {0003-4916},
    url          = {https://www.sciencedirect.com/science/article/pii/0003491681902426}
}

@article{Rein:1987cb,
    title        = {{Angular Distribution in Neutrino Induced Single Pion Production Processes}},
    author       = {Rein, D.},
    year         = 1987,
    journal      = {Zeitschrift für Physik C},
    volume       = 35,
    pages        = {43--64},
    doi          = {10.1007/BF01561054}
}

@article{Graczyk:2009qm,
    title        = {{$C^{A}_5$ axial form factor from bubble chamber experiments}},
    author       = {Graczyk, K. M. and Kielczewska, D. and Przewlocki, P. and Sobczyk, J. T.},
    year         = 2009,
    journal      = {Physical Review D},
    volume       = 80,
    pages        = {093001},
    doi          = {10.1103/PhysRevD.80.093001},
    eprint       = {0908.2175},
    archiveprefix = {arXiv},
    primaryclass = {hep-ph}
}

@article{Kuzmin:2006dh,
    title        = {{Axial masses in quasielastic neutrino scattering and single-pion neutrino production on nucleons and nuclei}},
    author       = {Kuzmin, Konstantin S. and Lyubushkin, Vladimir V. and Naumov, Vadim A.},
    year         = 2006,
    journal      = {Acta Physica Polonica B},
    volume       = 37,
    pages        = {2337--2348},
    editor       = {Graczyk, K. M. and Sobczyk, J. T.},
    eprint       = {hep-ph/0606184},
    archiveprefix = {arXiv}
}

@article{RuizSimo:2016rtu,
    title        = {{Relativistic model of 2p-2h meson exchange currents in (anti)neutrino scattering}},
    author       = {Ruiz Simo, I. and Amaro, J. E. and Barbaro, M. B. and De Pace, A. and Caballero, J. A. and Donnelly, T. W.},
    year         = 2017,
    journal      = {Journal of Physics G},
    volume       = 44,
    number       = 6,
    pages        = {065105},
    doi          = {10.1088/1361-6471/aa6a06},
    eprint       = {1604.08423},
    archiveprefix = {arXiv},
    primaryclass = {nucl-th}
}

@article{PASCHALIS2020135110,
    title        = {Nucleon-nucleon correlations and the single-particle strength in atomic nuclei},
    author       = {S. Paschalis and M. Petri and A.O. Macchiavelli and O. Hen and E. Piasetzky},
    year         = 2020,
    journal      = {Physics Letters B},
    volume       = 800,
    pages        = 135110,
    doi          = {https://doi.org/10.1016/j.physletb.2019.135110},
    issn         = {0370-2693},
    url          = {https://www.sciencedirect.com/science/article/pii/S0370269319308329}
}

@article{RPA-Default,
    title        = {{Inclusive quasielastic charged-current neutrino-nucleus reactions}},
    author       = {Nieves,J. and Amaro, J. E. and Valverde, M.},
    year         = 2004,
    journal      = {Physical Review C},
    volume       = 70,
    pages        = {055503},
    editor       = {},
    eprint       = {hep-ph/0606184},
    archiveprefix = {arXiv}
}

@article{Llewellyn-Smith,
    title        = {{Neutrino reactions at accelerator energies}},
    author       = {C.H. Llewellyn Smith},
    year         = 1971,
    journal      = {Physics Reports},
    volume       = 3,
    pages        = 5,
    editor       = {},
    eprint       = {},
    archiveprefix = {}
}

@article{NEUT_update,
    title        = {{The NEUT neutrino interaction simulation program library}},
    author       = {Hayato, Y. and Pickering, L.},
    year         = 2021,
    journal      = {European Physical Journal Special Topics},
    volume       = 230,
    pages        = {4469--4481},
    editor       = {},
    eprint       = {},
    archiveprefix = {}
}

@article{GENIE,
    title        = {{The GENIE neutrino Monte Carlo generator}},
    author       = {Andreopoulos, C. and others},
    year         = 2010,
    journal      = {Nuclear Instruments and Methods in Physics Research},
    volume       = 614,
    pages        = {87--104},
    editor       = {},
    eprint       = {},
    archiveprefix = {}
}

@article{SuperKamiokande,
    title        = {{The JHF-Kamioka neutrino project}},
    author       = {Itow, Y. and others},
    year         = 2001,
    journal      = {arxiv:0106019v1},
    volume       = {},
    number       = {},
    pages        = {},
    editor       = {},
    eprint       = {},
    archiveprefix = {}
}

@article{HyperKamiokande,
    title        = {{Hyper-Kamiokande Design Report}},
    author       = {Abe, K. and others},
    journal      = {arXiv:1805.04163},
    year         = {2018},
    doi          = {https://doi.org/10.48550/ arXiv.1805.04163},
    collaboration = {The Hyper-Kamiokande Collaboration},
    archiveprefix = {hep-ex}
}

@article{DUNE,
    title        = {{report number FERMILAB-PUB-20-025-ND}},
    author       = {Abi, B. and others},
    year         = {2020},
    journal      = {arXiv:2002.03005},
    collaboration = {The DUNE Collaboration},
    archiveprefix = {hep-ex}
}

@article{PMNS,
    title        = {{Remarks on the Unified Model of Elementary Particles}},
    author       = {Maki, Z. and Nakagawa, M. and Sakata, S.},
    year         = 1962,
    journal      = {Progress of Theoretical Physics},
    volume       = 28,
    number       = 5,
    pages        = {870--880},
    doi          = {10.1016/0003-4916(81)90242-6},
    reportnumber = {PITHA-80-10}
}

@article{Nova,
    title        = {{The NOvA Experiment}},
    author       = {Habig, A.},
    year         = 2012,
    journal      = {Nuclear Physics B},
    volume       = {229-232},
    number       = 5,
    pages        = 460,
    doi          = {https://doi.org/10.1016/j.nuclphysbps.2012.09.097},
    collaboration = {the NOvA collaboration}
}

@article{MINUIT2,
    title        = {{MINUIT-a system for function minimization and analysis of the parameter errors and correlations}},
    author       = {James, F. and Roos, M.},
    year         = 1984,
    journal      = {Commputer Physics Communications},
    volume       = 10,
    number       = 6,
    pages        = {343--367},
    doi          = {10.1016/0010-4655(75)90039-9}
}

@article{PhysRevD.101.112004,
    title        = {Simultaneous measurement of the muon neutrino charged-current cross section on oxygen and carbon without pions in the final state at {T}2{K}},
    author       = {Abe, K. and others},
    year         = 2020,
    month        = {Jun},
    journal      = {Physical Review D},
    publisher    = {American Physical Society},
    volume       = 101,
    pages        = 112004,
    doi          = {10.1103/PhysRevD.101.112004},
    url          = {https://link.aps.org/doi/10.1103/PhysRevD.101.112004},
    collaboration = {T2K Collaboration},
    issue        = 11,
    numpages     = 32
}

@article{GENIE_FSI,
    title        = {{Final State Interactions in GENIE}},
    author       = {S.A. Dytman and A.S. Meyer},
    year         = 2011,
    journal      = {AIP Conference Proceedings},
    volume       = 1405,
    number       = {},
    pages        = {213--218},
    doi          = {}
}

@article{ABE2012211,
    title        = {Measurements of the {T}2{K} neutrino beam properties using the {INGRID} on-axis near detector},
    author       = {Abe, K. and others},
    year         = 2012,
    journal      = {Nuclear Instruments and Methods in Physics Research Section A: Accelerators, Spectrometers, Detectors and Associated Equipment},
    volume       = 694,
    pages        = {211--223},
    doi          = {https://doi.org/10.1016/j.nima.2012.03.023},
    issn         = {0168-9002},
    url          = {https://www.sciencedirect.com/science/article/pii/S0168900212002987}
}

@article{Abgrall:2019EPJC.79.2,
    title        = {{Measurements of $\pi^{\pm}, K^{\pm}$ and proton double differential yields from the surface of the T2K replica target for incoming 31 GeV/c protons with the NA61/SHINE spectrometer at the CERN SPS}},
    author       = {Abgrall, N. and others},
    year         = 2019,
    journal      = {European Physics Journal C},
    volume       = 79,
    number       = 2,
    pages        = 100,
    doi          = {10.1140/epjc/s10052-019-6583-0}
}

@article{Abgrall:2016EPJC.76.11,
    title        = {{Measurements of $\pi^{\pm}$ differential yields from the surface of the T2K replica target for incoming 31 GeV/c protons with the NA61/SHINE spectrometer at the CERN SPS}},
    author       = {Abgrall, N. and others},
    year         = 2016,
    journal      = {The European Physics Journal C},
    volume       = 76,
    number       = 11,
    pages        = 617,
    doi          = {10.1140/epjc/s10052-016-4440-y}
}

@article{Abgrall:2016EPJC.76.2,
    title        = {{Measurements of $\pi^{\pm}, K^{\pm}, K^{0}_{s}, \Lambda$ and proton production in proton–carbon interactions at 31 GeV/c with the NA61/SHINE spectrometer at the CERN SPS}},
    author       = {Abgrall, N. and others},
    year         = 2016,
    journal      = {The European Physics Journal C},
    volume       = 76,
    number       = 2,
    pages        = 84,
    doi          = {10.1140/epjc/s10052-016-3898-y}
}

@article{PhysRevD.102.074012,
    title        = {Parametrization and applications of the low-${Q}^{2}$ nucleon vector form factors},
    author       = {Borah, Kaushik and Hill, Richard J. and Lee, Gabriel and Tomalak, Oleksandr},
    year         = 2020,
    month        = {Oct},
    journal      = {Physical Review D},
    publisher    = {American Physical Society},
    volume       = 102,
    pages        = {074012},
    doi          = {10.1103/PhysRevD.102.074012},
    url          = {https://link.aps.org/doi/10.1103/PhysRevD.102.074012},
    issue        = 7,
    numpages     = 16
}

@article{NITTA2004147,
    title        = {The {K}2{K} {S}ci{B}ar detector},
    author       = {Nitta, K. and others},
    year         = 2004,
    journal      = {Nuclear Instruments and Methods in Physics Research Section A: Accelerators, Spectrometers, Detectors and Associated Equipment},
    volume       = 535,
    number       = 1,
    pages        = {147--151},
    doi          = {https://doi.org/10.1016/j.nima.2004.07.272},
    issn         = {0168-9002},
    url          = {https://www.sciencedirect.com/science/article/pii/S0168900204015918},
    note         = {Proceedings of the 10th International Vienna Conference on Instrumentation}
}

@article{AGOSTINELLI2003250,
    title = {Geant4—a simulation toolkit},
    journal = {Nuclear Instruments and Methods in Physics Research Section A: Accelerators, Spectrometers, Detectors and Associated Equipment},
    volume = {506},
    number = {3},
    pages = {250-303},
    year = {2003},
    issn = {0168-9002},
    doi = {https://doi.org/10.1016/S0168-9002(03)01368-8},
    url = {https://www.sciencedirect.com/science/article/pii/S0168900203013688},
    author = {S. Agostinelli and others}
}

@article{Canetti_2012,
   title={Matter and antimatter in the universe},
   volume={14},
   ISSN={1367-2630},
   url={http://dx.doi.org/10.1088/1367-2630/14/9/095012},
   DOI={10.1088/1367-2630/14/9/095012},
   number={9},
   journal={New Journal of Physics},
   publisher={IOP Publishing},
   author={Canetti, Laurent and Drewes, Marco and Shaposhnikov, Mikhail},
   year={2012},
   month=sep,
   pages={095012}
}

@article{PhysRevD.91.112002,
  title = {Measurement of the ${\nu}_{\mu}$ charged current quasielastic cross section on carbon with the {T}2{K} on-axis neutrino beam},
  author = {Abe, K. and others},
  collaboration = {The T2K Collaboration},
  journal = {Phys. Rev. D},
  volume = {91},
  issue = {11},
  pages = {112002},
  numpages = {17},
  year = {2015},
  month = {Jun},
  publisher = {American Physical Society},
  doi = {10.1103/PhysRevD.91.112002},
  url = {https://link.aps.org/doi/10.1103/PhysRevD.91.112002}
}

\section{Data Release}

A data release from the T2K experiment for the results found in this paper can be found here: https://doi.org/10.5281/zenodo.16949979. 

\end{document}